\documentclass[prx,aps,twocolumn,showpacs,amsmath,amssymb,superscriptaddress,floatfix]{revtex4-2}

\usepackage{graphicx}
\usepackage[export]{adjustbox}
\usepackage{dcolumn}
\usepackage{upgreek}
\usepackage{bm,dsfont}
\usepackage{amssymb,mathtools}
\usepackage{booktabs}
\usepackage{color}
\usepackage{microtype}
\usepackage[T1]{fontenc}
\usepackage{lmodern} 
\usepackage[shortlabels]{enumitem}
\usepackage{changes}
\usepackage{physics}

\allowdisplaybreaks
\usepackage{placeins}
\usepackage[colorlinks,plainpages=false,pdfpagemode=UseNone,pdfstartview=FitBH]{hyperref}
\hypersetup{
colorlinks = true,
linkcolor = [rgb]{0.87,0.18,0.22},
citecolor = [rgb]{0.0,0.60,0.32},
urlcolor = [rgb]{0.20, 0.30, 0.54}}

\usepackage{sidecap}

\renewcommand{\tablename}{Tab.}
\makeatletter\renewcommand{\fnum@table}[1]{\tablename~\thetable.}\makeatother

\DeclarePairedDelimiter\ceil{\lceil}{\rceil}

\newcommand{\ii}{\mathrm{i}}
\renewcommand{\Re}{\operatorname{Re}}
\renewcommand{\Im}{\operatorname{Im}}
\newcommand{\stfac}[1]{\langle\!\langle#1\rangle\!\rangle}

\newcommand{\h}[1]{H_\mathrm{#1}}
\newcommand{\sro}{Sr$_2$RuO$_4$}

\newcommand{\gb}{G_\mathrm{b}(z)}
\newcommand{\gi}{G(z)}
\newcommand{\sig}{\Sigma(z)}

\usepackage{tikz}
\usetikzlibrary{math,positioning}
\usetikzlibrary{external}
\usepackage{tikz-network}

\definecolor{seq1}{HTML}{E4F1F7}
\definecolor{seq2}{HTML}{C5E1EF}
\definecolor{seq3}{HTML}{9EC9E2}
\definecolor{seq4}{HTML}{6CB0D6}
\definecolor{seq5}{HTML}{3C93C2}
\definecolor{seq6}{HTML}{226E9C}

\definecolor{div1}{HTML}{009392}
\definecolor{div2}{HTML}{39B185}
\definecolor{div3}{HTML}{9CCB86}
\definecolor{div4}{HTML}{E9E29C}
\definecolor{div5}{HTML}{EEB479}
\definecolor{div6}{HTML}{E88471}
\definecolor{div7}{HTML}{CF597E}

\definecolor{qual1}{HTML}{FF1F5B}
\definecolor{qual2}{HTML}{00CD6C}
\definecolor{qual3}{HTML}{009ADE}
\definecolor{qual4}{HTML}{AF58BA}
\definecolor{qual5}{HTML}{FFC61E}
\definecolor{qual6}{HTML}{F28522}

\newcommand{\includeTikz}[2]{
	\includegraphics[scale=1, valign=c, raise=#1 pt]{tensors/#2}
}

\definecolor{citecolor}{rgb}{0.0,0.60,0.32}

\begin{document}

\title{Tree tensor-network real-time multiorbital impurity Solver: Spin-orbit coupling and correlation functions in Sr$_2$RuO$_4$}

\author{X.~Cao}
\affiliation{Institute for Theoretical Physics, Heidelberg University, Philosophenweg 19, 69120 Heidelberg, Germany}
\affiliation{Center for Computational Quantum Physics, Flatiron Institute, 162 Fifth Avenue, New York, New York 10010, USA}

\author{Y.~Lu}
\email{yilu@nju.edu.cn}
\affiliation{Institute for Theoretical Physics, Heidelberg University, Philosophenweg 19, 69120 Heidelberg, Germany}
\affiliation{National Laboratory of Solid State Microstructures and Department of Physics, Nanjing University, Nanjing 210093, China}
\affiliation{Collaborative Innovation Center of Advanced Microstructures, Nanjing University, Nanjing 210093, China}

\author{P.~Hansmann}
\affiliation{Department of Physics, University of Erlangen-Nuremberg, 91058 Erlangen, Germany}

\author{M.~W.~Haverkort}
\affiliation{Institute for Theoretical Physics, Heidelberg University, Philosophenweg 19, 69120 Heidelberg, Germany}

\date{\today}

\begin{abstract}
	We present a tree tensor-network impurity solver suited for general multiorbital systems. The network is constructed to efficiently capture the entanglement structure and symmetry of an impurity problem. The solver works directly on the real-time/frequency axis and generates spectral functions with energy-independent resolution of the order of one percent of the correlated bandwidth. Combined with an optimized representation of the impurity bath, it efficiently solves self-consistent dynamical mean-field equations and calculates various dynamical correlation functions for systems with off-diagonal Green's functions. For the archetypal correlated Hund's metal Sr$_2$RuO$_4$, we show that both the low-energy quasiparticle spectra related to the van Hove singularity and the high-energy atomic multiplet excitations can be faithfully resolved. In particular, we show that while the spin-orbit coupling has only minor effects on the orbital-diagonal one-particle spectral functions, it has a more profound impact on the low-energy spin and orbital response functions.

\end{abstract}

\maketitle

\section{Introduction}\label{sec:intro}

The field of strongly correlated materials constitutes an important part of condensed matter physics. The interplay of the electrons' spin, charge, and orbital degrees of freedom leads to rich phase diagrams and emergent phenomena beyond the grasp of a single-particle theory. At low temperatures, the quasielectron mass in these materials could become thousands of times heavier than the bare one~\cite{Stewart1984,Hewson1997}. Metals can turn into insulators~\cite{Imada1998} exhibiting various forms of magnetism, which may yet become superconducting upon doping~\cite{Lee2006,Keimer2015}. To predict or even understand the properties of these materials poses strong challenges due to their vast diversity of energy scales. Nevertheless, our understanding of these quantum many-body systems has progressed steadily, owing to the advancement of both theoretical and experimental tools. In particular, the invention of dynamical mean-field theory (DMFT)~\cite{Metzner1989,Georges1996} has been an important step that allows us to understand key aspects of correlations over the full interaction range from non-interacting particles to the atomic limit within a single framework. Due to significant advances in numerical techniques, its merger with \emph{ab initio} methods such as density functional theory (DFT+DMFT) is today an accurate quantitative method for material-realistic computations~\cite{Kotliar2004,Georges2013}. On the experimental side, the development of various spectroscopic techniques has provided valuable information on the correlated electronic structure over a wide energy range~\cite{Damascelli2003,deGroot2008}, complementing those available to low-energy measurements such as transport. Core-level spectroscopy~\cite{deGroot2008}, for example, has proved invaluable in determining the interaction parameters in transition metal oxides~\cite{Bocquet1992} that help to establish their now standard classification~\cite{Zaanen1985}. In addition, recent advances in the instrumentation of resonant x-ray scattering have enabled the measurements of diverse correlation functions with unprecedented momentum and energy resolution~\cite{Ament2011}, providing new insights into the complex ordering phenomena and dynamics in correlated materials.

Recent years have seen growing efforts to bring a closer marriage between the two fronts, which, alas, faces many hurdles. At the root of the most acute difficulties is the availability of robust algorithms, or rather the lack thereof, for treating the DMFT auxiliary impurity problems. Quantum Monte Carlo (QMC) solvers~\cite{Gull2011,Georges1992,Ulmke1995,Rubtsov2005,Werner2006,Werner2006b} are the most diffuse for solving multi-orbital models. However, the intrinsic fermionic sign problem of QMC severely hinders their application in many material-realistic scenarios or relevant temperature ranges. The problem is particularly severe for materials with low local symmetry (e.g., triclinic or monoclinic local point groups) or for those with nonnegligible spin-orbit coupling (SOC), where the local one-particle Green's function cannot be diagonalized simultaneously for all frequencies on a single basis set of the fermionic degrees of freedom. This leads to the need for solvers that can explicitly treat off-diagonal Green's functions. Similar needs also arise from real-space cluster extensions of DMFT (cellular DMFT)~\cite{Kotliar2001}. Another limitation of QMC solvers is that obtaining real-frequency spectra for direct comparison with experiments involves an ill-conditioned inversion problem prone to uncertainties due to the statistical noise~\cite{Jarrell1996,Lu2017}. Exact diagonalization (ED)~\cite{Caffarel1994,Sangiovanni2006,Capone2007,Koch2008,Zgid2012,Lin2013,Lu2014}, on the other hand, is capable of calculating various spectral functions directly on real axes regardless of the system symmetry~\cite{Haverkort2012,Haverkort2014,Hariki2017,Hariki2020}, although with limited energy resolution (coarse bath discretization) and/or number of orbitals due to its exponential scaling of computation cost in system size. Another real-axis method is the numerical renormalization group (NRG)~\cite{Wilson1975,Bulla2008,Bulla1999,Bulla2011,Bulla2005,Pruschke2000}, which has exponentially good energy resolution for low-energy spectra. However, its resolution at high energies is less than optimal due to its logarithmic bath discretization. In addition, its application to multiorbital systems is usually limited to models with high orbital and/or interaction symmetry~\cite{Stadler2015,Stadler2019,Kugler2020}. The real-space density-matrix renormalization group (DMRG)~\cite{White1992,White1993,Schollwock2005,Hallberg2006,Garcia2004,Raas2004,Karski2005} with its matrix-product-state (MPS) reincarnation~\cite{Schollwock2011,Wolf2014,Wolf2015,Ganahl2014,Ganahl2015} requires no specific bath discretization, and has arguably the best uniform resolution on the full frequency axis. Similar to the case of NRG, attempts to apply DMRG to multiorbital systems have so far only seen success in two-orbital model calculations~\cite{Ganahl2015,Holzner2010}, or materials with high symmetry~\cite{Bauernfeind2017}.

In this paper, we introduce a tree tensor-network (TTN)~\cite{Murg2010,Shi2006} based solver for solving general multiorbital impurity problems on the real-time/frequency axis. As a generalization of the linear network of MPS, the TTN has a more flexible spatial topology, which allows for efficient representation of a multiorbital system without mapping it onto a one-dimensional chain, a process that inevitably introduces artificial long-range entanglement that is detrimental to DMRG methods. We describe the basic properties of the proposed TTN states and outline how they are applied to solving multiorbital impurity problems in Sec.~\ref{sec:solver}. Some technical aspects pertaining to this TTN are provided in Appendix~\ref{app:ttps}. After presenting the main ideas of the TTN, we discuss another important aspect of the real-axis DMFT calculations in Sec.~\ref{sec:dmft}, describing a numerically stable loop construction by expressing Green's functions in different matrix forms. Details of the transformations and arithmetics of the Green's functions are gathered in Appendix~\ref{app:gf} for further reference. In Sec.~\ref{sec:geometry}, we apply the TTN solver to a standard three-orbital impurity model and conduct a systematic evaluation of its numerical performance. In particular, the effects of different bath geometries (Fig.~\ref{fig:geo}) are closely examined. Following this model study, Sec.~\ref{sec:sro} presents the results obtained on a realistic correlated material, \sro, which is among the most studied quantum materials~\cite{Mackenzie2003,Mackenzie2017} and is the subject of many recent theoretical works~\cite{Haverkort2008,Mravlje2011,Mravlje2016,Zhang2016,Kim2018,Tamai2019,Linden2020,Kugler2020,Lee2020}. We first benchmark our calculations against the available results from previous studies, then proceed to present and discuss new results, especially those obtained with SOC, which poses a severely limiting difficulty for QMC solvers and have so far remained completely elusive for real-axis methods.

\section{Tree Tensor-Network Impurity Solver}\label{sec:solver}

\begin{figure}[tb]
  \includegraphics{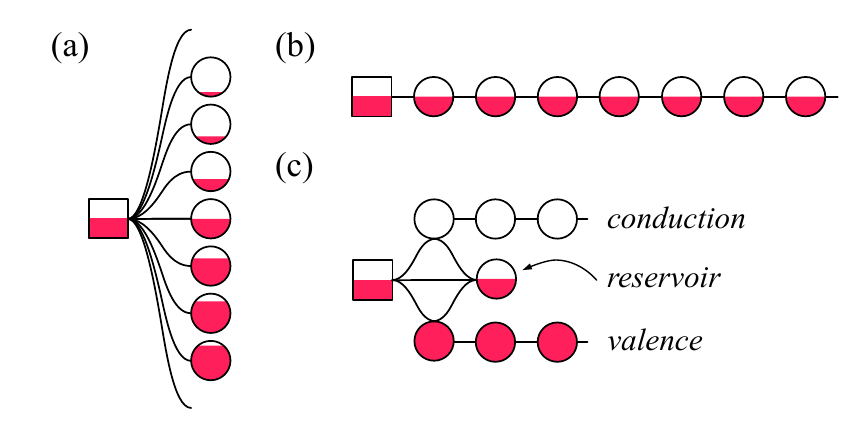}
  \caption{\label{fig:geo} Graphical representation of an impurity model in the (a) star, (b) chain, or (c) natural-orbital geometry. The interacting impurity site is represented by a square and the bath is represented by circles. Each site generally has fraction occupation, depicted by partially filled symbols. The hybridization between two sites is represented by solid lines. For later convenience, we term the two chains of bath sites in (c) as ``conduction'' and ``valence'', which are almost empty and filled, respectively. The ``reservoir'' site complements the occupation of the impurity site~\cite{Lu2014,Lu2019}.}
\end{figure}

A typical Anderson impurity Hamiltonian $\h{imp}$ reads

\begin{equation}\label{eq:imp}
  \begin{split}
    \h{imp} & = \h{loc} + \h{bath}, \\
    \h{loc} & = \sum_{\{\tau\}} \epsilon_{\tau_1\tau_2} c^\dag_{\tau_1} c_{\tau_2} + \sum_{\{\tau\}} U_{\tau_1\tau_2\tau_3\tau_4} c^\dag_{\tau_1} c^\dag_{\tau_2} c_{\tau_4} c_{\tau_3},\\
    \h{bath} & = \sum_{\kappa_1,\kappa_2} \epsilon_{\kappa_1\kappa_2} c^\dag_{\kappa_1} c_{\kappa_2} + \sum_{\tau,\kappa} V_{\tau\kappa} c^\dag_\tau c_\kappa + \mathrm{H.c.},
  \end{split}
\end{equation}
which comprises a locally interacting impurity described by $\h{loc}$ and its coupling to a non-interacting bath $\h{bath}$. $\tau$ ($\kappa$) is a collection of quantum numbers denoting the impurity (bath) degrees of freedom of the fermionic operators $c^{(\dag)}$. In addition to the Coulomb interaction, the impurity site typically hosts various one-body effects such as  low-symmetry local crystal field and SOC. The bath part of the Hamiltonian $\h{bath}$ here is given in a discretized form, whose parameters $\epsilon_{\kappa \kappa'}$ and $V_{\tau\kappa}$ are determined by the bath geometry as well as the specific discretization scheme. The graphical representation of the single-orbital version of Hamiltonian \eqref{eq:imp} is given in Fig.~\ref{fig:geo}, which depicts three possible bath geometries that can be transformed between one another via unitary transformations. While the star and chain geometries are commonly seen throughout the literature, the natural-orbital (NO) one has been explored only recently~\cite{Lu2014,Lu2019}. We note that although mathematically equivalent, different geometries lead to very different entanglement properties and, consequently, markedly different computation cost~\cite{Wolf2014,Lu2019,Kohn2021}. A conscious choice of the bath geometry, therefore, may prove pivotal for solving complex multiorbital impurity problems. A detailed discussion of the different geometries will be presented in Sec.~\ref{sec:geometry}.

\subsection{Tree Tensor-Network Representation of the Impurity Model}\label{sec:solver:ttn}
Consider a discrete many-body Hamiltonian such as the impurity model~\eqref{eq:imp} consisting of a large number $N$ of sites, each associated with a local Hilbert space spanned by a canonical basis set $\mathcal{H}_i=\mathrm{span} \{|\sigma_i\rangle\}$ of dimension $d_i$. Although not necessary, we assume the same $d$ for all sites for ease of notation. For a typical Hubbard-type site in the impurity model, for example, $d=4$ with $\sigma_i\in\{0,\uparrow,\downarrow,\uparrow\downarrow \}$. A corresponding general many-body wave function is then a state vector
\begin{equation}\label{eq:mps1}
	|\psi\rangle = \sum_{\sigma_1\cdots\sigma_N} c^{\sigma_1\cdots\sigma_N} |\sigma_1\cdots\sigma_N\rangle \equiv \sum_{\sigma_1\cdots\sigma_N} c^{\vec{\sigma}} |\vec{\sigma}\rangle
\end{equation}
in the exponentially large $d^N$-dimensional Hilbert space $\mathcal{H}=\otimes_i\mathcal{H}_i$. The coefficients $c^{\vec{\sigma}}$ can be viewed as a rank-$N$ tensor, which can be factorized into a set of smaller ones via a series of singular value decomposition (SVD)~\cite{Schollwock2011,Orus2014}, such that
\begin{equation}\label{eq:mps2}
	c^{\vec{\sigma}} = \sum_{e_1 \cdots e_E}M^{\sigma_{v_1}}_{l_{v_1}}M^{\sigma_{v_2}}_{l_{v_2}} \cdots M^{\sigma_{v_V}}_{l_{v_V}},
\end{equation}
where the tensors $M$ can be viewed as a set of $V$ vertices/nodes connected by a set of $E$ edges/bonds in a finite graph. Note that in general $V \geq N$, where the equality holds only if each tensor is associated with a physical site. For an auxiliary tensor not representing a physical site, its physical index $\sigma_{v_i}$ can be mapped to a dummy vacuum state or simply removed. The index $l_{v_i}$ collects all edges $\{e_j\}$ linked to the vertex $v_i$. For brevity we shall suppress the summation over these auxiliary edge indices hereafter when no confusion arises. Similarly, the Hamiltonian
\begin{equation}\label{eq:mpo1}
		H = \sum_{\substack{\sigma_1\cdots\sigma_N\\
		\sigma'_1\cdots\sigma'_N}} O_{\vec{\sigma}}^{\vec{\sigma}'} \dyad*{\vec{\sigma}'}{\vec{\sigma}}
\end{equation}
can also be expressed as a tensor network with the same geometry of the states as
\begin{equation}\label{eq:mpo2}
	O_{\vec{\sigma}}^{\vec{\sigma}'} = W^{\sigma'_{v_1}}_{\sigma_{v_1}}W^{\sigma'_{v_2}}_{\sigma_{v_2}} \cdots W^{\sigma'_{v_V}}_{\sigma_{v_V}}.
\end{equation}
One prominent example of the tensor network is the MPS (or tensor train~\cite{Oseledets2011}), a linear tensor network that enjoys tremendous success in modeling low-dimensional correlated systems, especially in the context of DMRG~\cite{Schollwock2011}. In this case, one finds an approximate compressed rewriting of~\eqref{eq:mps1} in the form of \eqref{eq:mps2} with the dimension of each bond limited below an upper bound $m$ by keeping only the largest singular values in the SVD. The standard quantum mechanical operations are carried out using tensor arithmetics, and the ground state can then be found variationally with the tensors playing the role of variational parameters. The single-orbital impurity models depicted in Fig.~\ref{fig:geo} can be mapped into the exact same linear geometry as the MPS, and efficiently solved using DMRG~\cite{Wolf2014,Wolf2015,Ganahl2014}. For multiorbital models, however, it requires braiding multiple chains into a single one, which inevitably introduces long-range interactions and entanglement that greatly increase the computation complexity. The MPS-DMRG solutions on the real-axis are therefore mostly limited to no more than two orbitals~\cite{Ganahl2015,Holzner2010}. Recently, the fork tensor-product state was proposed for solving multiorbital impurity problems~\cite{Holzner2010,Bauernfeind2017}. There, the bath degrees of freedom belonging to different orbitals are spatially separated into different chains, where only the tensors associated to the impurity sites at the end of each chain are linearly connected. Such a network effectively avoids the artificial long-range interaction that is detrimental to MPS when dealing with multiorbital systems, thus overcoming the previous limitations in orbital numbers~\cite{Bauernfeind2017}. It is however not yet clear how it performs for systems with orbital-off-diagonal interactions.

\begin{figure}
	\includeTikz{0}{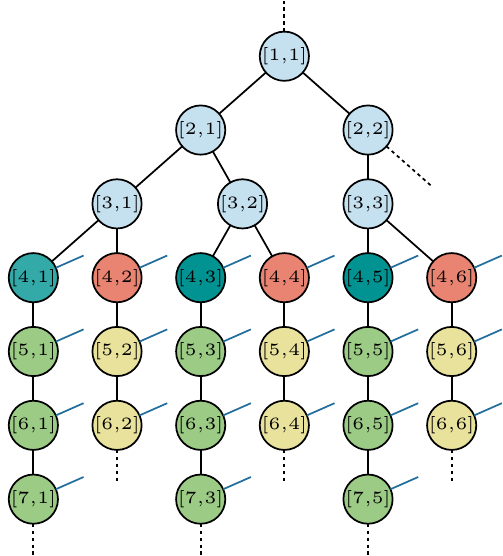}
	\caption{Sketch of the proposed TTN. A tensor on the $j-$th branch in the $i-$th layer (starting from the root as the first layer) is labeled as $[i,\,j]$. The tensors in the first three layers here are auxiliary tensors without physical indices (blue lines). The dummy bonds are represented by dashed lines. For star and chain geometries, this tree represents a six-orbital model.
	The impurity tensors are placed in the fourth layer, and the corresponding bath tensors are placed on each following branch. For the natural-orbital geometry, the fourth-layer tensors in the odd (even) branch represents the impurity (reservoir) orbital, followed by the conduction (valence) bath chains in later layers. This tree then represents a three-orbital model.}
	\label{fig:ttps}
\end{figure}

The fork tensor-product state belongs to a more general class of TTN states~\cite{Murg2010,Shi2006}. In this section, we propose another tree structure as shown in Fig.~\ref{fig:ttps}. A set of auxiliary tensors is structured as a binary tree of depth $\Delta=\ceil*{\log_2 N_o}$ for star and chain geometries, where $N_o$ is the number of orbitals. The natural-orbital geometry requires one more layer due to its double-chain structure. The maximal distance between a pair of impurity orbitals $2\Delta$ thus scales logarithmically with $N_o$, as opposed to linear scaling of the fork structure. In addition, the tree structure is highly symmetric, with each orbital equally distanced from the root. This closely resembles the symmetry of the impurity problem, which usually has certain rotational symmetries depending on its underlying point group. We thus expect such a tree geometry to efficiently capture the entanglement structure of multiorbital problems.

The acyclic structure of the tree means that cutting any bond between two tensors results in a bipartition of the network, whereby subsequent Schmidt decomposition and compression can be performed. Therefore, state-of-the-art DMRG ground-state search and various time-evolution algorithms developed for MPS~\cite{Schollwock2011,Paeckel2019} can be straightforwardly implemented for a general TTN state~\cite{Lubich2013,Schroeder2017,Bauernfeind2020}. We hence omit most of the technical details here, and only outline some aspects specific to the proposed TTN structure of Fig.~\ref{fig:ttps} in Appendix~\ref{app:ttps}.

\subsection{Green's Functions}\label{sec:solver:gf}

The key objective of an impurity solver is to compute the retarded local Green's function on the impurity site, which in the time domain reads
\begin{equation}\label{eq:gretard}
  \begin{split}
    G_{\tau_i\tau_j}(t) & = -\ii \uptheta(t) \ev*{\comm*{c_{\tau_i}(t)}{c_{\tau_j}^\dag(0)}\!_+}_0 \\
    & = \uptheta(t) [G_{\tau_i\tau_j}^>(t)-G_{\tau_i\tau_j}^<(t)],
  \end{split}
\end{equation}
where the greater and lesser Green's functions are given as
\begin{equation}\label{eq:ggl}
  \begin{split}
    G_{\tau_i\tau_j}^>(t) & = -\ii \ev*{c_{\tau_i}(t) c_{\tau_j}^\dag(0)}_0 = -\ii \ev*{c_{\tau_i} e^{-\ii (\h{imp}-E_0)t} c_{\tau_j}^\dag}_0, \\
    G_{\tau_i\tau_j}^<(t) & = \ii \ev*{c_{\tau_j}^\dag(0) c_{\tau_i}(t)}_0 = \ii \ev*{c_{\tau_j}^\dag e^{\ii (\h{imp}-E_0)t} c_{\tau_i}}_0.
  \end{split}
\end{equation}
Here, $\ev*{\cdot}_0$ denotes the expectation value in the ground state and $E_0$ is the ground state energy. Time-translation invariance is assumed here, as we are concerned with the equilibrium case in this paper. Its Fourier-transformed counterpart $G_{\tau_i\tau_j}(z=\omega+\ii 0^+) = \int \dd t e^{\ii z t} G_{\tau_i\tau_j}(t)$ is analytic in the upper half-plane.

In the context of tensor networks, there are several methods available for directly obtaining real-frequency spectral functions~\cite{Dargel2012,Holzner2011,Ganahl2014,Nocera2016}. In this paper, we opt for first computing $G^{\gtrless}(t)$ in Eq.~\eqref{eq:ggl} and subsequently Fourier transforming them into the frequency domain. When combined with linear prediction for obtaining time-series data in the long-time limit~\cite{Barthel2009}, such an approach has proven to be capable of obtaining spectral functions with high resolution. For this purpose, we choose the time-dependent variational principle (TDVP)~\cite{Haegeman2011,Haegeman2016,Lubich2013,Bauernfeind2020} to time evolve the TTN state. TDVP has the advantage over other algorithms such as time evolving block decimation, as it can deal with more complicated Hamiltonians that include long-range interactions, which are commonly encountered in complex multiorbital impurity models. In addition, a recent benchmark study~\cite{Paeckel2019} shows that TDVP excels in terms of both accuracy and efficiency among all the popular time-evolution algorithms for tensor-network states.

\section{Multiorbital Dynamical Mean-Field Theory}\label{sec:dmft}

Once equipped with an appropriate impurity solver, the remaining task, for our purpose here, is to construct DMFT loops that are numerically stable and minimize the error propagation between consecutive iterations. Real-axis methods often rely on numerical expression for Green's functions that involves broadening of a finite set of poles, and naive arithmetic operations on these approximate objects may contribute to additional inaccuracies of the spectra and self-energies (see discussions in, e.g., Ref.~\cite{Bulla1998}). In this section, we detail a DMFT loop construction based on the transformation between different matrix representations of Green's functions as described in Appendix~\ref{app:gf}. We emphasize here that such a loop construction and the arithmetic operations of the Green's functions within require no broadening of the poles. This guarantees that the spectral-weight sum rule is obeyed, thus eliminating the necessity for renormalization of spectral weights between iterations.

\begin{enumerate}[(1),wide, nosep]
  \item A single DMFT iteration starts with a given bath Green's function (Weiss function) $G_b(z)$, usually obtained from the previous iteration, in the form
  \begin{equation}\label{eq:dmft01}
    \gb = (z-\alpha^b_{0}-\sum_{i=1}^{N_b} {\beta^b_i}^\dag \frac{1}{z-\alpha^b_{i}} {\beta^b_i})^{-1},
  \end{equation}
  where the $d\!\times\!d$ matrices $\alpha^b_i$ and $\beta^b_i$ ($i\geq 1$) specify the onsite energies of the bath sites and their hybridization with the impurity with onsite energy $\alpha^b_0$. The number of bath sites $N_b$ usually needs to be kept $\gtrsim \mathcal{O}(10^2)$ to achieve good energy resolution. The frequency variable $z=\omega+\ii 0^+$ corresponds to the retarded Green's function. The above equation is equivalently expressed using the hybridization function
  \begin{equation}\label{eq:dmft01b}
    \Delta(z) = \mu + \alpha^b_{0} + \sum_{i=1}^{N_b} {\beta^b_i}^\dag \frac{1}{z-\alpha^b_{i}} {\beta^b_i},
  \end{equation}
  where $\mu$ is the chemical potential determined from the previous iteration. When no $G_b(z)$ is known, one can also start with the non-interacting lattice Green's function $G_0(z)$ given by band-structure calculations, which is usually given as a list of $N_0$ block poles specified by their energies $\epsilon^b_{i}$ and weights $w^b_{i}$,
  \begin{equation}\label{eq:dmft02}
    \gb = \sum_{i=1}^{N_0} \frac{w^b_{i}}{z-\epsilon^b_{i}}.
  \end{equation}
  It can be conveniently transformed into the form of Eq.~\eqref{eq:dmft01} as described in Appendix~\ref{app:gf}.

  \item With the determined bath parameters, one can construct the impurity Hamiltonian \eqref{eq:imp} and solve it using the TTN solver proposed in the previous section. The real-frequency Green's function $\gi$ is then obtained by calculating the real-time impurity Green's function using time-evolution algorithms such as TDVP, which is transformed into the form
  \begin{equation}\label{eq:dmft03}
    \gi = (z-\alpha_0-\sum_{i=1}^{N_p} {\beta_i}^\dag \frac{1}{z-\alpha_i} \beta_i)^{-1}.
  \end{equation}

  \item The impurity self-energy $\sig$ is then computed as
  \begin{equation}\label{eq:dmft04}
    \begin{split}
      \sig & = \gb^{-1} - \gi^{-1} \\
           & = (\alpha_0-\alpha^b_0) + \sum_{i=1}^{N_p} {\beta_i}^\dag \frac{1}{z-\alpha_i} \beta_i - \sum_{i=1}^{N_b} {\beta^b_i}^\dag \frac{1}{z-\alpha^b_i} \beta^b_i\\
           & = (\alpha_0-\alpha^b_0) + \sum_{i=1}^{dN_p} \frac{\tilde w_{i}}{z- \tilde \epsilon_{i}} - \sum_{i=1}^{dN_b} \frac{\tilde w^b_{i}}{z-\tilde \epsilon^b_{i}} \\
           & = \alpha^s_0 + \sum_{i=1}^{N_s} \frac{w^s_{i}}{z-\epsilon^s_{i}},
    \end{split}
  \end{equation}
  where in the last two lines, two lists of poles are obtained by eigendecomposition of matrices $\alpha_i$ and $\alpha^b_i$, which are subsequently combined into a new one. In the current implementation, the poles are merged such that their combined weight and first moment are conserved.

  \item Under the DMFT approximation whereby the lattice self-energy $\Sigma_{\mathrm{lat}}(\vb{k},z)$ is identified as $\sig$, the local Green's function is calculated as
	\begin{equation}\label{eq:dmft05}
		\begin{split}
			G_\mathrm{loc}(z) & = \int_\mathrm{B.Z.} \frac{\dd \vb{k}}{(2\pi)^3} \frac{1}{z + \mu - \epsilon_{\vb{k}} - \sig} \\
			& \equiv G_0(z+\mu-\sig),
		\end{split}
  \end{equation}
  where the chemical potential $\mu$ is updated to give the correct filling of $G_\mathrm{loc}(z)$. The integrated non-interacting Green's function $G_0(z)=1/(2\pi)^3\int_\mathrm{B.Z.} \dd \vb{k} (z - \epsilon_{\vb{k}})^{-1}$ reads
  \begin{equation}\label{eq:dmft06}
    G_0(z) = (z-\alpha^0_0-\sum_{i=1}^{N_0} {\beta^0_i}^\dag \frac{1}{z-\alpha^0_i} \beta^0_i)^{-1},
  \end{equation}
  where we have set its chemical potential to zero.
  For a given $\mu$, we then obtain
  \begin{equation}\label{eq:dmft07}
    \begin{split}
      G_\mathrm{loc}(z) = &\left(z+\mu-\alpha^s_0-\alpha^0_0-\sum_{i=1}^{N_s} \frac{w^s_{i}}{z-\epsilon^s_{i}}\right.\\
      & \left.-\sum_{i=1}^{N_0} {\beta^0_i}^\dag \frac{1}{z+\mu-\alpha^s_0-\alpha^0_0-\sum_{i=1}^{N_s} \frac{w^s_{i}}{z-\epsilon^s_{i}}} \beta^0_i \right)^{-1}.
    \end{split}
  \end{equation}
  Each summand in the second summation in Eq.~\eqref{eq:dmft07} has the form of a Green's function and can be transformed into a list of poles. Eventually, we obtain
  \begin{equation}\label{eq:dmft08}
    G_\mathrm{loc}(z) = (z-\alpha^l_0-\sum_{i=1}^{N_l} {\beta^l_i}^\dag \frac{1}{z-\alpha^l_i} \beta^l_i)^{-1}
  \end{equation}
  by combining all the lists of poles, where the matrices $\alpha^l_i$ and $\beta^l_i$ implicitly depend on $\mu$.

  \item To complete the DMFT iteration, a new bath Green's function is computed as
  \begin{equation}
    \begin{split}
      G_b^\prime(z) & = (G^{-1}_\mathrm{loc}(z) + \sig)^{-1} \\
      & = (z-\alpha^l_0 + \alpha^s_0 -\sum_{i=1}^{N_l} {\beta^l_i}^\dag \frac{1}{z-\alpha^l_i} \beta^l_i + \sum_{i=1}^{N_s} \frac{w^s_{i}}{z-\epsilon^s_{i}})^{-1} \\
      & = (z-{\alpha^{b'}_0} - \sum_{i=1}^{{N_L}} \frac{\tilde w^{b'}_{i}}{z-\tilde \epsilon^{b'}_{i}})^{-1}.
    \end{split}
  \end{equation}
  Note that $\gb$ here is usually kept quasi-continuous with large $N_L \gtrsim \mathcal{O}(10^3)$. Before given as input for the next iteration as in Eq.~\eqref{eq:dmft01}, a further discretization needs to be performed, which may follow different schemes. Similar to other real-axis methods, choices include merging poles between logarithmically or linearly spaced frequency nodes, or such that the resultant weight of each pole ($\Tr w^b$) is equal (``equal-weight'' discretization)~\cite{deVega2015}.
\end{enumerate}

The above steps are repeated until the Green's functions and self-energies are converged.

\section{Three-orbital model with different bath geometries}\label{sec:geometry}

In this section, we test the efficiency of the proposed TTN solver using a three-orbital impurity model, and in particular, benchmark its performance when the impurity is coupled to the same bath represented by the different geometries in Fig.~\ref{fig:geo}. The purpose of such an experiment is two fold. One is to obtain a systematic overview of the effects of bath representations on the computation complexity, which, while being critical for solving complex multiorbital problems, has so far been scarcely studied only in the context of a single-orbital model~\cite{Wolf2014,Kohn2021}. The other is to serve as a guidance for optimizing the bath geometry in the next section, where the closely related exemplary material \sro{} is studied.

\begin{figure}[tb]
  \includegraphics{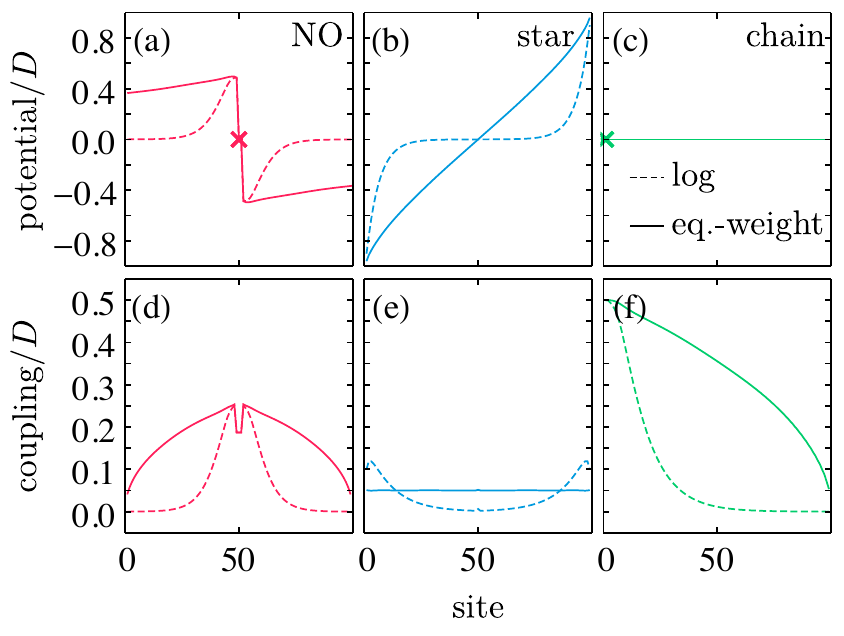}
  \caption{\label{fig:bath_para} (a)--(c) The on-site potential of each site and (d)-(f) its coupling to the impurity (for star geometry) or the next site (for NO and chain geometries) for $N_b=99$. For the NO geometry, the sites are ordered from the end of the conduction chain to the end of the valence chain via the impurity and reservoir sites (see Fig.~\ref{fig:geo}). The location of the impurity site is marked by a cross in (a) and (c) for the NO and chain geometries. For the NO geometry, the sites on the left (right) of the impurity correspond to the conduction (valence) chain. The logarithmic discretization is performed with the ``NRG discretization parameter'' $\Lambda=1.2$~\cite{Bulla2008}.}
\end{figure}

The model we studied is $O(3)_{orbital} \times SU(2)_{spin}$ symmetric and has only intraorbital hoppings. Each impurity orbital is separately coupled to the bath with an identical semielliptic spectral function
\begin{equation}
  -\frac{1}{\pi}\Im \Delta(\omega+\ii 0^+) = \frac{2}{\pi D}\sqrt{1-(\frac{\omega}{D})^2},
\end{equation}
where $D$ is the half bandwidth.
The Coulomb interaction part takes the rotationally invariant Kanamori form for a $t_{2g}$ system~\cite{Georges2013}
\begin{equation}\label{eq:kanamori}
  H_\mathrm{K} = \frac{1}{2}(U-3J) N(N-1)+\frac{5}{2}J N-2J\mathbf{S}^2 - \frac{1}{2}J\mathbf{L}^2,
\end{equation}
where $\mathbf{L}$ and $\mathbf{S}$ are the total orbital and spin momentum operators, respectively. Throughout this section, the impurity occupation is fixed at half filling with total number $\ev*{N}=3.0$. We compare the ground-state properties as well as time-evolution performances with different $U$ values and fixed $J=U/6$. For completeness, we also test different discretization schemes for the bath, using the same setting with $N_b=99$ bath sites per orbital. The bath parameters for different geometries and discretization schemes are shown in Fig.~\ref{fig:bath_para}. Note that the linear discretization has qualitatively the same performance as the equal-weight one, and its data are omitted hereafter for clarity. All calculations in this section are performed under the global $U(1)_\mathrm{charge}$ and $U(1)_{S_z}$ symmetries.

\subsection{Ground state}

\begin{figure}[tb]
  \includegraphics{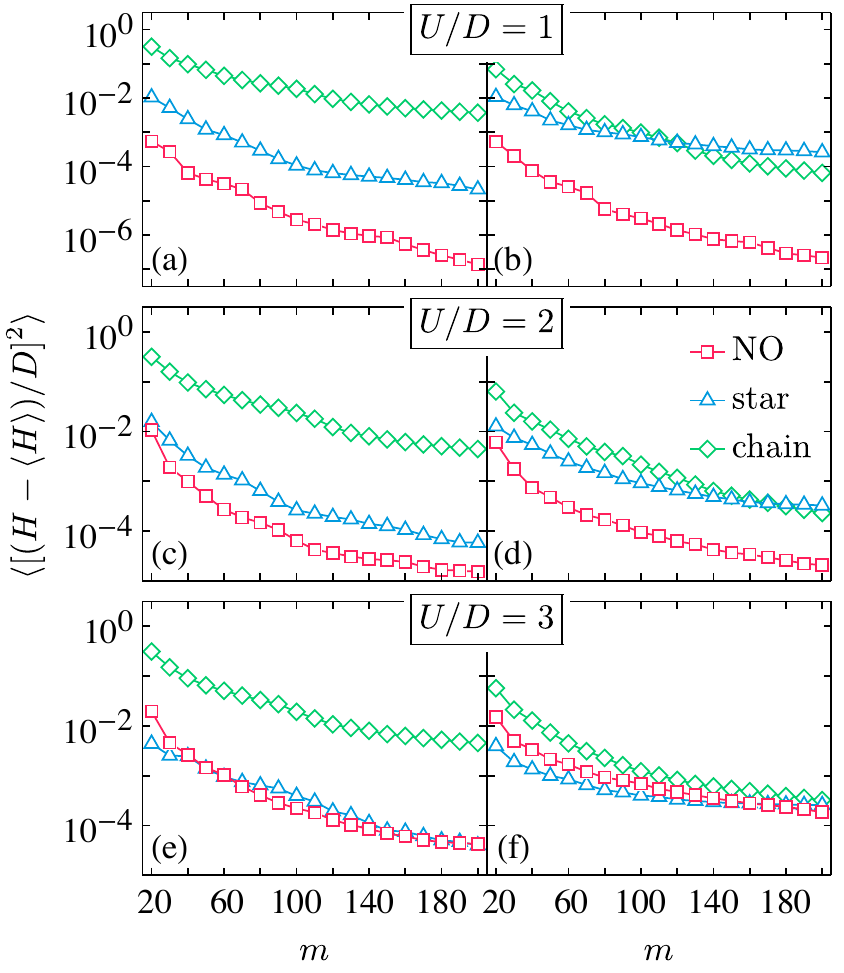}
  \caption{\label{fig:gs3orb} Ground-state variance $\ev*{[H-\ev*{H})/D]^2}$ of the three-orbital model with $N_b=99$ as a function of maximal bond dimension $m$ for (a),(b) $U/D=1$, (c),(d)$=2$, and (e),(f)$=3$. The left (right) column shows results for the bath with equal-weight (logarithmic) discretization.}
\end{figure}

Figure~\ref{fig:gs3orb} shows the ground-state energy variance of the three-orbital model for different bath geometries with varying interaction strength. Compared to the star and chain geometries, the NO geometry has overall better convergence properties within the tested interaction range, regardless of the bath discretization scheme. For $U/D\leq2$ in particular, its ground-state variance is up to more than four orders of magnitude lower than the others for a given $m$, and it achieves a comparable variance with less than half the bond dimension required by the star or chain geometry. In addition, one observes that the NO geometry has a relatively steady performance under different discretization schemes, while that of the other two geometries varies. Among the star and chain geometries, we find that the star geometry has a better performance when all energy scales are evenly represented using the equal-weight (or linear) discretization, in agreement with previous findings~\cite{Wolf2014}. However, its advantage shrinks for discretization schemes that emphasize the low-energy part. For the logarithmically discretized bath tested here, it shows comparable or even worse performance than the chain for large $m$ values.

\begin{figure}[tb]
  \includegraphics[width=\linewidth]{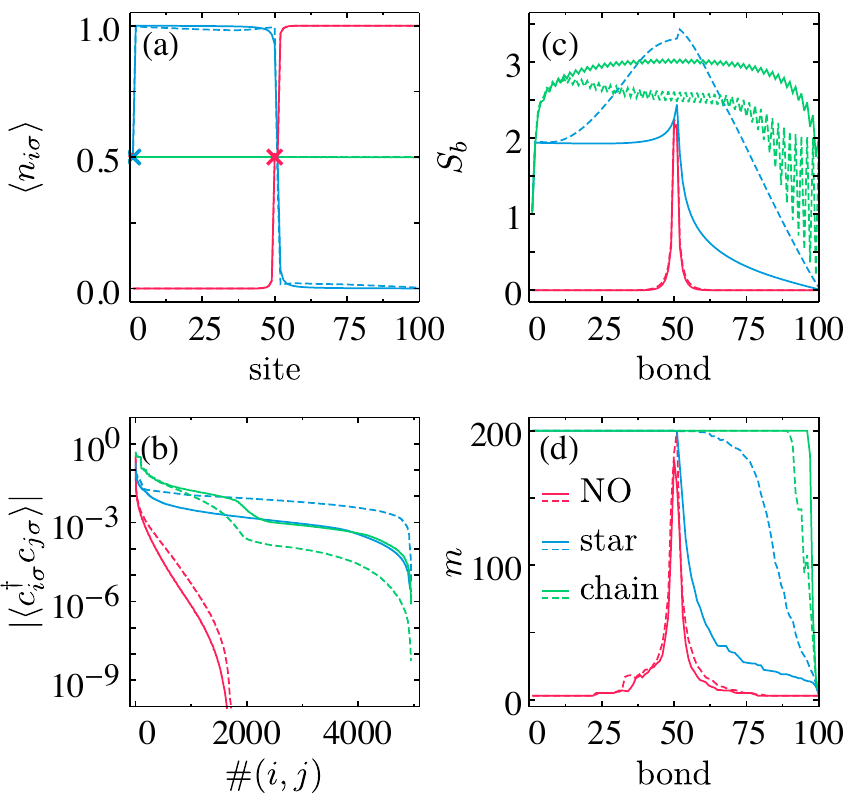}
  \caption{\label{fig:gs3orb_stat} (a) The diagonal and (b) off-diagonal ($i<j$) parts of the density matrix $\ev*{c^\dag_{i\sigma}c_{j\sigma}}$ (spin $\sigma=\uparrow\!/\!\downarrow$) of the ground state for the three-orbital model with $U/D=2$. The location of the impurity site in each geometry is marked by a cross in (a). (c) Bond entanglement entropy $S_b$ and (d) bond dimension distribution of the ground-state wave function with $m=200$ and truncation weight $w_t=10^{-12}$. The solid (dashed) lines correspond to equal-weight (logarithmic) discretization in all panels.}
\end{figure}

\begin{figure*}[t]
  \includegraphics[width=\textwidth]{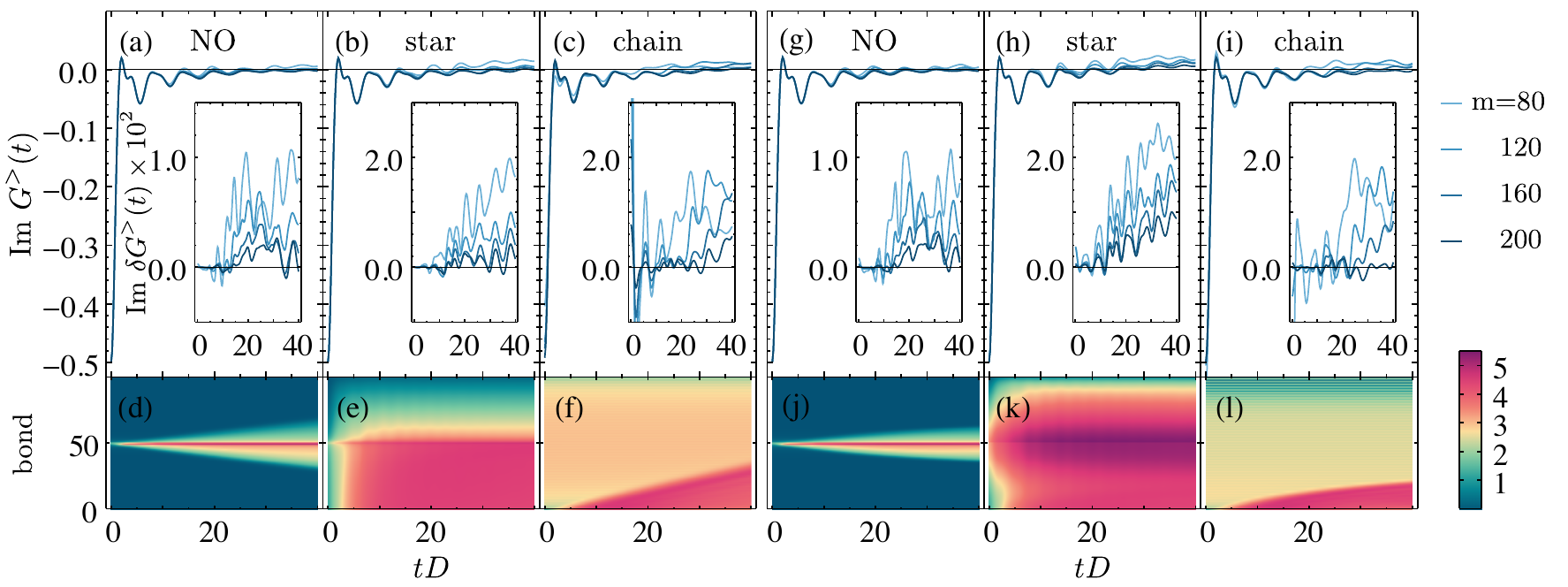}
  \caption{\label{fig:evo3orb} Time evolution of the imaginary part of $G^>(t)$ (upper panels) and bond entanglement entropies $S_b$ (lower panels) calculated in the NO, star, or chain bath geometry for $U/D=2$. Results for (a)--(f) equal-weight and (g)--(l) logarithmic discretization. For each bath geometry and discretization, $G^>(t)$ are calculated using the maximal bond dimensions $m=80$, 120, 160, and 200. The insets show their deviation $\delta G^>(t)$ from a reference $G_\mathrm{ref}^>(t)$ calculated using $m=300$.}
\end{figure*}

We can gain more insight into the different convergence behaviors by examining the ground-state charge density and entanglement distribution with different bath geometries. As shown in Fig.~\ref{fig:gs3orb_stat}(a), for the NO and star geometries, the onsite density distribution essentially follows a step function, where sites with positive (negative) potential (see Fig.~\ref{fig:bath_para}) are almost fully empty (filled). While the NO geometry shows no apparent difference between discretization schemes, for the star geometry, the deviation of the density from integers is more spread out in the whole system with logarithmic discretization. For the chain geometry, on the other hand, the density is evenly distributed among all sites, seemingly indicating a worst-case scenario with a large degree of delocalization, which is usually associated with high entanglement~\cite{Wolf2014}. Here, it is instructive to consider a periodic free fermion model, whose ground-state density matrix has only integer diagonal elements in the optimal momentum-space representation, but has identical fractional diagonal elements and pronounced off-diagonal ones in the highly entangled real-space representation. Figure~\ref{fig:gs3orb_stat}(b) shows the sorted amplitudes of all off-diagonal elements of the density matrix. For the NO geometry, only a small fraction of the elements shows nonvanishing values. By contrast, the star and chain geometries show more widely distributed off-diagonal elements that strongly depend on the bath discretization. As already indicated by the ground-state variance in Fig.~\ref{fig:gs3orb}, the star geometry is more entangled using logarithmic discretization. This can be understood, as in this case the majority of the bath sites are closely positioned around zero potential, which facilitates charge fluctuations between the impurity and bath, or among the bath sites. The opposite is true for the chain geometry. Under logarithmic discretization, the later bath sites in the chain, by design, are only exponentially weakly coupled to the system, resulting in a more localized entanglement close to the impurity.

The above intuitive arguments are corroborated by the distribution of bond entanglement entropies $S_b$ shown in Fig.~\ref{fig:gs3orb_stat}(c). For the NO geometry, the entropies are sharply localized around the impurity site and show little discretization dependence. Comparatively, for the star or chain geometry, the entropies are significantly larger and more spread out, with the former (latter) favoring the equal-weight (logarithmic) discretization. The distribution of the entropies is directly mirrored by the bond dimension of the ground-state wave function in Fig.~\ref{fig:gs3orb_stat}(d). While the ground state of the NO geometry can be accurately described with bond dimensions smaller than 200 given a truncation error of $w_t=10^{-12}$ per bond, both the star and chain geometries would require larger $m$ to reach the same level of accuracy.

Based on the above observations, we conclude that among the three geometries, the NO geometry has the optimal performance overall in obtaining the most accurate ground-state description (Fig.~\ref{fig:gs3orb}) with the least computational complexity (Fig.~\ref{fig:gs3orb_stat}).

\subsection{Time evolution}

To understand how different bath geometries perform under time evolution and how that is affected by the discretization scheme, we further compute the greater Green’s function $G^>(t)$ [Eq.~\eqref{eq:ggl}] for one of the impurity orbitals in different settings.

Figures~\ref{fig:evo3orb}(a)--\ref{fig:evo3orb}(c) show the imaginary part of $G^>(t)$ in different bath geometries with equal-weight discretization at $U/D=2$. With a small $m=80$, although with different magnitudes, results for all geometries deviate noticeably from zero in the long-time limit, indicating a relatively poor quality of data. Since an exact result is not available for such an interacting model, we use a $G_\mathrm{ref}^>(t)$ calculated with $m=300$ as a reference to evaluate the convergence behavior with respect to $m$ in each geometry. The insets of Figs.~\ref{fig:evo3orb}(a)--\ref{fig:evo3orb}(c) show the deviation $\delta G^>(t)$ from the reference, which, as expected, decreases with increasing $m$ values. We observe that for a given $m$, $\delta G^>(t)$ is smaller in the NO geometry than in the others by over a factor of two. The different performances can be explained by inspecting the corresponding evolution of bond entanglement entropies $S_b$ shown in Figs.~\ref{fig:evo3orb}(d)--\ref{fig:evo3orb}(f). Similar to the ground-state results in Fig.~\ref{fig:gs3orb_stat}(c), the entropies are centered around the impurity site in the NO geometry. While they propagate further into the chains at later times, the entropies stay relatively localized with moderate values below 3.0. In the star geometry, despite a relatively well behaved distribution at $t=0$, the entropies increase rapidly with time, and saturate after $tD \sim 5$ at a value around 4.5 in the first half chain. The chain geometry has a widespread initial entropy distribution, which explains its relatively large errors in $G^>(t)$ already at early times. Figures~\ref{fig:evo3orb}(g)--\ref{fig:evo3orb}(l) show the same observables as in Figs.~\ref{fig:evo3orb}(a)--\ref{fig:evo3orb}(f), which are instead obtained using the logarithmically discretized bath. Qualitatively, the same observations are made here. We only note that similar to the ground-state case, while the performance of NO is essentially independent of discretization schemes, that of the star or chain varies.

\begin{figure}[t]
  \includegraphics[width=\linewidth]{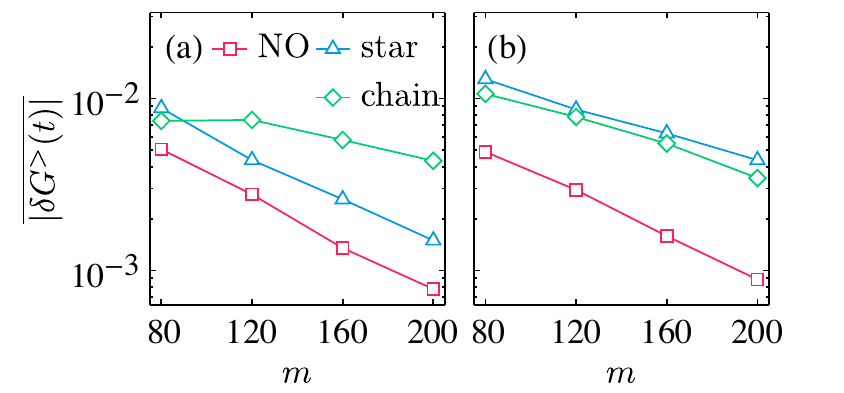}
  \caption{\label{fig:evo3orb_diffnorm} Average norm of $\delta G^>(t)$ calculated using different $m$ values over the time interval $tD \in [0,\,40]$ for (a) equal-weight and (b) logarithmic bath discretization (see Fig.~\ref{fig:evo3orb} and text).}
\end{figure}

Finally, to summarize the time-evolution results in different geometries, we show the average norm of $\delta G^>(t)$ over the full calculated time interval in Fig.~\ref{fig:evo3orb_diffnorm}. For all cases, the accuracy of $G^>(t)$ converges exponentially with increasing $m$. In general, the NO geometry has smaller errors than the star or chain for a given $m$, as well as a higher convergence rate with increasing $m$. Depending on the discretization scheme, for the largest $m=200$ tested here, the deviation is about a factor of three smaller for the NO than for the star. We note that similar results are also obtained for $U/D=1$ and 3.

\section{DMFT of \texorpdfstring{S\MakeLowercase{r}}{Sr}$_2$\texorpdfstring{R\MakeLowercase{u}}{Ru}O$_4$}\label{sec:sro}

After the above case study of the three-orbital impurity model and detailed comparison of the computational complexity and accuracy using different bath settings, we now turn our attention to a material-realistic application of the TTN solver to the DMFT calculation of the archetypal Hund's metal \sro{}~\cite{Mackenzie2003,Mackenzie2017}, and study, in particular, the effects of SOC on its one and two-particle observables. While previous DMFT works have proven successful in explaining many aspects of experiments such as mass renormalization~\cite{Mravlje2011,Tamai2019} and static magnetic responses~\cite{Strand2019} despite the neglect of SOC in the calculation, SOC has always been an awkward obstacle and its role for many observables has remained a matter of speculation.

The low-energy physics of \sro{} is dictated by the Ru 4$d$-$t_{2g}$ orbitals hybridized with the O 2$p$~\cite{Haverkort2008}. Similar to previous work~\cite{Mravlje2011,Mravlje2016,Zhang2016,Kim2018,Tamai2019,Linden2020,Kugler2020}, we construct the minimal one-body tight-binding Hamiltonian $\h{TB}$ using the three maximally localized $t_{2g}$ orbitals obtained from the non-SOC DFT Kohn-Sham orbitals~\cite{wien2k_a,wien2k_b,wannier90}. The full interacting Hamiltonian is then defined as $H=\h{TB}+\h{K}$ by adding the Coulomb term $\h{K}$~\eqref{eq:kanamori} with $U=2.3$ and $J=0.4$~\cite{Kim2018,Strand2019,Tamai2019,Linden2020,Kugler2020}. The unit of energy/frequency is the electron volt (eV) hereafter, unless otherwise noted. To include the effects of SOC, a local term is added as
\begin{equation}\label{eq:soc}
  \h{so}=\ii \frac{\lambda}{2} \sum_{m m' m''}\epsilon_{m m' m''} \sum_{\sigma\sigma'} c^\dag_{m\sigma} c_{m'\sigma'} \tau^{m''}_{\sigma\sigma'},
\end{equation}
where $\vb*{\epsilon}$ is the completely antisymmetric tensor and $\vb*{\tau}$ is the Pauli vector. The operators here are defined in the restricted $t_{2g}$ orbital-spin basis $(m,\sigma) \in \{yz(1), xz(2), xy(3)\}\otimes \{\uparrow, \downarrow\}$. We choose an isotropic coupling constant $\lambda=0.11$~\cite{Haverkort2008}.

The DMFT calculations are performed with $N_b=149$ bath sites for each impurity orbital. The large $N_b$ gives a faithful quasi-continuous representation of the bath, and also allows time evolution to reach for sufficiently long time, both of which are crucial for obtaining accurate impurity Green's functions. Based on the observations in Sec.~\ref{sec:geometry}, the NO bath geometry is adopted to minimize the computation cost. It is known that spatially separating the spin up and down degrees of freedom into different chains can improve the computational efficiency~\cite{Saberi2008,Ganahl2015,Bauernfeind2017}, a trick, however, that can no longer be applied in the presence of SOC. For a more general evaluation of numerical performances, we therefore choose not to separate the spin channels even for the calculations without SOC. For the SOC case, the calculations are performed in the eigenbasis of the SOC operator~\eqref{eq:soc} ($J$ basis), which minimizes the number of off-diagonal Green's functions and also allows us to take advantage of the global $U(1)$ symmetry associated with the quantum number $J_z$. In addition, orbitals with off-diagonal Green's functions are placed on neighboring branches to further minimize long-range entanglement. For both calculations with and without SOC, we keep a truncation weight $w_t=10^{-14}$ for the ground state and $10^{-9}$ for the time evolution. The maximal bond dimension $m$ is set to $200$ ($600$) and $240$ ($720$) for the auxiliary (site) tensors for calculations without and with SOC, respectively. All dynamic quantities are calculated in the time domain up to $t=40$ eV$^{-1}$ with a time step of $\Delta t=0.1$ eV$^{-1}$ and Fourier transformed into the frequency domain after performing the linear prediction.

\subsection{Single-particle spectra, self-energies, and Fermi surfaces}\label{sec:sro:spectra}

\begin{figure}[t]
  \includegraphics[width=\linewidth]{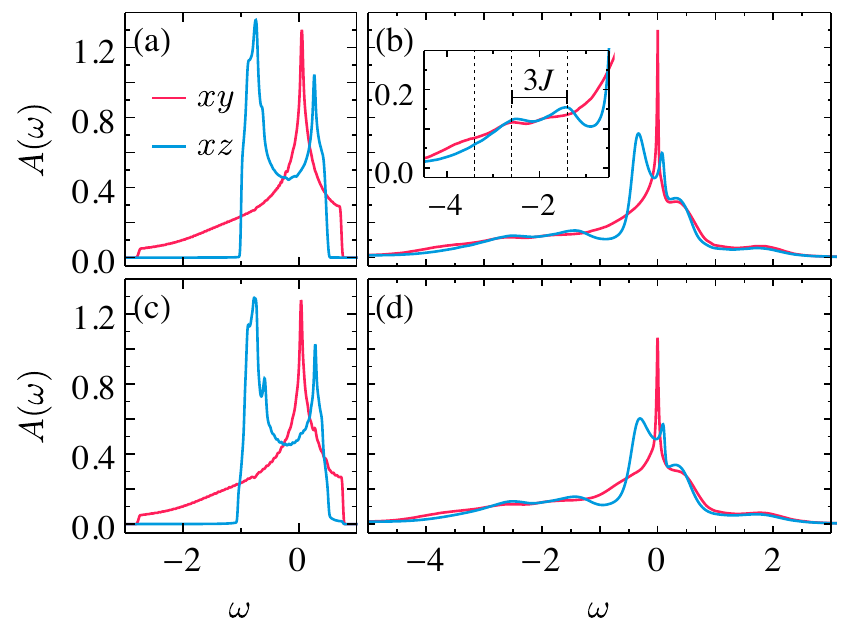}
  \caption{\label{fig:dosA} Non-interacting (left panels) and DMFT (right panels) local spectral functions for the \sro{} calculated (a),(b) without and (c),(d) with SOC. The inset in (b) shows the details of the lower Hubbard bands, where the dashed lines mark the location of atomic multiplet excitations.}
\end{figure}

\emph{Results without SOC.} Figure~\ref{fig:dosA}(a) shows the noninteracting spectral functions (density of states) of the \sro{} without SOC. The quasi-two-dimensional $xy$ band shows a van Hove singularity slightly above the Fermi energy. Its spectral function has a long tail below the Fermi energy and has a relatively larger bandwidth than the $xz$ (or symmetrically equivalent $yz$) band, which shows typical quasi-one-dimensional features with singularities located at the edge of the bands. These spectral features, albeit significantly renormalized, remain visible in the low-energy Fermi-liquid quasiparticle spectra for the interacting case, as shown in Fig.~\ref{fig:dosA}(b), within the range of $\abs{\omega} \lesssim 1.0$. In addition, a side peak at $\omega \simeq 0.4$ can be identified in the spectra for both orbitals, which is associated to the screening of orbital degrees of freedom~\cite{Stadler2015,Stadler2019}.

For both the $xy$ and $xz$ bands, the DMFT spectra manifest Hubbard bands at high energies ($\abs{\omega}\gtrsim$ 1.0) induced by the Coulomb interaction. Two well-defined substructures, appearing as vestiges of the atomic multiplet excitations, can be identified in the $xz$ lower Hubbard band around $-2.6$ and $-1.4$ with an energy separation of $\simeq 3J$. We note that these features were not resolved in earlier QMC~\cite{Deng2019} or NRG~\cite{Kugler2020} studies, which have relatively limited resolving powers at high energies. Neglecting the crystal field, the $N=4$ ground state of the atomic $t_{2g}$ system \eqref{eq:kanamori} has the quantum numbers $(S,L)=(1,1)$. The electron removal final states with $N=3$ have three different symmetry sectors, i.e., $(3/2,0)$, $(1/2, 2)$, and $(1/2, 1)$, ordered from low to high energy, with energy differences $3J$ and $2J$ between them, respectively. Following this atomic analysis, an additional structure is thus expected around $-3.4$~eV. However, simple numerical analysis shows that this transition has a considerably lower amplitude compared with the other two, which is further reduced for the $xz$ orbital by the inclusion of a crystal field, $\epsilon_{\mathrm{CF}}=\epsilon_{xz}-\epsilon_{xy}$ comparable to the DFT bare value ($\simeq$ 0.1) found in \sro. This could explain its apparent absence in the $xz$ spectrum in Fig.~\ref{fig:dosA}. Similar structures can be identified on the $xy$ spectrum with a small shift to higher excitation energies. Compared to the $xz$ band, the multiplet peaks are less pronounced due to the large bare bandwidth of the $xy$ band, with the exception of the peak at $-3.6$, whose amplitude is enhanced by the crystal field in contrast to the $xz$.

\begin{figure}[t]
  \includegraphics[width=\linewidth]{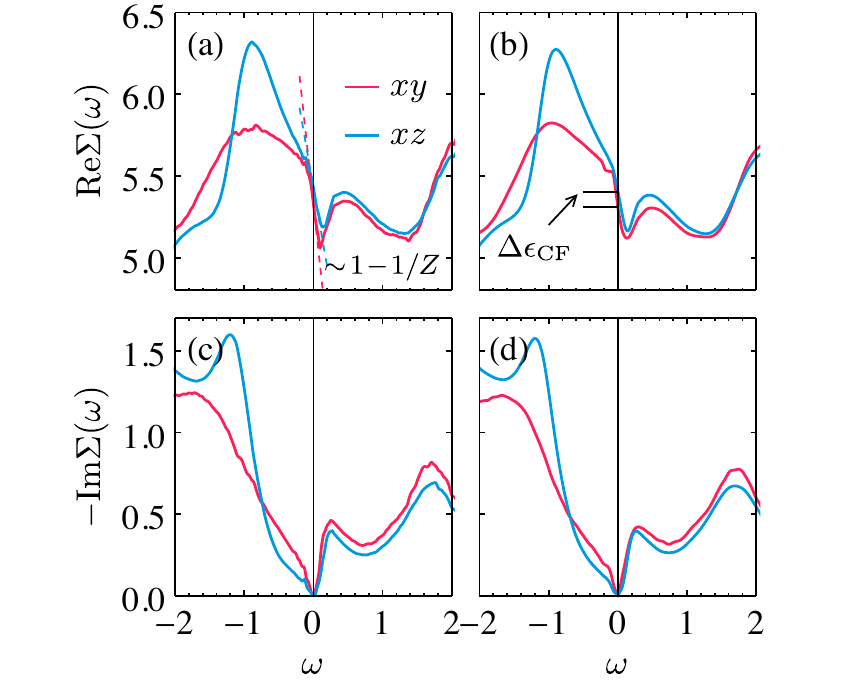}
  \caption{\label{fig:sigma} Real (upper panels) and imaginary (lower panels) parts of the self-energy for the \sro{} (a),(c) without and (b),(d) with SOC. Linear fits $\Re \Sigma(\omega) \simeq \Re \Sigma(0) + (1-1/Z)\omega$ are shown in (a) for both the $xy$ and $xz$ bands around $\omega=0$. The difference of $\Re \Sigma(0)$ shown in (c) corresponds to a correlation induced crystal field change.}
\end{figure}

The effect of correlation can be more quantitatively assessed by examining the self-energy in Fig.~\ref{fig:sigma}. As shown in Fig.~\ref{fig:sigma}(a), the real part of the self-energy $\Re \Sigma(\omega)$ is linear within a small energy window close to the Fermi energy, consistent with the expected Fermi-liquid ground state of \sro. The slope $\partial_\omega \Re \Sigma|_{\omega=0}$ is steeper for the $xy$ orbital than for the $xz$, signaling a stronger renormalization effect for the former despite its larger bare bandwidth. This seemingly surprising observation is attributed to the relatively smaller spectral weight of the hybridization function $\Delta(\omega)$ for the $xy$ orbital in the vicinity of Fermi energy due to the van Hove singularity~\cite{Mravlje2011,Kugler2020}. The calculated quasiparticle weights $Z=(1-\partial_\omega \Re \Sigma|_{\omega=0})^{-1}$ are $Z_{xy}\approx0.20$ and $Z_{xz}\approx0.29$, in good agreement with experiments~\cite{Mackenzie2003,Tamai2019} and previous calculations~\cite{Mravlje2011,Kim2018,Kugler2020}. The orbital differentiated correlation effect can also be seen in the corresponding (negative) imaginary part of the self-energy $-\Im \Sigma(\omega)$ in Fig.~\ref{fig:sigma}(c), where the quasiparticle excitation in the $xy$ band has a relatively shorter life-time close to the Fermi energy than the $xz$ band. The correlation also leads to an enhancement of the crystal field~\cite{Zhang2016} $\Delta \epsilon_{\mathrm{CF}}=\Re \Sigma_{xz}(0)-\Re \Sigma_{xy}(0) \approx 0.09$, a change comparable to its bare value of $\simeq 0.1$. Consequently, the charge densities are more evenly distributed between the $xy$ and $xz/yz$ orbitals, with orbital population $(n_{xy}, n_{xz})\approx(1.29,1.35)$, in contrast to the more polarized DFT values $(1.22, 1.39)$. For comparison, experiments find an almost uniform distribution $(1.33, 1.33)$~\cite{Mackenzie2003}.

Finally, we note here that in addition to the quantities discussed above, the general line shape of the self-energy with all its key features, including the two linear regions separated at $\omega \simeq -0.05$ for $\Re \Sigma_{xz}(\omega)$ and the retraction of normalization with $\partial_\omega \Re \Sigma_{xy/xz}>0$ for $\omega$ between $0.2$ and $0.4$, is in good agreement with previous results~\cite{Kim2018,Kugler2020}.

\emph{Results with SOC.} The energy scale of SOC is only a few percent of the bandwidth, therefore, no significant effects on the line shape of the non-interacting spectral function are expected, except for momenta where the bands are (near) degenerate~\cite{Haverkort2008}. As shown in Fig.~\ref{fig:dosA}(c), the most notable changes are spotted around $\omega \simeq \pm 0.5$, where fine structures emerge on both spectra due to the SOC induced hybridization at the $xy$ and $xz/yz$ band crossing. In addition, the edges of the $xz/yz$ band are noticeably ``smeared out'' over an energy interval of the order of the SOC due to the lifted degeneracy. Not surprisingly, the correlated spectra in Fig.~\ref{fig:dosA}(d) show no qualitative differences compared to those in Fig.~\ref{fig:dosA}(b), except for small changes on the detailed line shape of the quasiparticle spectra which can be traced back to the changes in the bare spectra in Fig.~\ref{fig:dosA}(c). Correspondingly, the self-energy in Figs.~\ref{fig:sigma}(b) and~\ref{fig:sigma}(d) is essentially the same as their counterparts without SOC. Similar to Ref.~\cite{Kim2018}, we find the quasiparticle weights $Z_{xy}$ and $Z_{xz}$ unchanged within an error of $0.03$ when compared to the values without SOC.

\begin{figure}[t]
  \includegraphics[width=\linewidth]{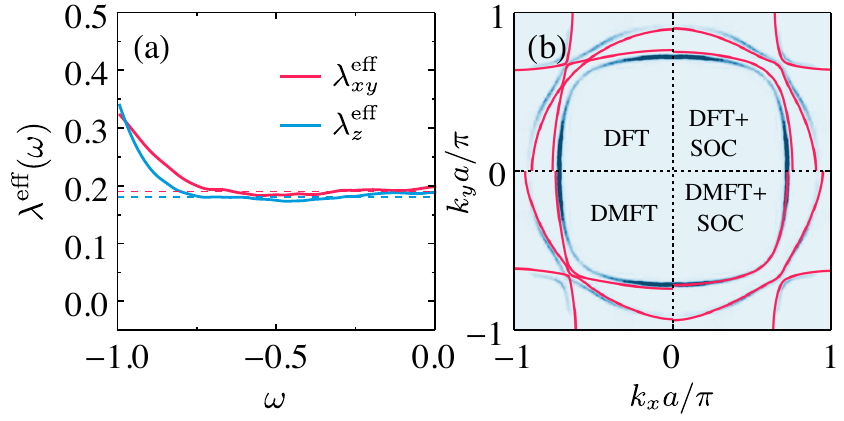}
  \caption{\label{fig:lambda_FS} (a) Effective in-plane and out-of-plane SOC constants $\lambda^\mathrm{eff}(\omega)$ near the Fermi energy (see text). The dashed lines show their average values within the energy interval $[-0.8,0.0]$ at 0.19 and 0.18, respectively. (b) Summary of Fermi surface plots at $k_z c = 0.4\pi$ obtained from DFT and DMFT calculations with or without SOC in comparison with angle-resolved photoemission spectroscopy (ARPES) results (blue) from Ref.~\cite{Tamai2019}. The experimental data are symmetrized here assuming $D_4$ symmetry.}
\end{figure}

The unremarkable effects of SOC observed so far validate earlier calculations that neglected SOC~\cite{Mravlje2011,Mravlje2016,Tamai2019}, especially results pertaining to the the orbital-diagonal quantities. However, it does not mean that SOC is inconsequential to the physical properties of \sro. In terms of single-particle observables, it was shown early on~\cite{Haverkort2008} that the inclusion of SOC is crucial for correctly describing the experimental Fermi surface. Once the Coulomb interactions are included, the off-diagonal components of the self-energy strongly modify the bare values of SOC~\cite{Zhang2016}. The frequency-dependent effective SOC, given as $\lambda^\mathrm{eff}_{xy}(\omega)=\lambda-2\Im \Sigma_{yz\uparrow,xz\uparrow}(\omega)$ and $\lambda^\mathrm{eff}_{z}(\omega)=\lambda+2\Im \Sigma_{xz\uparrow,xy\downarrow}(\omega)$, is plotted in Fig.~\ref{fig:lambda_FS}(a). The isotropic atomic SOC now acquires some anisotropy between the in-plane and out-of-plane values due to the lower-symmetry local point group $D_{4h}$ of the Ru ion. At $\omega=0$, the values $\lambda^\mathrm{eff}_{xy} \approx \lambda^\mathrm{eff}_z\approx 0.20$, which is about twice the bare value $0.11$. We note that due to the van Hove singularity in the close proximity of the Fermi energy, the details of the Green's functions and consequently the self-energies at low energy (especially for $\abs{\omega}\lesssim 0.05$) show moderate variations (up to about 20\%) depending on the details of the calculation. This is limited by the intrinsic difficulties of capturing long-time dynamics using time-evolution algorithms and linear prediction. Nonetheless, similar to Ref.~\cite{Kim2018}, we find that $\lambda^\mathrm{eff}_{xy/z}$ are essentially energy independent within a region below the Fermi energy, albeit with a narrower energy range of $\sim 0.8$. The average values within this region are $\bar \lambda^\mathrm{eff}_{xy}=0.19$ and $\bar \lambda^\mathrm{eff}_{z}=0.18$. We summarize the effects of SOC by plotting the Fermi surfaces obtained from DFT and DMFT calculations with or without SOC in Fig.~\ref{fig:lambda_FS}(b). Overall, the DMFT+SOC result has the best agreement with experiment. Compared to the DMFT result without SOC, the SOC lifts the degeneracies at the crossing along the Brillouin zone diagonals, giving rise to well-separated sheets as observed in experiment. As a result, the $k_z$ modulation of the Fermi surface is strongly suppressed (not shown), in qualitative agreement with findings in quantum oscillation experiments~\cite{Bergemann2000}. In addition, the SOC further balances the orbital occupation with $(n_{xy}, n_{xz})\approx(1.33,1.33)$, leading to correctly sized Fermi surface pockets~\cite{Bergemann2000,Mackenzie2003}.

\begin{figure}[t]
  \includegraphics[width=\linewidth]{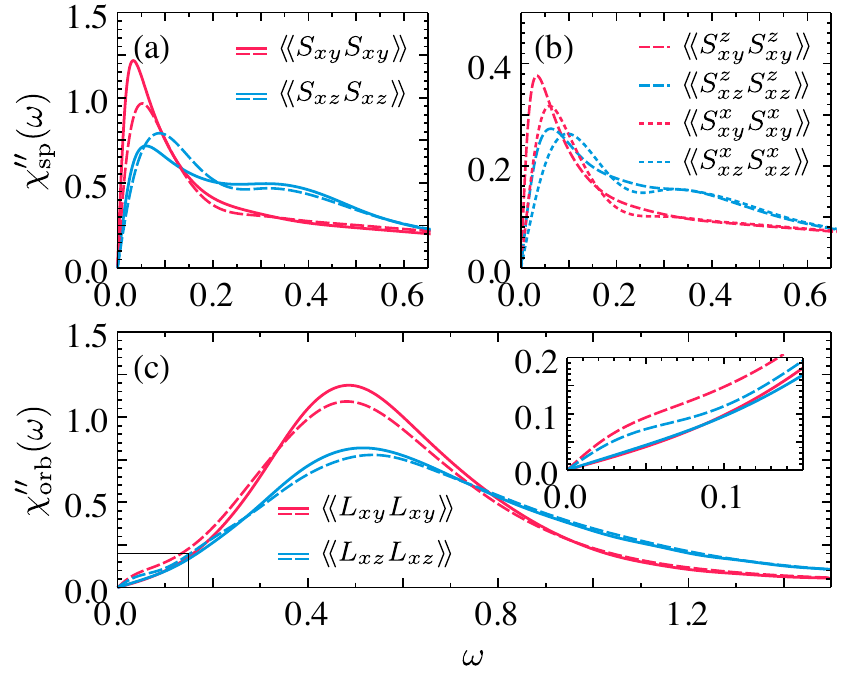}
	\caption{\label{fig:chi} (a) Total and (b) component-resolved spin $\chi''_{\mathrm{sp}}(\omega)$ and (c) orbital $\chi''_{\mathrm{orb}}(\omega)$ structure factors of the $xy$ and $xz$ orbitals without (solid lines) and with (dashed lines) SOC. The inset in (c) shows the details of the low-energy part of $\chi''_{\mathrm{orb}}(\omega)$. The total spin structure factor for orbital $m$ is obtained by summing over the contributions of all spin components $\stfac{S_m S_m}=\sum_\alpha \stfac{S^\alpha_m S^\alpha_m}$.}
\end{figure}

\subsection{Spin and orbital susceptibilities}\label{sec:sro:chi}

In addition to the single-particle quantities discussed above, we further study the importance of SOC on the dynamics of \sro, which is believed to hold keys to understanding the long-standing puzzle of its superconductivity~\cite{Mackenzie2003,Mackenzie2017}. To this end, we calculate the dynamical spin and orbital susceptibilities,
\begin{equation}
  \chi_{m m^\prime}(t) = -\ii \uptheta(t)\ev*{\comm*{O_m(t)}{O_{m^\prime}(0)}_-}_0,
\end{equation}
where the operators $O_m$ are either the spin operators $S^\alpha_{m} = \frac{1}{2}\sum_{\sigma\sigma^\prime} c^\dag_{m\sigma} \tau^\alpha_{\sigma\sigma^\prime} c_{m\sigma^\prime}$ or the orbital ones $L_m=\ii \sum_{m'm''\sigma} \epsilon_{mm'm''} c^\dag_{m'\sigma} c_{m''\sigma}$. One can define dynamic structure factors $\stfac{O_m O_{m^\prime}}\equiv\chi''(\omega)\equiv-1 /\pi \Im \chi(\omega)$ as the imaginary part of the Fourier-transformed susceptibilities $\chi(\omega)=\int \dd t\,e^{\ii \omega t} \chi(t)$.

Figure~\ref{fig:chi} shows the orbital-resolved spin and orbital structure factors. The spectra show different characteristic energy scales, with the maxima of spin responses positioned below 0.1 and those of orbital around 0.5. For the spin responses in Fig.~\ref{fig:chi}(a), in the absence of SOC, the spectral weight of the $xy$ orbital is more concentrated at lower energy than that of $xz$, consistent with previous results~\cite{Kugler2020}. Upon the inclusion of SOC, the spin response maxima for both orbitals are noticeably pushed up in energy due to the coupling to higher-energy orbital responses. Such changes of the characteristic energies are significant enough to be observed in spectroscopic experiments such as resonant inelastic x-ray scattering, which can measure orbital-resolved correlation functions~\cite{Kim2017,Suzuki2019}. Another immediate consequence of SOC is the breaking of the $SU(2)$ spin-rotation symmetry. As shown in Fig.~\ref{fig:chi}(b), a clear anisotropy is observed between the in-plane and out-of-plane components of $\chi''_{\mathrm{sp}}(\omega)$ for both orbitals. The more pronounced fluctuations of the out-of-plane component are in qualitative agreement with the observed anisotropy in \sro{} in inelastic neutron scattering experiments~\cite{Braden2002a} and the $c$-axis Ising magnetic anisotropy in Ti-substituted \sro~\cite{Braden2002b}. In Fig.~\ref{fig:chi}(c), the orbital structure factors show changes compatible with those of the spin responses in Fig.~\ref{fig:chi}(a). Part of the spectral weight is transferred to lower energy in the presence of SOC for both orbitals, appearing as small ``shoulders'' around the characteristic energy of the spin responses.

\begin{figure}[t]
  \includegraphics[width=\linewidth]{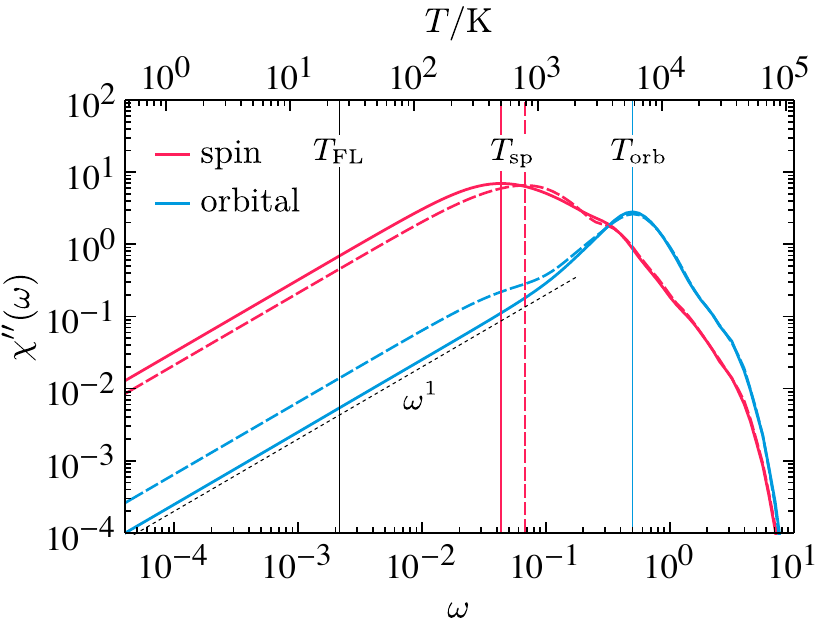}
	\caption{\label{fig:chilog} Total spin and orbital structure factors of the $t_{2g}$ orbitals without (solid lines) and with (dashed lines) SOC. The spectra are obtained by summing over contributions of all orbital and spin components: ${\chi^\mathrm{tot}_\mathrm{sp}}''(\omega)=\sum_{\alpha} \stfac{S^\alpha S^\alpha}$ with $S^\alpha=\sum_m S_m^\alpha$, and ${\chi^\mathrm{tot}_\mathrm{orb}}''(\omega)=\sum_{m} \stfac{L_m L_m}$. The spin and orbital Kondo scales are marked by vertical lines. The experimentally estimated Fermi-liquid temperature $T_\mathrm{FL}\approx 25$ K~\cite{Mackenzie2003} is also marked for comparison. The guide-to-the-eye black dotted line indicates the Fermi-liquid linear power law.}
\end{figure}

The logarithmic plot in Fig.~\ref{fig:chilog} reveals more information of the effects of SOC on the spin and orbital susceptibilities. As already seen in Fig.~\ref{fig:chi}, the spin and orbital responses show distinct energy scales. Their respective Kondo scales $T_\mathrm{sp}$ and $T_\mathrm{orb}$, defined here as the maxima of the spectra~\cite{Stadler2015,Stadler2019}, differ by about one order of magnitude. Without SOC, the system goes through a two-stage screening process before entering the Fermi-liquid ground state below the Fermi-liquid energy scale $T_\mathrm{FL} \approx 25$ K~\cite{Mackenzie2003}, signified by a linear behavior of the susceptibilities $\chi'' \sim \omega$. The screening of the spin degrees of freedom ($T_\mathrm{sp} \simeq 5 \times 10^2$ K) happens after the orbital ones are fully screened below $10^3$ K where $\chi''_\mathrm{orb}$ enters a linear region~\cite{Kugler2020}. This so-called spin-orbital separation is one of the main features characterizing the Hund's metal~\cite{Georges2013,Stadler2015,Stadler2019,Deng2019} nature of \sro. On the other hand, SOC strongly couples these two otherwise separated screening processes, resulting in spectra weight transfer to higher (lower) energies for the spin (orbital) susceptibility. $\chi''_\mathrm{orb}$ now shows a clear plateau at $T_\mathrm{sp}$, which means that, as opposed to the case without SOC, the orbital degrees of freedom are \emph{not} fully decoupled from the spin dynamics until both of them are fully screened below $T_\mathrm{FL}$. This may have important implications for constructing low-energy models for \sro{}, especially those concerning its superconductivity~\cite{Ng2000,Scaffidi2014}. Future work on momentum-resolved dynamical spin and orbital spin susceptibilities~\cite{Strand2019} including SOC may be interesting.

\section{Conclusions and Outlook}

In conclusion, we have presented a multiorbital impurity solver based on TTN. The acyclic structure of the network allows for straightforward implementation of SVD-based methods for ground-state search as well as real-time evolution calculations for obtaining dynamical quantities, which are well established and have proven remarkably successful in the MPS context. The flexible tree structure can adapt to impurity Hamiltonians with different orbital numbers and bath geometries. Using a three-orbital model, we have shown that, among the most common bath geometries, the natural-orbital geometry has the best performance overall for both ground-state and time-evolution calculations over a large parameter space.

The combined advantage of the tree tensor network and the natural-orbital bath geometry allows for efficient and accurate solution of general multiorbital impurity problems. As a benchmark, we performed real-axis DMFT calculations of \sro{} both without and with SOC, the latter of which has remained elusive due to its computational complexity. In the low-energy quasiparticle region, our real-frequency single-particle spectra and self-energies are in good agreement with previous results obtained using NRG and QMC solvers that are known for high accuracies close to zero frequency. Moreover, unlike NRG and QMC, the uniform resolution of the presented solver on the entire frequency axis also makes it possible to identify atomic multiplet excitations in the high-energy Hubbard bands, which are important for determining interaction parameters of the system and which are not yet resolved in previous studies. We have also shown that the system has distinct in-plane and out-of-plane spin responses when SOC is present, which are closely related to the experimentally observed magnetic anisotropy. In addition, the spin and orbital degrees of freedom are strongly coupled throughout a wide energy range above the Fermi-liquid energy scale, thus modifying the complete spin-orbital-separation picture in the absence of SOC.

The presented TTN solver opens up several interesting possibilities for future work. Many well-known materials such as VO$_2$, V$_2$O$_3$, or Fe$_3$O$_4$ have low-symmetry local point-group symmetries at the transition metal sites despite their high-symmetry crystal systems, where the structure distortion plays an important role in determining their properties such as the metal-insulator transition. One can now revisit these classic materials, explicitly including the off-diagonal part of the local Green's function. Another direct application for the presented solver is the cellular DMFT, which recently has been revisited and extended by cluster-size extrapolation schemes on the one- and two-particle levels~\cite{Klett2020,Schaefer2021}. While on the square lattice at half filling relatively large cluster sizes of up to $9^2$ sites could be reached, a triangular lattice geometry turned out to be hardly feasible even for cluster sizes $<10$ sites with current QMC solvers~\cite{Wietek2021}.

\begin{acknowledgments}
	We thank Daniel Bauernfeind for useful discussions. This work is partially supported by Deutsche Forschungsgemeinschaft (DFG) under Germany’s Excellence Strategy EXC2181/1-390900948 (the Heidelberg STRUCTURES Excellence Cluster).
\end{acknowledgments}

\appendix
\counterwithin{figure}{section}

\section{Canonical forms, DMRG, and time evolution of TTN}\label{app:ttps}

The basic building blocks of the proposed TTN (Fig.~\ref{fig:ttps}) are a tensor:
\begin{equation}
 M^{\sigma[i,j]}_{e_a,e_b,e_r}\equiv\includeTikz{0}{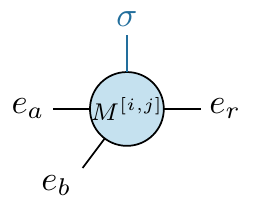}\, ,
\end{equation}
where the bonds of the tensor, i.e., $e_r$, $e_a$, and $e_b$, connect to tensors at its root in the upper layer, and on its left $a$ and right $b$ child branches in the next layer, respectively. The rank-3 auxiliary tensors in the root layers of the tree and rank-3 physical tensors on the leaves can then be expressed by setting the physical index $\sigma$ or bond index $e_b$ as a dummy one. Similar to MPS, we can encode the symmetry information of the underlying systems into the site tensors. Currently, Abelian $U(1)$ symmetries are implemented.

Assuming the same bond dimension $m$ for all bonds, the auxiliary tensors have $m^3$ parameters, while the physical site tensors have $dm^2$. For typical $m\sim\mathcal{O}(10^2)$ and $d \sim \mathcal{O}(1)$, the time cost of SVD on an auxiliary tensor is then $\mathcal{O}(m^4)$, much greater than the cost $\mathcal{O}(dm^3)$ needed for a site tensor. In addition, two-site local optimization schemes of DMRG and TDVP now cost $\mathcal{O}(m^5)$ on auxiliary tensors. To achieve a balance between accuracy and computational costs, we adopt ($i$) the single-site DMRG algorithm with subspace expansion for the ground-state search, and ($ii$) a hybrid TDVP for time evolution, where the auxiliary tensors are updated by a single-site TDVP and the site tensors by the two-site variant.

\subsection{Canonical forms and orthogonality center}\label{app:ttps:forms}

Similar to MPS, the tensors in the TTN need to be brought into canonical forms to facilitate efficient computation. Given a pair of connected tensors, one can always normalize one towards the other by SVD
\begin{equation}\label{eq:norm}
	\begin{split}
		\includeTikz{0}{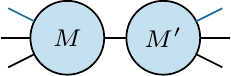} & = \includeTikz{0}{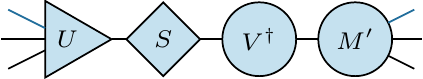} \\
		& = \includeTikz{0}{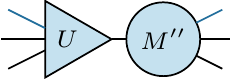}\,,
	\end{split}
\end{equation}
where the first site tensor $M$ is now left-normalized satisfying $U^\dag U=1$. Depending on along which bond \emph{of $M$} the normalization is performed, we term the site tensor $r$, $a$, or $b$ normalized. As the tree is acyclic, given a node $v$, any other node in the network is connected to it by a unique simple path. One can define $v$ as the orthogonality center after iteratively normalizing all tensors towards it starting from the most distant ones. The state site tensors of its $a$, $b$, and $r$ subtrees can be contracted into single tensors $A^{\vec{\sigma}_{v,a}}_{m_a}$, $B^{\vec{\sigma}_{v,b}}_{m_b}$, and $R^{\vec{\sigma}_{v,r}}_{m_r}$. Here, $\vec{\sigma}_{v,t}$ ($t$=$a$, $b$, $r$) collects all physical indices in the $t$ subtree of node $v$. We can define similar contraction for a TTN operator of each sub-tree as $O^{{\vec{\sigma}'}_{v,t}}_{\vec{\sigma}_{v,t} w_t}$. It is then possible to write
\begin{equation}\label{eq:expval}
	\ev*{H}{\psi}=\includeTikz{-15}{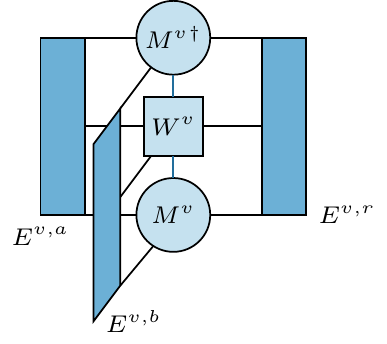}\,,
\end{equation}
where the environment tensors $E^{v,t}$ associated to, e.g., $a$ bond, are
\begin{equation}
	E^{v,a}_{m_a,w_a,m_a} = \sum_{\substack{\vec{\sigma}_{v,a} {\vec{\sigma}'}_{v,a}}}
	A^{\vec{\sigma}_{v,a}}_{m_a} O^{{\vec{\sigma}'}_{v,a}}_{\vec{\sigma}_{v,a} w_a} A_{{\vec{\sigma}'}_{v,a}, m_a}^\dag.
\end{equation}
In the single-site DMRG, the ground-state search is then performed by sweeping over the full system multiple times and variationally updating the local state tensors $M^v$ in each step to find the extreme value of Eq.~\eqref{eq:expval}.

\subsection{Subspace expansion for single-site DMRG}

While the two-site DMRG is able to distribute particles throughout the system and partially circumvents the convergence issues of its single-site variant, it is more time consuming, especially for networks containing higher rank tensors with large bond dimensions. In this paper, we adopt a single-site algorithm supplied with subspace expansion~\cite{Hubig2015}, which allows dynamical increase of the otherwise fixed bond dimension during each tensor update. As an example, the $e_r$ bond of the tensor $M^v$ of size $(d, m_a, m_b, m_r)$ is expanded as $M^v \rightarrow \tilde M^v = \begin{pmatrix}M^v & P^v\end{pmatrix}$ with $P^v$ of size $(d, m_a, m_b, w_r m_r)$ obtained using available tensors
\begin{equation}
	P^v = \alpha E^{v,a} M^{v} W^{v} E^{v,b}.
\end{equation}
Correspondingly, the tensor $M^{v'}$ connected to $M^v$ via the $e_r$ bond is be expanded by a zero matrix $M^{v'} \rightarrow \tilde{M}^{v'}=\begin{pmatrix}
	M^{v'}\\
	0
\end{pmatrix}$ to match the dimension of $M^v$. After such an expansion, $\tilde M^v$ is truncated with respect to the $e_r$ bond to $\tilde{m} \geq m$. The scalar mixing factor $\alpha$ controls the amplitude of perturbation to the state and eventually scales down to zero to reach convergence.

\subsection{Time evolution}

The TTN is time evolved by TDVP, which is originally derived from the Dirac-Frenkel variational principle for MPS~\cite{Haegeman2011,Haegeman2016} and later extended to TTN~\cite{Lubich2013,Schroeder2017,Bauernfeind2020}. Its main idea is to project the action of $H$ to the tangent space of the state manifold with a fixed bond dimension by introducing a tangent space projector $P_{T_{|\psi\rangle}}$. With such a projection, the modified time-dependent Schr\"odinger equation written as
\begin{equation}\label{eq:tdse}
	\frac{d}{dt}|\psi(t)\rangle = -\ii P_{T_{|\psi\rangle}} H |\psi(t)\rangle,
\end{equation}
after a Lie-Trotter splitting of the tangent space projector, resolves into a system of equation of motion for each TTN tensor $M^{v}$, and center matrix $C^{v}$ which is obtained by a SVD of $M^{v}(t+\Delta t)$ with respect to its $e_r$ bond. The single-site local update is accomplished by
\begin{equation}\label{eq:tdvp}
  \begin{split}
M^{v}(t+\Delta t) = e^{-i\Delta t H^{v}_{(1)}} M^{v}(t), \\
C^{v}(t+\Delta t) = e^{i\Delta t H^{v}_{(0)}} C^{v}(t).
  \end{split}
\end{equation}
Here, $H^{v}_{(1)}$ and $H^{v}_{(0)}$ are the single-site and zero-site effective Hamiltonians on node $v$:
\begin{equation}\label{eq:heffa}
  \hat{H}^{v}_{(1)}=\includeTikz{-18}{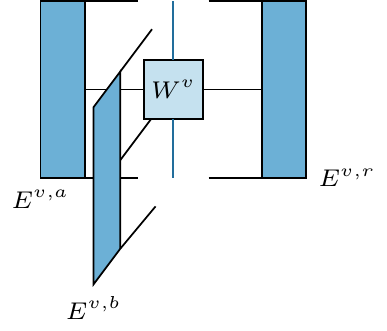}\,,
\end{equation}
and
\begin{equation}
	\hat{H}^{v}_{(0)}=\includeTikz{-2}{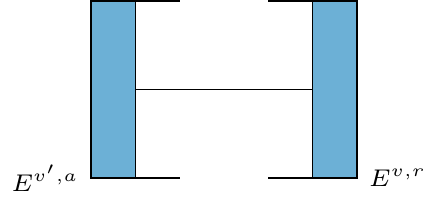}\,.
\end{equation}
The environment tensors $E^{v}$ are constructed iteratively during the TDVP sweep. The above single-site scheme with a leading computational cost of $\mathcal{O}(m^4)$ is used to time evolve all the auxiliary tensors. For site tensors having just two bonds, a two-site variant is adopted, in which two site tensors are updated simultaneously with a cost of $\mathcal{O}(m^4)$. In this case, the bond dimension is adjusted dynamically to contain the entanglement growth during time evolution.

\begin{figure}[tb]
	\centering
	\includegraphics[width=0.5\textwidth]{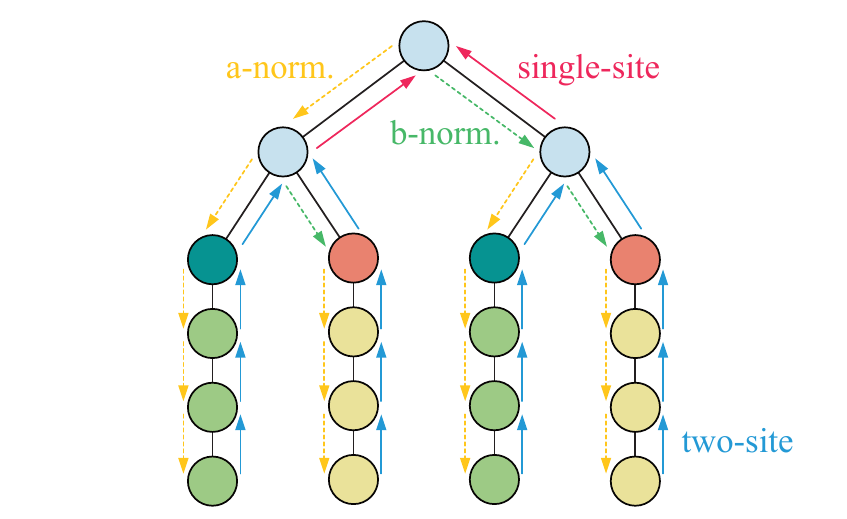}
	\caption{Sketch of a single TDVP sweep. The solid (dashed) arrows indicate time-evolution (normalization) operations (see text).}
	\label{fig:tdvp}
\end{figure}

A TDVP sweep which evolves the state from $\ket{\psi(t)}$ to $\ket{\psi(t+\Delta t)}$ is illustrated in Fig.~\ref{fig:tdvp}. The state is first brought in to a canonical form by setting the root node ($[1,1]$) as the orthogonality center. The time evolutions are carried out by visiting each node in an anticlockwise manner. By performing operations on each node as depicted in Fig.~\ref{fig:tdvp}, each visited node during the sweep is guaranteed to be the orthogonality center. The ``$a$($b$)-normalized'' operations are defined in Eq.~\eqref{eq:norm}. The ``single-site update'' represents time evolving the site tensor one step forward, followed by a backward time evolution of the center tensor according to Eq.~\eqref{eq:tdvp}. The ``two-site update'' updates the visited node and its root node simultaneously with the two-site effective Hamiltonian, followed by a backward time evolution of its root node. One can further split the first-order TDVP sweep into a second order integrator with Trotter error of the order of $\mathcal{O}(\Delta t^3)$ by two symmetric TDVP sweeps that each evolves the state half time step $\Delta t/2$ forward.

\section{Matrix Representation of Green's functions}\label{app:gf}

In the spectral representation, a Green's function can be expressed as a series of discrete block poles
\begin{equation}\label{eq:gl}
  G_L(z) = \sum_{i=1}^{N_L} \frac{w_i}{z-\epsilon_i},
\end{equation}
where $w_i$, a $d\!\times\!d$ dimension matrix with $d$ the number of impurity spinful orbitals, is the weight of the pole at energy $\epsilon_i$. The matrices $w_i$ satisfy the normalization condition $\sum_i \Tr w_i = d$. The number of poles $N_L$ should be chosen such that details of the Green's function are well resolved, which is typically of the order of a few hundreds. The Green's function can be equivalently represented as the resolvent of an arrowhead matrix
\begin{equation}\label{eq:ha}
  H_A = \left( \begin{array}{ccccc} u_0    & v_1^\dag    & v_2^\dag      & v_3^\dag     & \cdots \\
    v_1    & u_1    & 0      & 0      & \cdots \\
    v_2    & 0      & u_2    & 0      & \cdots \\
    v_3    & 0      & 0      & u_3    & \cdots \\
    \vdots & \vdots & \vdots & \vdots & \ddots \end{array} \right),
\end{equation}
with matrix elements $u_i$ and $v_i$ ($i \leq N$) of dimension $d\!\times\!d$. This leads to the expression
\begin{equation}\label{eq:ga}
    G_A(z) = (z - H_A)^{-1}_d = \frac{1}{z-u_0-\sum_{i=1}^N v_i^\dag\frac{1}{z-u_i}v_i},
\end{equation}
where the subscript $d$ denotes the leading $d\!\times\!d$ block. For numerical convenience, the summation in Eq.~\eqref{eq:ga} is sometimes equivalently expressed as a list of poles $\sum_{j=1}^{dN} \frac{w^u_j}{z-\epsilon^u_j}$ by eigendecomposition of the matrices $u_i$.

One can alternatively transform $H_A$ into a block tridiagonal form using the block Lanczos or Householder method,
\begin{equation}\label{eq:ht}
  H_T = \left( \begin{array}{ccccc} a_0    & b_1^\dag    & 0      & 0      & \cdots \\
    b_1    & a_1    & b_2^\dag    & 0      & \cdots \\
    0      & b_2    & a_2    & b_3^\dag    & \cdots \\
    0      & 0      & b_3    & a_3    & \cdots \\
    \vdots & \vdots & \vdots & \vdots & \ddots \end{array} \right),
\end{equation}
with the Green's function given as
\begin{equation}\label{eq:gt}
  G_T(z) = (z - H_T)^{-1}_d = \cfrac{1}{z-a_0-b_1^\dag \cfrac{1}{z-a_1-b_2^\dag \cfrac{1}{\cdots}b_2}b_1}.
\end{equation}

We dub the above different forms of the Green's function as L, A, and T forms, respectively. Following Ref.~\cite{Lu2014}, the transformation between these different forms can be formulated using the numerically exact procedures detailed below.

%
%

\subsection{From T, A forms to L form}\label{app:gf:toL}

This transformation can be performed by computing the eigensystem of $H_T$ or $H_A$. $G_L(z)$ is then given by identifying $\epsilon_i$ as the eigenvalues and $w_{i} = \phi_{i} \phi^\dag_{i}$ with $\phi_i$ the first $d$ coefficients of the corresponding right eigenvectors. The total number of poles, $N_L=d(N+1)$, which could be reduced by merging poles with degenerate or nearly degenerate $\epsilon_i$.

\subsection{From T-form to A-form}\label{app:gf:TtoA}

The equivalence between $G_T(z)$ \eqref{eq:gt} and $G_A(z)$ \eqref{eq:ga} requires $a_0=u_0$ and
\begin{equation*}
  b_1^\dag \cfrac{1}{z-a_1-b_2^\dag \cfrac{1}{\cdots}b_2}b_1 = \sum_{i=1}^N v_i^\dag\frac{1}{z-u_i}v_i.
\end{equation*}
The continued fraction on the left hand side can be viewed as a Green's function. Following Appendix~\ref{app:gf:toL}, it can be expressed as a list of poles
\begin{equation*}
  b_1^\dag \cfrac{1}{z-a_1-b_2^\dag\cfrac{1}{\cdots}b_2} b_1 = b_1^\dag \left( \sum_{j=1}^{dN} \frac{\tilde \phi_j \tilde \phi_j^\dag}{z-\tilde \epsilon_j} \right) b_1,
\end{equation*}
where similarly, $\tilde \epsilon_j$ and $\tilde \phi_j$ are the eigenvalues and first $d$ elements of the corresponding eigenvectors of the submatrix $H_T(0;0)$. Comparing the two equations above, we can then identify
\begin{equation*}
  \begin{split}
    u_i & = \mathrm{diag}\{\tilde \epsilon_{d (i-1)}, \tilde \epsilon_{d (i-1)+1}, \dots, \tilde \epsilon_{d i-1}\} \\
    v_i & = \{\tilde \phi_{d(i-1)}, \tilde \phi_{d (i-1)+1}, \dots, \tilde \phi_{d i-1}\}^\dag b_1.
  \end{split}
\end{equation*}
Note that in a DMFT loop, when the bath Green's function is expressed in the A-form, the above $u_i$ and $v_i$ matrices directly determine the bath parameters for the star geometry (see Sec.~\ref{sec:intro}). For a given number of bath sites $N_b$, it is then desirable to first merge the $dN$ poles into $N_b$ poles
\begin{equation*}
  b_1^\dag \left( \sum_{j=1}^{dN} \frac{\tilde \phi_j \tilde \phi_j^\dag}{z-\tilde \epsilon_j} \right) b_1 \longrightarrow \sum_{i=1}^{N_b} \frac{\tilde w'_i}{z-\tilde \epsilon'_i}.
\end{equation*}
The new set of matrices is then given by $u_i = \tilde \epsilon'_i \mathds{1}_d$, and $v_i$ by some decomposition of the positive-definite Hermitian matrices $\tilde w'_i$ such that $\tilde w'_i = v_i^\dag v_i$. This can be directly done by the Cholesky decomposition, in which case $v_i$ is an upper triangular matrix. Alternatively, one can also perform the eigendecomposition of $\tilde w'_i$ such that $\tilde w'_i=V_i \Lambda_i V^\dag_i$. Here $V_i$ is the square $d\!\times\!d$ matrix whose columns are the eigenvectors of $\tilde w'_i$, and $\Lambda_i$ is the diagonal matrix whose diagonal elements are the corresponding eigenvalues. In this case, one can define Hermitian matrices $v_i=V_i \sqrt{\Lambda_i}V^\dag_i$.

\subsection{From L-form to T-form}\label{app:gf:LtoT}

The transformation from Eq.~\eqref{eq:gl} to \eqref{eq:gt} equates to finding $a_i$ and $b_i$ that satisfy
\begin{equation*}
  \sum_{i=1}^{N_L} \frac{w_i}{z-\epsilon_i} = \cfrac{1}{z-a_0-b_1^\dag \cfrac{1}{z-a_1-b_2^\dag \cfrac{1}{\cdots}b_2}b_1}.
\end{equation*}
We again find a decomposition $w_i = v_i^\dag v_i$. The above equation can then be equivalently expressed as
\begin{equation*}
\cfrac{1}{z-\sum_i v_i^\dag\frac{1}{z-\epsilon_i}v_i} = \cfrac{1}{z - \cfrac{1}{z-a_0-b_1^\dag\cfrac{1}{z-a_1-b_2^\dag\cfrac{1}{\cdots}b_2}b_1}}.
\end{equation*}
The left hand side has the form of $G_A(z)$, and $a_i$ and $b_i$ can then be found by tridiagonalizing the corresponding matrix $H_A$.


\bibliographystyle{apsrev4-2}
\bibliography{multi_solver.bib}

\begin{thebibliography}{102}%
\makeatletter
\providecommand \@ifxundefined [1]{%
 \@ifx{#1\undefined}
}%
\providecommand \@ifnum [1]{%
 \ifnum #1\expandafter \@firstoftwo
 \else \expandafter \@secondoftwo
 \fi
}%
\providecommand \@ifx [1]{%
 \ifx #1\expandafter \@firstoftwo
 \else \expandafter \@secondoftwo
 \fi
}%
\providecommand \natexlab [1]{#1}%
\providecommand \enquote  [1]{``#1''}%
\providecommand \bibnamefont  [1]{#1}%
\providecommand \bibfnamefont [1]{#1}%
\providecommand \citenamefont [1]{#1}%
\providecommand \href@noop [0]{\@secondoftwo}%
\providecommand \href [0]{\begingroup \@sanitize@url \@href}%
\providecommand \@href[1]{\@@startlink{#1}\@@href}%
\providecommand \@@href[1]{\endgroup#1\@@endlink}%
\providecommand \@sanitize@url [0]{\catcode `\\12\catcode `\$12\catcode
  `\&12\catcode `\#12\catcode `\^12\catcode `\_12\catcode `\%12\relax}%
\providecommand \@@startlink[1]{}%
\providecommand \@@endlink[0]{}%
\providecommand \url  [0]{\begingroup\@sanitize@url \@url }%
\providecommand \@url [1]{\endgroup\@href {#1}{\urlprefix }}%
\providecommand \urlprefix  [0]{URL }%
\providecommand \Eprint [0]{\href }%
\providecommand \doibase [0]{https://doi.org/}%
\providecommand \selectlanguage [0]{\@gobble}%
\providecommand \bibinfo  [0]{\@secondoftwo}%
\providecommand \bibfield  [0]{\@secondoftwo}%
\providecommand \translation [1]{[#1]}%
\providecommand \BibitemOpen [0]{}%
\providecommand \bibitemStop [0]{}%
\providecommand \bibitemNoStop [0]{.\EOS\space}%
\providecommand \EOS [0]{\spacefactor3000\relax}%
\providecommand \BibitemShut  [1]{\csname bibitem#1\endcsname}%
\let\auto@bib@innerbib\@empty
\bibitem [{\citenamefont {Stewart}(1984)}]{Stewart1984}%
  \BibitemOpen
  \bibfield  {author} {\bibinfo {author} {\bibfnamefont {G.~R.}\ \bibnamefont
  {Stewart}},\ }\href {https://doi.org/10.1103/RevModPhys.56.755} {\bibfield
  {journal} {\bibinfo  {journal} {Rev. Mod. Phys.}\ }\textbf {\bibinfo {volume}
  {56}},\ \bibinfo {pages} {755} (\bibinfo {year} {1984})}\BibitemShut
  {NoStop}%
\bibitem [{\citenamefont {Hewson}(1997)}]{Hewson1997}%
  \BibitemOpen
  \bibfield  {author} {\bibinfo {author} {\bibfnamefont {A.~C.}\ \bibnamefont
  {Hewson}},\ }\href@noop {} {\emph {\bibinfo {title} {The Kondo Problem to
  Heavy Fermions}}},\ Vol.~\bibinfo {volume} {2}\ (\bibinfo  {publisher}
  {Cambridge University Press, Cambridge},\ \bibinfo {year} {1997})\BibitemShut
  {NoStop}%
\bibitem [{\citenamefont {Imada}\ \emph {et~al.}(1998)\citenamefont {Imada},
  \citenamefont {Fujimori},\ and\ \citenamefont {Tokura}}]{Imada1998}%
  \BibitemOpen
  \bibfield  {author} {\bibinfo {author} {\bibfnamefont {M.}~\bibnamefont
  {Imada}}, \bibinfo {author} {\bibfnamefont {A.}~\bibnamefont {Fujimori}},\
  and\ \bibinfo {author} {\bibfnamefont {Y.}~\bibnamefont {Tokura}},\ }\href
  {https://doi.org/10.1103/RevModPhys.70.1039} {\bibfield  {journal} {\bibinfo
  {journal} {Rev. Mod. Phys.}\ }\textbf {\bibinfo {volume} {70}},\ \bibinfo
  {pages} {1039} (\bibinfo {year} {1998})}\BibitemShut {NoStop}%
\bibitem [{\citenamefont {Lee}\ \emph {et~al.}(2006)\citenamefont {Lee},
  \citenamefont {Nagaosa},\ and\ \citenamefont {Wen}}]{Lee2006}%
  \BibitemOpen
  \bibfield  {author} {\bibinfo {author} {\bibfnamefont {P.~A.}\ \bibnamefont
  {Lee}}, \bibinfo {author} {\bibfnamefont {N.}~\bibnamefont {Nagaosa}},\ and\
  \bibinfo {author} {\bibfnamefont {X.-G.}\ \bibnamefont {Wen}},\ }\href
  {https://doi.org/10.1103/RevModPhys.78.17} {\bibfield  {journal} {\bibinfo
  {journal} {Rev. Mod. Phys.}\ }\textbf {\bibinfo {volume} {78}},\ \bibinfo
  {pages} {17} (\bibinfo {year} {2006})}\BibitemShut {NoStop}%
\bibitem [{\citenamefont {Keimer}\ \emph {et~al.}(2015)\citenamefont {Keimer},
  \citenamefont {Kivelson}, \citenamefont {Norman}, \citenamefont {Uchida},\
  and\ \citenamefont {Zaanen}}]{Keimer2015}%
  \BibitemOpen
  \bibfield  {author} {\bibinfo {author} {\bibfnamefont {B.}~\bibnamefont
  {Keimer}}, \bibinfo {author} {\bibfnamefont {S.~A.}\ \bibnamefont
  {Kivelson}}, \bibinfo {author} {\bibfnamefont {M.~R.}\ \bibnamefont
  {Norman}}, \bibinfo {author} {\bibfnamefont {S.}~\bibnamefont {Uchida}},\
  and\ \bibinfo {author} {\bibfnamefont {J.}~\bibnamefont {Zaanen}},\ }\href
  {https://doi.org/10.1038/nature14165} {\bibfield  {journal} {\bibinfo
  {journal} {Nature}\ }\textbf {\bibinfo {volume} {518}},\ \bibinfo {pages}
  {179} (\bibinfo {year} {2015})}\BibitemShut {NoStop}%
\bibitem [{\citenamefont {Metzner}\ and\ \citenamefont
  {Vollhardt}(1989)}]{Metzner1989}%
  \BibitemOpen
  \bibfield  {author} {\bibinfo {author} {\bibfnamefont {W.}~\bibnamefont
  {Metzner}}\ and\ \bibinfo {author} {\bibfnamefont {D.}~\bibnamefont
  {Vollhardt}},\ }\href {https://doi.org/10.1103/PhysRevLett.62.324} {\bibfield
   {journal} {\bibinfo  {journal} {Phys. Rev. Lett.}\ }\textbf {\bibinfo
  {volume} {62}},\ \bibinfo {pages} {324} (\bibinfo {year} {1989})}\BibitemShut
  {NoStop}%
\bibitem [{\citenamefont {Georges}\ \emph {et~al.}(1996)\citenamefont
  {Georges}, \citenamefont {Kotliar}, \citenamefont {Krauth},\ and\
  \citenamefont {Rozenberg}}]{Georges1996}%
  \BibitemOpen
  \bibfield  {author} {\bibinfo {author} {\bibfnamefont {A.}~\bibnamefont
  {Georges}}, \bibinfo {author} {\bibfnamefont {G.}~\bibnamefont {Kotliar}},
  \bibinfo {author} {\bibfnamefont {W.}~\bibnamefont {Krauth}},\ and\ \bibinfo
  {author} {\bibfnamefont {M.~J.}\ \bibnamefont {Rozenberg}},\ }\href
  {https://doi.org/10.1103/RevModPhys.68.13} {\bibfield  {journal} {\bibinfo
  {journal} {Rev. Mod. Phys.}\ }\textbf {\bibinfo {volume} {68}},\ \bibinfo
  {pages} {13} (\bibinfo {year} {1996})}\BibitemShut {NoStop}%
\bibitem [{\citenamefont {Kotliar}\ and\ \citenamefont
  {Vollhardt}(2004)}]{Kotliar2004}%
  \BibitemOpen
  \bibfield  {author} {\bibinfo {author} {\bibfnamefont {G.}~\bibnamefont
  {Kotliar}}\ and\ \bibinfo {author} {\bibfnamefont {D.}~\bibnamefont
  {Vollhardt}},\ }\href {https://doi.org/10.1063/1.1712502} {\bibfield
  {journal} {\bibinfo  {journal} {Phys. Today}\ }\textbf {\bibinfo {volume}
  {57}},\ \bibinfo {pages} {53} (\bibinfo {year} {2004})}\BibitemShut {NoStop}%
\bibitem [{\citenamefont {Georges}\ \emph {et~al.}(2013)\citenamefont
  {Georges}, \citenamefont {Medici},\ and\ \citenamefont
  {Mravlje}}]{Georges2013}%
  \BibitemOpen
  \bibfield  {author} {\bibinfo {author} {\bibfnamefont {A.}~\bibnamefont
  {Georges}}, \bibinfo {author} {\bibfnamefont {L.~d.}\ \bibnamefont
  {Medici}},\ and\ \bibinfo {author} {\bibfnamefont {J.}~\bibnamefont
  {Mravlje}},\ }\href
  {https://doi.org/10.1146/annurev-conmatphys-020911-125045} {\bibfield
  {journal} {\bibinfo  {journal} {Annu. Rev. Condens. Matter Phys.}\ }\textbf
  {\bibinfo {volume} {4}},\ \bibinfo {pages} {137} (\bibinfo {year}
  {2013})}\BibitemShut {NoStop}%
\bibitem [{\citenamefont {Damascelli}\ \emph {et~al.}(2003)\citenamefont
  {Damascelli}, \citenamefont {Hussain},\ and\ \citenamefont
  {Shen}}]{Damascelli2003}%
  \BibitemOpen
  \bibfield  {author} {\bibinfo {author} {\bibfnamefont {A.}~\bibnamefont
  {Damascelli}}, \bibinfo {author} {\bibfnamefont {Z.}~\bibnamefont
  {Hussain}},\ and\ \bibinfo {author} {\bibfnamefont {Z.-X.}\ \bibnamefont
  {Shen}},\ }\href {https://doi.org/10.1103/RevModPhys.75.473} {\bibfield
  {journal} {\bibinfo  {journal} {Rev. Mod. Phys.}\ }\textbf {\bibinfo {volume}
  {75}},\ \bibinfo {pages} {473} (\bibinfo {year} {2003})}\BibitemShut
  {NoStop}%
\bibitem [{\citenamefont {De~Groot}\ and\ \citenamefont
  {Kotani}(2008)}]{deGroot2008}%
  \BibitemOpen
  \bibfield  {author} {\bibinfo {author} {\bibfnamefont {F.}~\bibnamefont
  {De~Groot}}\ and\ \bibinfo {author} {\bibfnamefont {A.}~\bibnamefont
  {Kotani}},\ }\href@noop {} {\emph {\bibinfo {title} {Core Level Spectroscopy
  of Solids}}}\ (\bibinfo  {publisher} {CRC Press, Boca Raton, FL},\ \bibinfo
  {year} {2008})\BibitemShut {NoStop}%
\bibitem [{\citenamefont {Bocquet}\ \emph {et~al.}(1992)\citenamefont
  {Bocquet}, \citenamefont {Mizokawa}, \citenamefont {Saitoh}, \citenamefont
  {Namatame},\ and\ \citenamefont {Fujimori}}]{Bocquet1992}%
  \BibitemOpen
  \bibfield  {author} {\bibinfo {author} {\bibfnamefont {A.~E.}\ \bibnamefont
  {Bocquet}}, \bibinfo {author} {\bibfnamefont {T.}~\bibnamefont {Mizokawa}},
  \bibinfo {author} {\bibfnamefont {T.}~\bibnamefont {Saitoh}}, \bibinfo
  {author} {\bibfnamefont {H.}~\bibnamefont {Namatame}},\ and\ \bibinfo
  {author} {\bibfnamefont {A.}~\bibnamefont {Fujimori}},\ }\href
  {https://doi.org/10.1103/PhysRevB.46.3771} {\bibfield  {journal} {\bibinfo
  {journal} {Phys. Rev. B}\ }\textbf {\bibinfo {volume} {46}},\ \bibinfo
  {pages} {3771} (\bibinfo {year} {1992})}\BibitemShut {NoStop}%
\bibitem [{\citenamefont {Zaanen}\ \emph {et~al.}(1985)\citenamefont {Zaanen},
  \citenamefont {Sawatzky},\ and\ \citenamefont {Allen}}]{Zaanen1985}%
  \BibitemOpen
  \bibfield  {author} {\bibinfo {author} {\bibfnamefont {J.}~\bibnamefont
  {Zaanen}}, \bibinfo {author} {\bibfnamefont {G.~A.}\ \bibnamefont
  {Sawatzky}},\ and\ \bibinfo {author} {\bibfnamefont {J.~W.}\ \bibnamefont
  {Allen}},\ }\href {https://doi.org/10.1103/PhysRevLett.55.418} {\bibfield
  {journal} {\bibinfo  {journal} {Phys. Rev. Lett.}\ }\textbf {\bibinfo
  {volume} {55}},\ \bibinfo {pages} {418} (\bibinfo {year} {1985})}\BibitemShut
  {NoStop}%
\bibitem [{\citenamefont {Ament}\ \emph {et~al.}(2011)\citenamefont {Ament},
  \citenamefont {van Veenendaal}, \citenamefont {Devereaux}, \citenamefont
  {Hill},\ and\ \citenamefont {van~den Brink}}]{Ament2011}%
  \BibitemOpen
  \bibfield  {author} {\bibinfo {author} {\bibfnamefont {L.~J.~P.}\
  \bibnamefont {Ament}}, \bibinfo {author} {\bibfnamefont {M.}~\bibnamefont
  {van Veenendaal}}, \bibinfo {author} {\bibfnamefont {T.~P.}\ \bibnamefont
  {Devereaux}}, \bibinfo {author} {\bibfnamefont {J.~P.}\ \bibnamefont
  {Hill}},\ and\ \bibinfo {author} {\bibfnamefont {J.}~\bibnamefont {van~den
  Brink}},\ }\href {https://doi.org/10.1103/RevModPhys.83.705} {\bibfield
  {journal} {\bibinfo  {journal} {Rev. Mod. Phys.}\ }\textbf {\bibinfo {volume}
  {83}},\ \bibinfo {pages} {705} (\bibinfo {year} {2011})}\BibitemShut
  {NoStop}%
\bibitem [{\citenamefont {Gull}\ \emph {et~al.}(2011)\citenamefont {Gull},
  \citenamefont {Millis}, \citenamefont {Lichtenstein}, \citenamefont
  {Rubtsov}, \citenamefont {Troyer},\ and\ \citenamefont {Werner}}]{Gull2011}%
  \BibitemOpen
  \bibfield  {author} {\bibinfo {author} {\bibfnamefont {E.}~\bibnamefont
  {Gull}}, \bibinfo {author} {\bibfnamefont {A.~J.}\ \bibnamefont {Millis}},
  \bibinfo {author} {\bibfnamefont {A.~I.}\ \bibnamefont {Lichtenstein}},
  \bibinfo {author} {\bibfnamefont {A.~N.}\ \bibnamefont {Rubtsov}}, \bibinfo
  {author} {\bibfnamefont {M.}~\bibnamefont {Troyer}},\ and\ \bibinfo {author}
  {\bibfnamefont {P.}~\bibnamefont {Werner}},\ }\href
  {https://doi.org/10.1103/RevModPhys.83.349} {\bibfield  {journal} {\bibinfo
  {journal} {Rev. Mod. Phys.}\ }\textbf {\bibinfo {volume} {83}},\ \bibinfo
  {pages} {349} (\bibinfo {year} {2011})}\BibitemShut {NoStop}%
\bibitem [{\citenamefont {Georges}\ and\ \citenamefont
  {Krauth}(1992)}]{Georges1992}%
  \BibitemOpen
  \bibfield  {author} {\bibinfo {author} {\bibfnamefont {A.}~\bibnamefont
  {Georges}}\ and\ \bibinfo {author} {\bibfnamefont {W.}~\bibnamefont
  {Krauth}},\ }\href {https://doi.org/10.1103/PhysRevLett.69.1240} {\bibfield
  {journal} {\bibinfo  {journal} {Phys. Rev. Lett.}\ }\textbf {\bibinfo
  {volume} {69}},\ \bibinfo {pages} {1240} (\bibinfo {year}
  {1992})}\BibitemShut {NoStop}%
\bibitem [{\citenamefont {Ulmke}\ \emph {et~al.}(1995)\citenamefont {Ulmke},
  \citenamefont {Jani\ifmmode~\check{s}\else \v{s}\fi{}},\ and\ \citenamefont
  {Vollhardt}}]{Ulmke1995}%
  \BibitemOpen
  \bibfield  {author} {\bibinfo {author} {\bibfnamefont {M.}~\bibnamefont
  {Ulmke}}, \bibinfo {author} {\bibfnamefont {V.}~\bibnamefont
  {Jani\ifmmode~\check{s}\else \v{s}\fi{}}},\ and\ \bibinfo {author}
  {\bibfnamefont {D.}~\bibnamefont {Vollhardt}},\ }\href
  {https://doi.org/10.1103/PhysRevB.51.10411} {\bibfield  {journal} {\bibinfo
  {journal} {Phys. Rev. B}\ }\textbf {\bibinfo {volume} {51}},\ \bibinfo
  {pages} {10411} (\bibinfo {year} {1995})}\BibitemShut {NoStop}%
\bibitem [{\citenamefont {Rubtsov}\ \emph {et~al.}(2005)\citenamefont
  {Rubtsov}, \citenamefont {Savkin},\ and\ \citenamefont
  {Lichtenstein}}]{Rubtsov2005}%
  \BibitemOpen
  \bibfield  {author} {\bibinfo {author} {\bibfnamefont {A.~N.}\ \bibnamefont
  {Rubtsov}}, \bibinfo {author} {\bibfnamefont {V.~V.}\ \bibnamefont
  {Savkin}},\ and\ \bibinfo {author} {\bibfnamefont {A.~I.}\ \bibnamefont
  {Lichtenstein}},\ }\href {https://doi.org/10.1103/PhysRevB.72.035122}
  {\bibfield  {journal} {\bibinfo  {journal} {Phys. Rev. B}\ }\textbf {\bibinfo
  {volume} {72}},\ \bibinfo {pages} {035122} (\bibinfo {year}
  {2005})}\BibitemShut {NoStop}%
\bibitem [{\citenamefont {Werner}\ \emph {et~al.}(2006)\citenamefont {Werner},
  \citenamefont {Comanac}, \citenamefont {de' Medici}, \citenamefont {Troyer},\
  and\ \citenamefont {Millis}}]{Werner2006}%
  \BibitemOpen
  \bibfield  {author} {\bibinfo {author} {\bibfnamefont {P.}~\bibnamefont
  {Werner}}, \bibinfo {author} {\bibfnamefont {A.}~\bibnamefont {Comanac}},
  \bibinfo {author} {\bibfnamefont {L.}~\bibnamefont {de' Medici}}, \bibinfo
  {author} {\bibfnamefont {M.}~\bibnamefont {Troyer}},\ and\ \bibinfo {author}
  {\bibfnamefont {A.~J.}\ \bibnamefont {Millis}},\ }\href
  {https://doi.org/10.1103/PhysRevLett.97.076405} {\bibfield  {journal}
  {\bibinfo  {journal} {Phys. Rev. Lett.}\ }\textbf {\bibinfo {volume} {97}},\
  \bibinfo {pages} {076405} (\bibinfo {year} {2006})}\BibitemShut {NoStop}%
\bibitem [{\citenamefont {Werner}\ and\ \citenamefont
  {Millis}(2006)}]{Werner2006b}%
  \BibitemOpen
  \bibfield  {author} {\bibinfo {author} {\bibfnamefont {P.}~\bibnamefont
  {Werner}}\ and\ \bibinfo {author} {\bibfnamefont {A.~J.}\ \bibnamefont
  {Millis}},\ }\href {https://doi.org/10.1103/PhysRevB.74.155107} {\bibfield
  {journal} {\bibinfo  {journal} {Phys. Rev. B}\ }\textbf {\bibinfo {volume}
  {74}},\ \bibinfo {pages} {155107} (\bibinfo {year} {2006})}\BibitemShut
  {NoStop}%
\bibitem [{\citenamefont {Kotliar}\ \emph {et~al.}(2001)\citenamefont
  {Kotliar}, \citenamefont {Savrasov}, \citenamefont {P\'alsson},\ and\
  \citenamefont {Biroli}}]{Kotliar2001}%
  \BibitemOpen
  \bibfield  {author} {\bibinfo {author} {\bibfnamefont {G.}~\bibnamefont
  {Kotliar}}, \bibinfo {author} {\bibfnamefont {S.~Y.}\ \bibnamefont
  {Savrasov}}, \bibinfo {author} {\bibfnamefont {G.}~\bibnamefont
  {P\'alsson}},\ and\ \bibinfo {author} {\bibfnamefont {G.}~\bibnamefont
  {Biroli}},\ }\href {https://doi.org/10.1103/PhysRevLett.87.186401} {\bibfield
   {journal} {\bibinfo  {journal} {Phys. Rev. Lett.}\ }\textbf {\bibinfo
  {volume} {87}},\ \bibinfo {pages} {186401} (\bibinfo {year}
  {2001})}\BibitemShut {NoStop}%
\bibitem [{\citenamefont {Jarrell}\ and\ \citenamefont
  {Gubernatis}(1996)}]{Jarrell1996}%
  \BibitemOpen
  \bibfield  {author} {\bibinfo {author} {\bibfnamefont {M.}~\bibnamefont
  {Jarrell}}\ and\ \bibinfo {author} {\bibfnamefont {J.~E.}\ \bibnamefont
  {Gubernatis}},\ }\href@noop {} {\bibfield  {journal} {\bibinfo  {journal}
  {Phys. Rep.}\ }\textbf {\bibinfo {volume} {269}},\ \bibinfo {pages} {133}
  (\bibinfo {year} {1996})}\BibitemShut {NoStop}%
\bibitem [{\citenamefont {Lu}\ and\ \citenamefont {Haverkort}(2017)}]{Lu2017}%
  \BibitemOpen
  \bibfield  {author} {\bibinfo {author} {\bibfnamefont {Y.}~\bibnamefont
  {Lu}}\ and\ \bibinfo {author} {\bibfnamefont {M.~W.}\ \bibnamefont
  {Haverkort}},\ }\href {https://doi.org/10.1140/epjst/e2017-70042-4}
  {\bibfield  {journal} {\bibinfo  {journal} {Eur. Phys. J. Spec. Top.}\
  }\textbf {\bibinfo {volume} {226}},\ \bibinfo {pages} {2549} (\bibinfo {year}
  {2017})}\BibitemShut {NoStop}%
\bibitem [{\citenamefont {Caffarel}\ and\ \citenamefont
  {Krauth}(1994)}]{Caffarel1994}%
  \BibitemOpen
  \bibfield  {author} {\bibinfo {author} {\bibfnamefont {M.}~\bibnamefont
  {Caffarel}}\ and\ \bibinfo {author} {\bibfnamefont {W.}~\bibnamefont
  {Krauth}},\ }\href {https://doi.org/10.1103/PhysRevLett.72.1545} {\bibfield
  {journal} {\bibinfo  {journal} {Phys. Rev. Lett.}\ }\textbf {\bibinfo
  {volume} {72}},\ \bibinfo {pages} {1545} (\bibinfo {year}
  {1994})}\BibitemShut {NoStop}%
\bibitem [{\citenamefont {Sangiovanni}\ \emph {et~al.}(2006)\citenamefont
  {Sangiovanni}, \citenamefont {Toschi}, \citenamefont {Koch}, \citenamefont
  {Held}, \citenamefont {Capone}, \citenamefont {Castellani}, \citenamefont
  {Gunnarsson}, \citenamefont {Mo}, \citenamefont {Allen}, \citenamefont {Kim},
  \citenamefont {Sekiyama}, \citenamefont {Yamasaki}, \citenamefont {Suga},\
  and\ \citenamefont {Metcalf}}]{Sangiovanni2006}%
  \BibitemOpen
  \bibfield  {author} {\bibinfo {author} {\bibfnamefont {G.}~\bibnamefont
  {Sangiovanni}}, \bibinfo {author} {\bibfnamefont {A.}~\bibnamefont {Toschi}},
  \bibinfo {author} {\bibfnamefont {E.}~\bibnamefont {Koch}}, \bibinfo {author}
  {\bibfnamefont {K.}~\bibnamefont {Held}}, \bibinfo {author} {\bibfnamefont
  {M.}~\bibnamefont {Capone}}, \bibinfo {author} {\bibfnamefont
  {C.}~\bibnamefont {Castellani}}, \bibinfo {author} {\bibfnamefont
  {O.}~\bibnamefont {Gunnarsson}}, \bibinfo {author} {\bibfnamefont {S.-K.}\
  \bibnamefont {Mo}}, \bibinfo {author} {\bibfnamefont {J.~W.}\ \bibnamefont
  {Allen}}, \bibinfo {author} {\bibfnamefont {H.-D.}\ \bibnamefont {Kim}},
  \bibinfo {author} {\bibfnamefont {A.}~\bibnamefont {Sekiyama}}, \bibinfo
  {author} {\bibfnamefont {A.}~\bibnamefont {Yamasaki}}, \bibinfo {author}
  {\bibfnamefont {S.}~\bibnamefont {Suga}},\ and\ \bibinfo {author}
  {\bibfnamefont {P.}~\bibnamefont {Metcalf}},\ }\href
  {https://doi.org/10.1103/PhysRevB.73.205121} {\bibfield  {journal} {\bibinfo
  {journal} {Phys. Rev. B}\ }\textbf {\bibinfo {volume} {73}},\ \bibinfo
  {pages} {205121} (\bibinfo {year} {2006})}\BibitemShut {NoStop}%
\bibitem [{\citenamefont {Capone}\ \emph {et~al.}(2007)\citenamefont {Capone},
  \citenamefont {de' Medici},\ and\ \citenamefont {Georges}}]{Capone2007}%
  \BibitemOpen
  \bibfield  {author} {\bibinfo {author} {\bibfnamefont {M.}~\bibnamefont
  {Capone}}, \bibinfo {author} {\bibfnamefont {L.}~\bibnamefont {de' Medici}},\
  and\ \bibinfo {author} {\bibfnamefont {A.}~\bibnamefont {Georges}},\ }\href
  {https://doi.org/10.1103/PhysRevB.76.245116} {\bibfield  {journal} {\bibinfo
  {journal} {Phys. Rev. B}\ }\textbf {\bibinfo {volume} {76}},\ \bibinfo
  {pages} {245116} (\bibinfo {year} {2007})}\BibitemShut {NoStop}%
\bibitem [{\citenamefont {Koch}\ \emph {et~al.}(2008)\citenamefont {Koch},
  \citenamefont {Sangiovanni},\ and\ \citenamefont {Gunnarsson}}]{Koch2008}%
  \BibitemOpen
  \bibfield  {author} {\bibinfo {author} {\bibfnamefont {E.}~\bibnamefont
  {Koch}}, \bibinfo {author} {\bibfnamefont {G.}~\bibnamefont {Sangiovanni}},\
  and\ \bibinfo {author} {\bibfnamefont {O.}~\bibnamefont {Gunnarsson}},\
  }\href {https://doi.org/10.1103/PhysRevB.78.115102} {\bibfield  {journal}
  {\bibinfo  {journal} {Phys. Rev. B}\ }\textbf {\bibinfo {volume} {78}},\
  \bibinfo {pages} {115102} (\bibinfo {year} {2008})}\BibitemShut {NoStop}%
\bibitem [{\citenamefont {Zgid}\ \emph {et~al.}(2012)\citenamefont {Zgid},
  \citenamefont {Gull},\ and\ \citenamefont {Chan}}]{Zgid2012}%
  \BibitemOpen
  \bibfield  {author} {\bibinfo {author} {\bibfnamefont {D.}~\bibnamefont
  {Zgid}}, \bibinfo {author} {\bibfnamefont {E.}~\bibnamefont {Gull}},\ and\
  \bibinfo {author} {\bibfnamefont {G.~K.-L.}\ \bibnamefont {Chan}},\ }\href
  {https://doi.org/10.1103/PhysRevB.86.165128} {\bibfield  {journal} {\bibinfo
  {journal} {Phys. Rev. B}\ }\textbf {\bibinfo {volume} {86}},\ \bibinfo
  {pages} {165128} (\bibinfo {year} {2012})}\BibitemShut {NoStop}%
\bibitem [{\citenamefont {Lin}\ and\ \citenamefont {Demkov}(2013)}]{Lin2013}%
  \BibitemOpen
  \bibfield  {author} {\bibinfo {author} {\bibfnamefont {C.}~\bibnamefont
  {Lin}}\ and\ \bibinfo {author} {\bibfnamefont {A.~A.}\ \bibnamefont
  {Demkov}},\ }\href {https://doi.org/10.1103/PhysRevB.88.035123} {\bibfield
  {journal} {\bibinfo  {journal} {Phys. Rev. B}\ }\textbf {\bibinfo {volume}
  {88}},\ \bibinfo {pages} {035123} (\bibinfo {year} {2013})}\BibitemShut
  {NoStop}%
\bibitem [{\citenamefont {Lu}\ \emph {et~al.}(2014)\citenamefont {Lu},
  \citenamefont {H\"oppner}, \citenamefont {Gunnarsson},\ and\ \citenamefont
  {Haverkort}}]{Lu2014}%
  \BibitemOpen
  \bibfield  {author} {\bibinfo {author} {\bibfnamefont {Y.}~\bibnamefont
  {Lu}}, \bibinfo {author} {\bibfnamefont {M.}~\bibnamefont {H\"oppner}},
  \bibinfo {author} {\bibfnamefont {O.}~\bibnamefont {Gunnarsson}},\ and\
  \bibinfo {author} {\bibfnamefont {M.~W.}\ \bibnamefont {Haverkort}},\ }\href
  {https://doi.org/10.1103/PhysRevB.90.085102} {\bibfield  {journal} {\bibinfo
  {journal} {Phys. Rev. B}\ }\textbf {\bibinfo {volume} {90}},\ \bibinfo
  {pages} {085102} (\bibinfo {year} {2014})}\BibitemShut {NoStop}%
\bibitem [{\citenamefont {Haverkort}\ \emph {et~al.}(2012)\citenamefont
  {Haverkort}, \citenamefont {Zwierzycki},\ and\ \citenamefont
  {Andersen}}]{Haverkort2012}%
  \BibitemOpen
  \bibfield  {author} {\bibinfo {author} {\bibfnamefont {M.~W.}\ \bibnamefont
  {Haverkort}}, \bibinfo {author} {\bibfnamefont {M.}~\bibnamefont
  {Zwierzycki}},\ and\ \bibinfo {author} {\bibfnamefont {O.~K.}\ \bibnamefont
  {Andersen}},\ }\href {https://doi.org/10.1103/PhysRevB.85.165113} {\bibfield
  {journal} {\bibinfo  {journal} {Phys. Rev. B}\ }\textbf {\bibinfo {volume}
  {85}},\ \bibinfo {pages} {165113} (\bibinfo {year} {2012})}\BibitemShut
  {NoStop}%
\bibitem [{\citenamefont {Haverkort}\ \emph {et~al.}(2014)\citenamefont
  {Haverkort}, \citenamefont {Sangiovanni}, \citenamefont {Hansmann},
  \citenamefont {Toschi}, \citenamefont {Lu},\ and\ \citenamefont
  {Macke}}]{Haverkort2014}%
  \BibitemOpen
  \bibfield  {author} {\bibinfo {author} {\bibfnamefont {M.~W.}\ \bibnamefont
  {Haverkort}}, \bibinfo {author} {\bibfnamefont {G.}~\bibnamefont
  {Sangiovanni}}, \bibinfo {author} {\bibfnamefont {P.}~\bibnamefont
  {Hansmann}}, \bibinfo {author} {\bibfnamefont {A.}~\bibnamefont {Toschi}},
  \bibinfo {author} {\bibfnamefont {Y.}~\bibnamefont {Lu}},\ and\ \bibinfo
  {author} {\bibfnamefont {S.}~\bibnamefont {Macke}},\ }\href
  {https://doi.org/10.1209/0295-5075/108/57004} {\bibfield  {journal} {\bibinfo
   {journal} {EPL}\ }\textbf {\bibinfo {volume} {108}},\ \bibinfo {pages}
  {57004} (\bibinfo {year} {2014})}\BibitemShut {NoStop}%
\bibitem [{\citenamefont {Hariki}\ \emph {et~al.}(2017)\citenamefont {Hariki},
  \citenamefont {Uozumi},\ and\ \citenamefont {Kune\ifmmode~\check{s}\else
  \v{s}\fi{}}}]{Hariki2017}%
  \BibitemOpen
  \bibfield  {author} {\bibinfo {author} {\bibfnamefont {A.}~\bibnamefont
  {Hariki}}, \bibinfo {author} {\bibfnamefont {T.}~\bibnamefont {Uozumi}},\
  and\ \bibinfo {author} {\bibfnamefont {J.}~\bibnamefont
  {Kune\ifmmode~\check{s}\else \v{s}\fi{}}},\ }\href
  {https://doi.org/10.1103/PhysRevB.96.045111} {\bibfield  {journal} {\bibinfo
  {journal} {Phys. Rev. B}\ }\textbf {\bibinfo {volume} {96}},\ \bibinfo
  {pages} {045111} (\bibinfo {year} {2017})}\BibitemShut {NoStop}%
\bibitem [{\citenamefont {Hariki}\ \emph {et~al.}(2020)\citenamefont {Hariki},
  \citenamefont {Winder}, \citenamefont {Uozumi},\ and\ \citenamefont
  {Kune\ifmmode~\check{s}\else \v{s}\fi{}}}]{Hariki2020}%
  \BibitemOpen
  \bibfield  {author} {\bibinfo {author} {\bibfnamefont {A.}~\bibnamefont
  {Hariki}}, \bibinfo {author} {\bibfnamefont {M.}~\bibnamefont {Winder}},
  \bibinfo {author} {\bibfnamefont {T.}~\bibnamefont {Uozumi}},\ and\ \bibinfo
  {author} {\bibfnamefont {J.}~\bibnamefont {Kune\ifmmode~\check{s}\else
  \v{s}\fi{}}},\ }\href {https://doi.org/10.1103/PhysRevB.101.115130}
  {\bibfield  {journal} {\bibinfo  {journal} {Phys. Rev. B}\ }\textbf {\bibinfo
  {volume} {101}},\ \bibinfo {pages} {115130} (\bibinfo {year}
  {2020})}\BibitemShut {NoStop}%
\bibitem [{\citenamefont {Wilson}(1975)}]{Wilson1975}%
  \BibitemOpen
  \bibfield  {author} {\bibinfo {author} {\bibfnamefont {K.~G.}\ \bibnamefont
  {Wilson}},\ }\href {https://doi.org/10.1103/RevModPhys.47.773} {\bibfield
  {journal} {\bibinfo  {journal} {Rev. Mod. Phys.}\ }\textbf {\bibinfo {volume}
  {47}},\ \bibinfo {pages} {773} (\bibinfo {year} {1975})}\BibitemShut
  {NoStop}%
\bibitem [{\citenamefont {Bulla}\ \emph {et~al.}(2008)\citenamefont {Bulla},
  \citenamefont {Costi},\ and\ \citenamefont {Pruschke}}]{Bulla2008}%
  \BibitemOpen
  \bibfield  {author} {\bibinfo {author} {\bibfnamefont {R.}~\bibnamefont
  {Bulla}}, \bibinfo {author} {\bibfnamefont {T.~A.}\ \bibnamefont {Costi}},\
  and\ \bibinfo {author} {\bibfnamefont {T.}~\bibnamefont {Pruschke}},\ }\href
  {https://doi.org/10.1103/RevModPhys.80.395} {\bibfield  {journal} {\bibinfo
  {journal} {Rev. Mod. Phys.}\ }\textbf {\bibinfo {volume} {80}},\ \bibinfo
  {pages} {395} (\bibinfo {year} {2008})}\BibitemShut {NoStop}%
\bibitem [{\citenamefont {Bulla}(1999)}]{Bulla1999}%
  \BibitemOpen
  \bibfield  {author} {\bibinfo {author} {\bibfnamefont {R.}~\bibnamefont
  {Bulla}},\ }\href {https://doi.org/10.1103/PhysRevLett.83.136} {\bibfield
  {journal} {\bibinfo  {journal} {Phys. Rev. Lett.}\ }\textbf {\bibinfo
  {volume} {83}},\ \bibinfo {pages} {136} (\bibinfo {year} {1999})}\BibitemShut
  {NoStop}%
\bibitem [{\citenamefont {Bulla}\ \emph {et~al.}(2001)\citenamefont {Bulla},
  \citenamefont {Costi},\ and\ \citenamefont {Vollhardt}}]{Bulla2011}%
  \BibitemOpen
  \bibfield  {author} {\bibinfo {author} {\bibfnamefont {R.}~\bibnamefont
  {Bulla}}, \bibinfo {author} {\bibfnamefont {T.~A.}\ \bibnamefont {Costi}},\
  and\ \bibinfo {author} {\bibfnamefont {D.}~\bibnamefont {Vollhardt}},\ }\href
  {https://doi.org/10.1103/PhysRevB.64.045103} {\bibfield  {journal} {\bibinfo
  {journal} {Phys. Rev. B}\ }\textbf {\bibinfo {volume} {64}},\ \bibinfo
  {pages} {045103} (\bibinfo {year} {2001})}\BibitemShut {NoStop}%
\bibitem [{\citenamefont {Bulla}\ \emph {et~al.}(2005)\citenamefont {Bulla},
  \citenamefont {Lee}, \citenamefont {Tong},\ and\ \citenamefont
  {Vojta}}]{Bulla2005}%
  \BibitemOpen
  \bibfield  {author} {\bibinfo {author} {\bibfnamefont {R.}~\bibnamefont
  {Bulla}}, \bibinfo {author} {\bibfnamefont {H.-J.}\ \bibnamefont {Lee}},
  \bibinfo {author} {\bibfnamefont {N.-H.}\ \bibnamefont {Tong}},\ and\
  \bibinfo {author} {\bibfnamefont {M.}~\bibnamefont {Vojta}},\ }\href
  {https://doi.org/10.1103/PhysRevB.71.045122} {\bibfield  {journal} {\bibinfo
  {journal} {Phys. Rev. B}\ }\textbf {\bibinfo {volume} {71}},\ \bibinfo
  {pages} {045122} (\bibinfo {year} {2005})}\BibitemShut {NoStop}%
\bibitem [{\citenamefont {Pruschke}\ \emph {et~al.}(2000)\citenamefont
  {Pruschke}, \citenamefont {Bulla},\ and\ \citenamefont
  {Jarrell}}]{Pruschke2000}%
  \BibitemOpen
  \bibfield  {author} {\bibinfo {author} {\bibfnamefont {T.}~\bibnamefont
  {Pruschke}}, \bibinfo {author} {\bibfnamefont {R.}~\bibnamefont {Bulla}},\
  and\ \bibinfo {author} {\bibfnamefont {M.}~\bibnamefont {Jarrell}},\ }\href
  {https://doi.org/10.1103/PhysRevB.61.12799} {\bibfield  {journal} {\bibinfo
  {journal} {Phys. Rev. B}\ }\textbf {\bibinfo {volume} {61}},\ \bibinfo
  {pages} {12799} (\bibinfo {year} {2000})}\BibitemShut {NoStop}%
\bibitem [{\citenamefont {Stadler}\ \emph {et~al.}(2015)\citenamefont
  {Stadler}, \citenamefont {Yin}, \citenamefont {von Delft}, \citenamefont
  {Kotliar},\ and\ \citenamefont {Weichselbaum}}]{Stadler2015}%
  \BibitemOpen
  \bibfield  {author} {\bibinfo {author} {\bibfnamefont {K.~M.}\ \bibnamefont
  {Stadler}}, \bibinfo {author} {\bibfnamefont {Z.~P.}\ \bibnamefont {Yin}},
  \bibinfo {author} {\bibfnamefont {J.}~\bibnamefont {von Delft}}, \bibinfo
  {author} {\bibfnamefont {G.}~\bibnamefont {Kotliar}},\ and\ \bibinfo {author}
  {\bibfnamefont {A.}~\bibnamefont {Weichselbaum}},\ }\href
  {https://doi.org/10.1103/PhysRevLett.115.136401} {\bibfield  {journal}
  {\bibinfo  {journal} {Phys. Rev. Lett.}\ }\textbf {\bibinfo {volume} {115}},\
  \bibinfo {pages} {136401} (\bibinfo {year} {2015})}\BibitemShut {NoStop}%
\bibitem [{\citenamefont {Stadler}\ \emph {et~al.}(2019)\citenamefont
  {Stadler}, \citenamefont {Kotliar}, \citenamefont {Weichselbaum},\ and\
  \citenamefont {{von Delft}}}]{Stadler2019}%
  \BibitemOpen
  \bibfield  {author} {\bibinfo {author} {\bibfnamefont {K.}~\bibnamefont
  {Stadler}}, \bibinfo {author} {\bibfnamefont {G.}~\bibnamefont {Kotliar}},
  \bibinfo {author} {\bibfnamefont {A.}~\bibnamefont {Weichselbaum}},\ and\
  \bibinfo {author} {\bibfnamefont {J.}~\bibnamefont {{von Delft}}},\ }\href
  {https://doi.org/https://doi.org/10.1016/j.aop.2018.10.017} {\bibfield
  {journal} {\bibinfo  {journal} {Ann. Phys.}\ }\textbf {\bibinfo {volume}
  {405}},\ \bibinfo {pages} {365 } (\bibinfo {year} {2019})}\BibitemShut
  {NoStop}%
\bibitem [{\citenamefont {Kugler}\ \emph {et~al.}(2020)\citenamefont {Kugler},
  \citenamefont {Zingl}, \citenamefont {Strand}, \citenamefont {Lee},
  \citenamefont {von Delft},\ and\ \citenamefont {Georges}}]{Kugler2020}%
  \BibitemOpen
  \bibfield  {author} {\bibinfo {author} {\bibfnamefont {F.~B.}\ \bibnamefont
  {Kugler}}, \bibinfo {author} {\bibfnamefont {M.}~\bibnamefont {Zingl}},
  \bibinfo {author} {\bibfnamefont {H.~U.~R.}\ \bibnamefont {Strand}}, \bibinfo
  {author} {\bibfnamefont {S.-S.~B.}\ \bibnamefont {Lee}}, \bibinfo {author}
  {\bibfnamefont {J.}~\bibnamefont {von Delft}},\ and\ \bibinfo {author}
  {\bibfnamefont {A.}~\bibnamefont {Georges}},\ }\href
  {https://doi.org/10.1103/PhysRevLett.124.016401} {\bibfield  {journal}
  {\bibinfo  {journal} {Phys. Rev. Lett.}\ }\textbf {\bibinfo {volume} {124}},\
  \bibinfo {pages} {016401} (\bibinfo {year} {2020})}\BibitemShut {NoStop}%
\bibitem [{\citenamefont {White}(1992)}]{White1992}%
  \BibitemOpen
  \bibfield  {author} {\bibinfo {author} {\bibfnamefont {S.~R.}\ \bibnamefont
  {White}},\ }\href {https://doi.org/10.1103/PhysRevLett.69.2863} {\bibfield
  {journal} {\bibinfo  {journal} {Phys. Rev. Lett.}\ }\textbf {\bibinfo
  {volume} {69}},\ \bibinfo {pages} {2863} (\bibinfo {year}
  {1992})}\BibitemShut {NoStop}%
\bibitem [{\citenamefont {White}(1993)}]{White1993}%
  \BibitemOpen
  \bibfield  {author} {\bibinfo {author} {\bibfnamefont {S.~R.}\ \bibnamefont
  {White}},\ }\href {https://doi.org/10.1103/PhysRevB.48.10345} {\bibfield
  {journal} {\bibinfo  {journal} {Phys. Rev. B}\ }\textbf {\bibinfo {volume}
  {48}},\ \bibinfo {pages} {10345} (\bibinfo {year} {1993})}\BibitemShut
  {NoStop}%
\bibitem [{\citenamefont {Schollw{\"o}ck}(2005)}]{Schollwock2005}%
  \BibitemOpen
  \bibfield  {author} {\bibinfo {author} {\bibfnamefont {U.}~\bibnamefont
  {Schollw{\"o}ck}},\ }\href {https://doi.org/10.1103/RevModPhys.77.259}
  {\bibfield  {journal} {\bibinfo  {journal} {Rev. Mod. Phys.}\ }\textbf
  {\bibinfo {volume} {77}},\ \bibinfo {pages} {259} (\bibinfo {year}
  {2005})}\BibitemShut {NoStop}%
\bibitem [{\citenamefont {Hallberg}(2006)}]{Hallberg2006}%
  \BibitemOpen
  \bibfield  {author} {\bibinfo {author} {\bibfnamefont {K.~A.}\ \bibnamefont
  {Hallberg}},\ }\href {https://doi.org/10.1080/00018730600766432} {\bibfield
  {journal} {\bibinfo  {journal} {Adv. Phys.}\ }\textbf {\bibinfo {volume}
  {55}},\ \bibinfo {pages} {477} (\bibinfo {year} {2006})}\BibitemShut
  {NoStop}%
\bibitem [{\citenamefont {Garc\'{\i}a}\ \emph {et~al.}(2004)\citenamefont
  {Garc\'{\i}a}, \citenamefont {Hallberg},\ and\ \citenamefont
  {Rozenberg}}]{Garcia2004}%
  \BibitemOpen
  \bibfield  {author} {\bibinfo {author} {\bibfnamefont {D.~J.}\ \bibnamefont
  {Garc\'{\i}a}}, \bibinfo {author} {\bibfnamefont {K.}~\bibnamefont
  {Hallberg}},\ and\ \bibinfo {author} {\bibfnamefont {M.~J.}\ \bibnamefont
  {Rozenberg}},\ }\href {https://doi.org/10.1103/PhysRevLett.93.246403}
  {\bibfield  {journal} {\bibinfo  {journal} {Phys. Rev. Lett.}\ }\textbf
  {\bibinfo {volume} {93}},\ \bibinfo {pages} {246403} (\bibinfo {year}
  {2004})}\BibitemShut {NoStop}%
\bibitem [{\citenamefont {Raas}\ \emph {et~al.}(2004)\citenamefont {Raas},
  \citenamefont {Uhrig},\ and\ \citenamefont {Anders}}]{Raas2004}%
  \BibitemOpen
  \bibfield  {author} {\bibinfo {author} {\bibfnamefont {C.}~\bibnamefont
  {Raas}}, \bibinfo {author} {\bibfnamefont {G.~S.}\ \bibnamefont {Uhrig}},\
  and\ \bibinfo {author} {\bibfnamefont {F.~B.}\ \bibnamefont {Anders}},\
  }\href {https://doi.org/10.1103/PhysRevB.69.041102} {\bibfield  {journal}
  {\bibinfo  {journal} {Phys. Rev. B}\ }\textbf {\bibinfo {volume} {69}},\
  \bibinfo {pages} {041102(R)} (\bibinfo {year} {2004})}\BibitemShut {NoStop}%
\bibitem [{\citenamefont {Karski}\ \emph {et~al.}(2005)\citenamefont {Karski},
  \citenamefont {Raas},\ and\ \citenamefont {Uhrig}}]{Karski2005}%
  \BibitemOpen
  \bibfield  {author} {\bibinfo {author} {\bibfnamefont {M.}~\bibnamefont
  {Karski}}, \bibinfo {author} {\bibfnamefont {C.}~\bibnamefont {Raas}},\ and\
  \bibinfo {author} {\bibfnamefont {G.~S.}\ \bibnamefont {Uhrig}},\ }\href
  {https://doi.org/10.1103/PhysRevB.72.113110} {\bibfield  {journal} {\bibinfo
  {journal} {Phys. Rev. B}\ }\textbf {\bibinfo {volume} {72}},\ \bibinfo
  {pages} {113110} (\bibinfo {year} {2005})}\BibitemShut {NoStop}%
\bibitem [{\citenamefont {Schollw\"ock}(2011)}]{Schollwock2011}%
  \BibitemOpen
  \bibfield  {author} {\bibinfo {author} {\bibfnamefont {U.}~\bibnamefont
  {Schollw\"ock}},\ }\href
  {https://doi.org/https://doi.org/10.1016/j.aop.2010.09.012} {\bibfield
  {journal} {\bibinfo  {journal} {Ann. Phys.}\ }\textbf {\bibinfo {volume}
  {326}},\ \bibinfo {pages} {96 } (\bibinfo {year} {2011})}\BibitemShut
  {NoStop}%
\bibitem [{\citenamefont {Wolf}\ \emph {et~al.}(2014)\citenamefont {Wolf},
  \citenamefont {McCulloch},\ and\ \citenamefont {Schollw\"ock}}]{Wolf2014}%
  \BibitemOpen
  \bibfield  {author} {\bibinfo {author} {\bibfnamefont {F.~A.}\ \bibnamefont
  {Wolf}}, \bibinfo {author} {\bibfnamefont {I.~P.}\ \bibnamefont
  {McCulloch}},\ and\ \bibinfo {author} {\bibfnamefont {U.}~\bibnamefont
  {Schollw\"ock}},\ }\href {https://doi.org/10.1103/PhysRevB.90.235131}
  {\bibfield  {journal} {\bibinfo  {journal} {Phys. Rev. B}\ }\textbf {\bibinfo
  {volume} {90}},\ \bibinfo {pages} {235131} (\bibinfo {year}
  {2014})}\BibitemShut {NoStop}%
\bibitem [{\citenamefont {Wolf}\ \emph {et~al.}(2015)\citenamefont {Wolf},
  \citenamefont {Go}, \citenamefont {McCulloch}, \citenamefont {Millis},\ and\
  \citenamefont {Schollw\"ock}}]{Wolf2015}%
  \BibitemOpen
  \bibfield  {author} {\bibinfo {author} {\bibfnamefont {F.~A.}\ \bibnamefont
  {Wolf}}, \bibinfo {author} {\bibfnamefont {A.}~\bibnamefont {Go}}, \bibinfo
  {author} {\bibfnamefont {I.~P.}\ \bibnamefont {McCulloch}}, \bibinfo {author}
  {\bibfnamefont {A.~J.}\ \bibnamefont {Millis}},\ and\ \bibinfo {author}
  {\bibfnamefont {U.}~\bibnamefont {Schollw\"ock}},\ }\href
  {https://doi.org/10.1103/PhysRevX.5.041032} {\bibfield  {journal} {\bibinfo
  {journal} {Phys. Rev. X}\ }\textbf {\bibinfo {volume} {5}},\ \bibinfo {pages}
  {041032} (\bibinfo {year} {2015})}\BibitemShut {NoStop}%
\bibitem [{\citenamefont {Ganahl}\ \emph {et~al.}(2014)\citenamefont {Ganahl},
  \citenamefont {Thunstr\"om}, \citenamefont {Verstraete}, \citenamefont
  {Held},\ and\ \citenamefont {Evertz}}]{Ganahl2014}%
  \BibitemOpen
  \bibfield  {author} {\bibinfo {author} {\bibfnamefont {M.}~\bibnamefont
  {Ganahl}}, \bibinfo {author} {\bibfnamefont {P.}~\bibnamefont {Thunstr\"om}},
  \bibinfo {author} {\bibfnamefont {F.}~\bibnamefont {Verstraete}}, \bibinfo
  {author} {\bibfnamefont {K.}~\bibnamefont {Held}},\ and\ \bibinfo {author}
  {\bibfnamefont {H.~G.}\ \bibnamefont {Evertz}},\ }\href
  {https://doi.org/10.1103/PhysRevB.90.045144} {\bibfield  {journal} {\bibinfo
  {journal} {Phys. Rev. B}\ }\textbf {\bibinfo {volume} {90}},\ \bibinfo
  {pages} {045144} (\bibinfo {year} {2014})}\BibitemShut {NoStop}%
\bibitem [{\citenamefont {Ganahl}\ \emph {et~al.}(2015)\citenamefont {Ganahl},
  \citenamefont {Aichhorn}, \citenamefont {Evertz}, \citenamefont
  {Thunstr\"om}, \citenamefont {Held},\ and\ \citenamefont
  {Verstraete}}]{Ganahl2015}%
  \BibitemOpen
  \bibfield  {author} {\bibinfo {author} {\bibfnamefont {M.}~\bibnamefont
  {Ganahl}}, \bibinfo {author} {\bibfnamefont {M.}~\bibnamefont {Aichhorn}},
  \bibinfo {author} {\bibfnamefont {H.~G.}\ \bibnamefont {Evertz}}, \bibinfo
  {author} {\bibfnamefont {P.}~\bibnamefont {Thunstr\"om}}, \bibinfo {author}
  {\bibfnamefont {K.}~\bibnamefont {Held}},\ and\ \bibinfo {author}
  {\bibfnamefont {F.}~\bibnamefont {Verstraete}},\ }\href
  {https://doi.org/10.1103/PhysRevB.92.155132} {\bibfield  {journal} {\bibinfo
  {journal} {Phys. Rev. B}\ }\textbf {\bibinfo {volume} {92}},\ \bibinfo
  {pages} {155132} (\bibinfo {year} {2015})}\BibitemShut {NoStop}%
\bibitem [{\citenamefont {Holzner}\ \emph {et~al.}(2010)\citenamefont
  {Holzner}, \citenamefont {Weichselbaum},\ and\ \citenamefont {von
  Delft}}]{Holzner2010}%
  \BibitemOpen
  \bibfield  {author} {\bibinfo {author} {\bibfnamefont {A.}~\bibnamefont
  {Holzner}}, \bibinfo {author} {\bibfnamefont {A.}~\bibnamefont
  {Weichselbaum}},\ and\ \bibinfo {author} {\bibfnamefont {J.}~\bibnamefont
  {von Delft}},\ }\href {https://doi.org/10.1103/PhysRevB.81.125126} {\bibfield
   {journal} {\bibinfo  {journal} {Phys. Rev. B}\ }\textbf {\bibinfo {volume}
  {81}},\ \bibinfo {pages} {125126} (\bibinfo {year} {2010})}\BibitemShut
  {NoStop}%
\bibitem [{\citenamefont {Bauernfeind}\ \emph {et~al.}(2017)\citenamefont
  {Bauernfeind}, \citenamefont {Zingl}, \citenamefont {Triebl}, \citenamefont
  {Aichhorn},\ and\ \citenamefont {Evertz}}]{Bauernfeind2017}%
  \BibitemOpen
  \bibfield  {author} {\bibinfo {author} {\bibfnamefont {D.}~\bibnamefont
  {Bauernfeind}}, \bibinfo {author} {\bibfnamefont {M.}~\bibnamefont {Zingl}},
  \bibinfo {author} {\bibfnamefont {R.}~\bibnamefont {Triebl}}, \bibinfo
  {author} {\bibfnamefont {M.}~\bibnamefont {Aichhorn}},\ and\ \bibinfo
  {author} {\bibfnamefont {H.~G.}\ \bibnamefont {Evertz}},\ }\href
  {https://doi.org/10.1103/PhysRevX.7.031013} {\bibfield  {journal} {\bibinfo
  {journal} {Phys. Rev. X}\ }\textbf {\bibinfo {volume} {7}},\ \bibinfo {pages}
  {031013} (\bibinfo {year} {2017})}\BibitemShut {NoStop}%
\bibitem [{\citenamefont {Murg}\ \emph {et~al.}(2010)\citenamefont {Murg},
  \citenamefont {Verstraete}, \citenamefont {Legeza},\ and\ \citenamefont
  {Noack}}]{Murg2010}%
  \BibitemOpen
  \bibfield  {author} {\bibinfo {author} {\bibfnamefont {V.}~\bibnamefont
  {Murg}}, \bibinfo {author} {\bibfnamefont {F.}~\bibnamefont {Verstraete}},
  \bibinfo {author} {\bibfnamefont {O.}~\bibnamefont {Legeza}},\ and\ \bibinfo
  {author} {\bibfnamefont {R.~M.}\ \bibnamefont {Noack}},\ }\href
  {https://doi.org/10.1103/PhysRevB.82.205105} {\bibfield  {journal} {\bibinfo
  {journal} {Phys. Rev. B}\ }\textbf {\bibinfo {volume} {82}},\ \bibinfo
  {pages} {205105} (\bibinfo {year} {2010})}\BibitemShut {NoStop}%
\bibitem [{\citenamefont {Shi}\ \emph {et~al.}(2006)\citenamefont {Shi},
  \citenamefont {Duan},\ and\ \citenamefont {Vidal}}]{Shi2006}%
  \BibitemOpen
  \bibfield  {author} {\bibinfo {author} {\bibfnamefont {Y.-Y.}\ \bibnamefont
  {Shi}}, \bibinfo {author} {\bibfnamefont {L.-M.}\ \bibnamefont {Duan}},\ and\
  \bibinfo {author} {\bibfnamefont {G.}~\bibnamefont {Vidal}},\ }\href
  {https://doi.org/10.1103/PhysRevA.74.022320} {\bibfield  {journal} {\bibinfo
  {journal} {Phys. Rev. A}\ }\textbf {\bibinfo {volume} {74}},\ \bibinfo
  {pages} {022320} (\bibinfo {year} {2006})}\BibitemShut {NoStop}%
\bibitem [{\citenamefont {Mackenzie}\ and\ \citenamefont
  {Maeno}(2003)}]{Mackenzie2003}%
  \BibitemOpen
  \bibfield  {author} {\bibinfo {author} {\bibfnamefont {A.~P.}\ \bibnamefont
  {Mackenzie}}\ and\ \bibinfo {author} {\bibfnamefont {Y.}~\bibnamefont
  {Maeno}},\ }\href {https://doi.org/10.1103/RevModPhys.75.657} {\bibfield
  {journal} {\bibinfo  {journal} {Rev. Mod. Phys.}\ }\textbf {\bibinfo {volume}
  {75}},\ \bibinfo {pages} {657} (\bibinfo {year} {2003})}\BibitemShut
  {NoStop}%
\bibitem [{\citenamefont {Mackenzie}\ \emph {et~al.}(2017)\citenamefont
  {Mackenzie}, \citenamefont {Scaffidi}, \citenamefont {Hicks},\ and\
  \citenamefont {Maeno}}]{Mackenzie2017}%
  \BibitemOpen
  \bibfield  {author} {\bibinfo {author} {\bibfnamefont {A.~P.}\ \bibnamefont
  {Mackenzie}}, \bibinfo {author} {\bibfnamefont {T.}~\bibnamefont {Scaffidi}},
  \bibinfo {author} {\bibfnamefont {C.~W.}\ \bibnamefont {Hicks}},\ and\
  \bibinfo {author} {\bibfnamefont {Y.}~\bibnamefont {Maeno}},\ }\href
  {https://doi.org/10.1038/s41535-017-0045-4} {\bibfield  {journal} {\bibinfo
  {journal} {npj Quantum Mater.}\ }\textbf {\bibinfo {volume} {2}},\ \bibinfo
  {pages} {40} (\bibinfo {year} {2017})}\BibitemShut {NoStop}%
\bibitem [{\citenamefont {Haverkort}\ \emph {et~al.}(2008)\citenamefont
  {Haverkort}, \citenamefont {Elfimov}, \citenamefont {Tjeng}, \citenamefont
  {Sawatzky},\ and\ \citenamefont {Damascelli}}]{Haverkort2008}%
  \BibitemOpen
  \bibfield  {author} {\bibinfo {author} {\bibfnamefont {M.~W.}\ \bibnamefont
  {Haverkort}}, \bibinfo {author} {\bibfnamefont {I.~S.}\ \bibnamefont
  {Elfimov}}, \bibinfo {author} {\bibfnamefont {L.~H.}\ \bibnamefont {Tjeng}},
  \bibinfo {author} {\bibfnamefont {G.~A.}\ \bibnamefont {Sawatzky}},\ and\
  \bibinfo {author} {\bibfnamefont {A.}~\bibnamefont {Damascelli}},\ }\href
  {https://doi.org/10.1103/PhysRevLett.101.026406} {\bibfield  {journal}
  {\bibinfo  {journal} {Phys. Rev. Lett.}\ }\textbf {\bibinfo {volume} {101}},\
  \bibinfo {pages} {026406} (\bibinfo {year} {2008})}\BibitemShut {NoStop}%
\bibitem [{\citenamefont {Mravlje}\ \emph {et~al.}(2011)\citenamefont
  {Mravlje}, \citenamefont {Aichhorn}, \citenamefont {Miyake}, \citenamefont
  {Haule}, \citenamefont {Kotliar},\ and\ \citenamefont
  {Georges}}]{Mravlje2011}%
  \BibitemOpen
  \bibfield  {author} {\bibinfo {author} {\bibfnamefont {J.}~\bibnamefont
  {Mravlje}}, \bibinfo {author} {\bibfnamefont {M.}~\bibnamefont {Aichhorn}},
  \bibinfo {author} {\bibfnamefont {T.}~\bibnamefont {Miyake}}, \bibinfo
  {author} {\bibfnamefont {K.}~\bibnamefont {Haule}}, \bibinfo {author}
  {\bibfnamefont {G.}~\bibnamefont {Kotliar}},\ and\ \bibinfo {author}
  {\bibfnamefont {A.}~\bibnamefont {Georges}},\ }\href
  {https://doi.org/10.1103/PhysRevLett.106.096401} {\bibfield  {journal}
  {\bibinfo  {journal} {Phys. Rev. Lett.}\ }\textbf {\bibinfo {volume} {106}},\
  \bibinfo {pages} {096401} (\bibinfo {year} {2011})}\BibitemShut {NoStop}%
\bibitem [{\citenamefont {Mravlje}\ and\ \citenamefont
  {Georges}(2016)}]{Mravlje2016}%
  \BibitemOpen
  \bibfield  {author} {\bibinfo {author} {\bibfnamefont {J.}~\bibnamefont
  {Mravlje}}\ and\ \bibinfo {author} {\bibfnamefont {A.}~\bibnamefont
  {Georges}},\ }\href {https://doi.org/10.1103/PhysRevLett.117.036401}
  {\bibfield  {journal} {\bibinfo  {journal} {Phys. Rev. Lett.}\ }\textbf
  {\bibinfo {volume} {117}},\ \bibinfo {pages} {036401} (\bibinfo {year}
  {2016})}\BibitemShut {NoStop}%
\bibitem [{\citenamefont {Zhang}\ \emph {et~al.}(2016)\citenamefont {Zhang},
  \citenamefont {Gorelov}, \citenamefont {Sarvestani},\ and\ \citenamefont
  {Pavarini}}]{Zhang2016}%
  \BibitemOpen
  \bibfield  {author} {\bibinfo {author} {\bibfnamefont {G.}~\bibnamefont
  {Zhang}}, \bibinfo {author} {\bibfnamefont {E.}~\bibnamefont {Gorelov}},
  \bibinfo {author} {\bibfnamefont {E.}~\bibnamefont {Sarvestani}},\ and\
  \bibinfo {author} {\bibfnamefont {E.}~\bibnamefont {Pavarini}},\ }\href
  {https://doi.org/10.1103/PhysRevLett.116.106402} {\bibfield  {journal}
  {\bibinfo  {journal} {Phys. Rev. Lett.}\ }\textbf {\bibinfo {volume} {116}},\
  \bibinfo {pages} {106402} (\bibinfo {year} {2016})}\BibitemShut {NoStop}%
\bibitem [{\citenamefont {Kim}\ \emph {et~al.}(2018)\citenamefont {Kim},
  \citenamefont {Mravlje}, \citenamefont {Ferrero}, \citenamefont {Parcollet},\
  and\ \citenamefont {Georges}}]{Kim2018}%
  \BibitemOpen
  \bibfield  {author} {\bibinfo {author} {\bibfnamefont {M.}~\bibnamefont
  {Kim}}, \bibinfo {author} {\bibfnamefont {J.}~\bibnamefont {Mravlje}},
  \bibinfo {author} {\bibfnamefont {M.}~\bibnamefont {Ferrero}}, \bibinfo
  {author} {\bibfnamefont {O.}~\bibnamefont {Parcollet}},\ and\ \bibinfo
  {author} {\bibfnamefont {A.}~\bibnamefont {Georges}},\ }\href
  {https://doi.org/10.1103/PhysRevLett.120.126401} {\bibfield  {journal}
  {\bibinfo  {journal} {Phys. Rev. Lett.}\ }\textbf {\bibinfo {volume} {120}},\
  \bibinfo {pages} {126401} (\bibinfo {year} {2018})}\BibitemShut {NoStop}%
\bibitem [{\citenamefont {Tamai}\ \emph {et~al.}(2019)\citenamefont {Tamai},
  \citenamefont {Zingl}, \citenamefont {Rozbicki}, \citenamefont {Cappelli},
  \citenamefont {Ricc\`o}, \citenamefont {de~la Torre}, \citenamefont
  {McKeown~Walker}, \citenamefont {Bruno}, \citenamefont {King}, \citenamefont
  {Meevasana}, \citenamefont {Shi}, \citenamefont
  {Radovi\ifmmode~\acute{c}\else \'{c}\fi{}}, \citenamefont {Plumb},
  \citenamefont {Gibbs}, \citenamefont {Mackenzie}, \citenamefont {Berthod},
  \citenamefont {Strand}, \citenamefont {Kim}, \citenamefont {Georges},\ and\
  \citenamefont {Baumberger}}]{Tamai2019}%
  \BibitemOpen
  \bibfield  {author} {\bibinfo {author} {\bibfnamefont {A.}~\bibnamefont
  {Tamai}}, \bibinfo {author} {\bibfnamefont {M.}~\bibnamefont {Zingl}},
  \bibinfo {author} {\bibfnamefont {E.}~\bibnamefont {Rozbicki}}, \bibinfo
  {author} {\bibfnamefont {E.}~\bibnamefont {Cappelli}}, \bibinfo {author}
  {\bibfnamefont {S.}~\bibnamefont {Ricc\`o}}, \bibinfo {author} {\bibfnamefont
  {A.}~\bibnamefont {de~la Torre}}, \bibinfo {author} {\bibfnamefont
  {S.}~\bibnamefont {McKeown~Walker}}, \bibinfo {author} {\bibfnamefont
  {F.~Y.}\ \bibnamefont {Bruno}}, \bibinfo {author} {\bibfnamefont {P.~D.~C.}\
  \bibnamefont {King}}, \bibinfo {author} {\bibfnamefont {W.}~\bibnamefont
  {Meevasana}}, \bibinfo {author} {\bibfnamefont {M.}~\bibnamefont {Shi}},
  \bibinfo {author} {\bibfnamefont {M.}~\bibnamefont
  {Radovi\ifmmode~\acute{c}\else \'{c}\fi{}}}, \bibinfo {author} {\bibfnamefont
  {N.~C.}\ \bibnamefont {Plumb}}, \bibinfo {author} {\bibfnamefont {A.~S.}\
  \bibnamefont {Gibbs}}, \bibinfo {author} {\bibfnamefont {A.~P.}\ \bibnamefont
  {Mackenzie}}, \bibinfo {author} {\bibfnamefont {C.}~\bibnamefont {Berthod}},
  \bibinfo {author} {\bibfnamefont {H.~U.~R.}\ \bibnamefont {Strand}}, \bibinfo
  {author} {\bibfnamefont {M.}~\bibnamefont {Kim}}, \bibinfo {author}
  {\bibfnamefont {A.}~\bibnamefont {Georges}},\ and\ \bibinfo {author}
  {\bibfnamefont {F.}~\bibnamefont {Baumberger}},\ }\href
  {https://doi.org/10.1103/PhysRevX.9.021048} {\bibfield  {journal} {\bibinfo
  {journal} {Phys. Rev. X}\ }\textbf {\bibinfo {volume} {9}},\ \bibinfo {pages}
  {021048} (\bibinfo {year} {2019})}\BibitemShut {NoStop}%
\bibitem [{\citenamefont {Linden}\ \emph {et~al.}(2020)\citenamefont {Linden},
  \citenamefont {Zingl}, \citenamefont {Hubig}, \citenamefont {Parcollet},\
  and\ \citenamefont {Schollw\"ock}}]{Linden2020}%
  \BibitemOpen
  \bibfield  {author} {\bibinfo {author} {\bibfnamefont {N.-O.}\ \bibnamefont
  {Linden}}, \bibinfo {author} {\bibfnamefont {M.}~\bibnamefont {Zingl}},
  \bibinfo {author} {\bibfnamefont {C.}~\bibnamefont {Hubig}}, \bibinfo
  {author} {\bibfnamefont {O.}~\bibnamefont {Parcollet}},\ and\ \bibinfo
  {author} {\bibfnamefont {U.}~\bibnamefont {Schollw\"ock}},\ }\href
  {https://doi.org/10.1103/PhysRevB.101.041101} {\bibfield  {journal} {\bibinfo
   {journal} {Phys. Rev. B}\ }\textbf {\bibinfo {volume} {101}},\ \bibinfo
  {pages} {041101(R)} (\bibinfo {year} {2020})}\BibitemShut {NoStop}%
\bibitem [{\citenamefont {Lee}\ \emph {et~al.}(2020)\citenamefont {Lee},
  \citenamefont {Kim},\ and\ \citenamefont {Go}}]{Lee2020}%
  \BibitemOpen
  \bibfield  {author} {\bibinfo {author} {\bibfnamefont {H.~J.}\ \bibnamefont
  {Lee}}, \bibinfo {author} {\bibfnamefont {C.~H.}\ \bibnamefont {Kim}},\ and\
  \bibinfo {author} {\bibfnamefont {A.}~\bibnamefont {Go}},\ }\href
  {https://doi.org/10.1103/PhysRevB.102.195115} {\bibfield  {journal} {\bibinfo
   {journal} {Phys. Rev. B}\ }\textbf {\bibinfo {volume} {102}},\ \bibinfo
  {pages} {195115} (\bibinfo {year} {2020})}\BibitemShut {NoStop}%
\bibitem [{\citenamefont {Lu}\ \emph {et~al.}(2019)\citenamefont {Lu},
  \citenamefont {Cao}, \citenamefont {Hansmann},\ and\ \citenamefont
  {Haverkort}}]{Lu2019}%
  \BibitemOpen
  \bibfield  {author} {\bibinfo {author} {\bibfnamefont {Y.}~\bibnamefont
  {Lu}}, \bibinfo {author} {\bibfnamefont {X.}~\bibnamefont {Cao}}, \bibinfo
  {author} {\bibfnamefont {P.}~\bibnamefont {Hansmann}},\ and\ \bibinfo
  {author} {\bibfnamefont {M.~W.}\ \bibnamefont {Haverkort}},\ }\href
  {https://doi.org/10.1103/PhysRevB.100.115134} {\bibfield  {journal} {\bibinfo
   {journal} {Phys. Rev. B}\ }\textbf {\bibinfo {volume} {100}},\ \bibinfo
  {pages} {115134} (\bibinfo {year} {2019})}\BibitemShut {NoStop}%
\bibitem [{\citenamefont {Kohn}\ and\ \citenamefont
  {Santoro}(2021)}]{Kohn2021}%
  \BibitemOpen
  \bibfield  {author} {\bibinfo {author} {\bibfnamefont {L.}~\bibnamefont
  {Kohn}}\ and\ \bibinfo {author} {\bibfnamefont {G.~E.}\ \bibnamefont
  {Santoro}},\ }\href {https://doi.org/10.1103/PhysRevB.104.014303} {\bibfield
  {journal} {\bibinfo  {journal} {Phys. Rev. B}\ }\textbf {\bibinfo {volume}
  {104}},\ \bibinfo {pages} {014303} (\bibinfo {year} {2021})}\BibitemShut
  {NoStop}%
\bibitem [{\citenamefont {Orús}(2014)}]{Orus2014}%
  \BibitemOpen
  \bibfield  {author} {\bibinfo {author} {\bibfnamefont {R.}~\bibnamefont
  {Orús}},\ }\href {https://doi.org/https://doi.org/10.1016/j.aop.2014.06.013}
  {\bibfield  {journal} {\bibinfo  {journal} {Ann. Phys.}\ }\textbf {\bibinfo
  {volume} {349}},\ \bibinfo {pages} {117 } (\bibinfo {year}
  {2014})}\BibitemShut {NoStop}%
\bibitem [{\citenamefont {Oseledets}(2011)}]{Oseledets2011}%
  \BibitemOpen
  \bibfield  {author} {\bibinfo {author} {\bibfnamefont {I.~V.}\ \bibnamefont
  {Oseledets}},\ }\href {https://doi.org/10.1137/090752286} {\bibfield
  {journal} {\bibinfo  {journal} {SIAM J. Sci. Comput.}\ }\textbf {\bibinfo
  {volume} {33}},\ \bibinfo {pages} {2295} (\bibinfo {year}
  {2011})}\BibitemShut {NoStop}%
\bibitem [{\citenamefont {Paeckel}\ \emph {et~al.}(2019)\citenamefont
  {Paeckel}, \citenamefont {K{\"o}hler}, \citenamefont {Swoboda}, \citenamefont
  {Manmana}, \citenamefont {Schollw{\"o}ck},\ and\ \citenamefont
  {Hubig}}]{Paeckel2019}%
  \BibitemOpen
  \bibfield  {author} {\bibinfo {author} {\bibfnamefont {S.}~\bibnamefont
  {Paeckel}}, \bibinfo {author} {\bibfnamefont {T.}~\bibnamefont {K{\"o}hler}},
  \bibinfo {author} {\bibfnamefont {A.}~\bibnamefont {Swoboda}}, \bibinfo
  {author} {\bibfnamefont {S.~R.}\ \bibnamefont {Manmana}}, \bibinfo {author}
  {\bibfnamefont {U.}~\bibnamefont {Schollw{\"o}ck}},\ and\ \bibinfo {author}
  {\bibfnamefont {C.}~\bibnamefont {Hubig}},\ }\href
  {https://doi.org/https://doi.org/10.1016/j.aop.2019.167998} {\bibfield
  {journal} {\bibinfo  {journal} {Ann. Phys.}\ }\textbf {\bibinfo {volume}
  {411}},\ \bibinfo {pages} {167998} (\bibinfo {year} {2019})}\BibitemShut
  {NoStop}%
\bibitem [{\citenamefont {Lubich}\ \emph {et~al.}(2013)\citenamefont {Lubich},
  \citenamefont {Rohwedder}, \citenamefont {Schneider},\ and\ \citenamefont
  {Vandereycken}}]{Lubich2013}%
  \BibitemOpen
  \bibfield  {author} {\bibinfo {author} {\bibfnamefont {C.}~\bibnamefont
  {Lubich}}, \bibinfo {author} {\bibfnamefont {T.}~\bibnamefont {Rohwedder}},
  \bibinfo {author} {\bibfnamefont {R.}~\bibnamefont {Schneider}},\ and\
  \bibinfo {author} {\bibfnamefont {B.}~\bibnamefont {Vandereycken}},\ }\href
  {https://doi.org/10.1137/120885723} {\bibfield  {journal} {\bibinfo
  {journal} {SIAM J. Matrix Anal. Appl.}\ }\textbf {\bibinfo {volume} {34}},\
  \bibinfo {pages} {470} (\bibinfo {year} {2013})}\BibitemShut {NoStop}%
\bibitem [{\citenamefont {Schröder}\ \emph {et~al.}()\citenamefont
  {Schröder}, \citenamefont {Turban}, \citenamefont {Musser}, \citenamefont
  {Hine},\ and\ \citenamefont {Chin}}]{Schroeder2017}%
  \BibitemOpen
  \bibfield  {author} {\bibinfo {author} {\bibfnamefont {F.~A. Y.~N.}\
  \bibnamefont {Schröder}}, \bibinfo {author} {\bibfnamefont {D.~H.~P.}\
  \bibnamefont {Turban}}, \bibinfo {author} {\bibfnamefont {A.~J.}\
  \bibnamefont {Musser}}, \bibinfo {author} {\bibfnamefont {N.~D.~M.}\
  \bibnamefont {Hine}},\ and\ \bibinfo {author} {\bibfnamefont {A.~W.}\
  \bibnamefont {Chin}},\ }\href@noop {} {}\Eprint
  {https://arxiv.org/abs/1710.01362} {arXiv:1710.01362} \BibitemShut {NoStop}%
\bibitem [{\citenamefont {Bauernfeind}\ and\ \citenamefont
  {Aichhorn}(2020)}]{Bauernfeind2020}%
  \BibitemOpen
  \bibfield  {author} {\bibinfo {author} {\bibfnamefont {D.}~\bibnamefont
  {Bauernfeind}}\ and\ \bibinfo {author} {\bibfnamefont {M.}~\bibnamefont
  {Aichhorn}},\ }\href {https://doi.org/10.21468/SciPostPhys.8.2.024}
  {\bibfield  {journal} {\bibinfo  {journal} {SciPost Phys.}\ }\textbf
  {\bibinfo {volume} {8}},\ \bibinfo {pages} {24} (\bibinfo {year}
  {2020})}\BibitemShut {NoStop}%
\bibitem [{\citenamefont {Dargel}\ \emph {et~al.}(2012)\citenamefont {Dargel},
  \citenamefont {W\"ollert}, \citenamefont {Honecker}, \citenamefont
  {McCulloch}, \citenamefont {Schollw\"ock},\ and\ \citenamefont
  {Pruschke}}]{Dargel2012}%
  \BibitemOpen
  \bibfield  {author} {\bibinfo {author} {\bibfnamefont {P.~E.}\ \bibnamefont
  {Dargel}}, \bibinfo {author} {\bibfnamefont {A.}~\bibnamefont {W\"ollert}},
  \bibinfo {author} {\bibfnamefont {A.}~\bibnamefont {Honecker}}, \bibinfo
  {author} {\bibfnamefont {I.~P.}\ \bibnamefont {McCulloch}}, \bibinfo {author}
  {\bibfnamefont {U.}~\bibnamefont {Schollw\"ock}},\ and\ \bibinfo {author}
  {\bibfnamefont {T.}~\bibnamefont {Pruschke}},\ }\href
  {https://doi.org/10.1103/PhysRevB.85.205119} {\bibfield  {journal} {\bibinfo
  {journal} {Phys. Rev. B}\ }\textbf {\bibinfo {volume} {85}},\ \bibinfo
  {pages} {205119} (\bibinfo {year} {2012})}\BibitemShut {NoStop}%
\bibitem [{\citenamefont {Holzner}\ \emph {et~al.}(2011)\citenamefont
  {Holzner}, \citenamefont {Weichselbaum}, \citenamefont {McCulloch},
  \citenamefont {Schollw\"ock},\ and\ \citenamefont {von Delft}}]{Holzner2011}%
  \BibitemOpen
  \bibfield  {author} {\bibinfo {author} {\bibfnamefont {A.}~\bibnamefont
  {Holzner}}, \bibinfo {author} {\bibfnamefont {A.}~\bibnamefont
  {Weichselbaum}}, \bibinfo {author} {\bibfnamefont {I.~P.}\ \bibnamefont
  {McCulloch}}, \bibinfo {author} {\bibfnamefont {U.}~\bibnamefont
  {Schollw\"ock}},\ and\ \bibinfo {author} {\bibfnamefont {J.}~\bibnamefont
  {von Delft}},\ }\href {https://doi.org/10.1103/PhysRevB.83.195115} {\bibfield
   {journal} {\bibinfo  {journal} {Phys. Rev. B}\ }\textbf {\bibinfo {volume}
  {83}},\ \bibinfo {pages} {195115} (\bibinfo {year} {2011})}\BibitemShut
  {NoStop}%
\bibitem [{\citenamefont {Nocera}\ and\ \citenamefont
  {Alvarez}(2016)}]{Nocera2016}%
  \BibitemOpen
  \bibfield  {author} {\bibinfo {author} {\bibfnamefont {A.}~\bibnamefont
  {Nocera}}\ and\ \bibinfo {author} {\bibfnamefont {G.}~\bibnamefont
  {Alvarez}},\ }\href {https://doi.org/10.1103/PhysRevE.94.053308} {\bibfield
  {journal} {\bibinfo  {journal} {Phys. Rev. E}\ }\textbf {\bibinfo {volume}
  {94}},\ \bibinfo {pages} {053308} (\bibinfo {year} {2016})}\BibitemShut
  {NoStop}%
\bibitem [{\citenamefont {Barthel}\ \emph {et~al.}(2009)\citenamefont
  {Barthel}, \citenamefont {Schollw\"ock},\ and\ \citenamefont
  {White}}]{Barthel2009}%
  \BibitemOpen
  \bibfield  {author} {\bibinfo {author} {\bibfnamefont {T.}~\bibnamefont
  {Barthel}}, \bibinfo {author} {\bibfnamefont {U.}~\bibnamefont
  {Schollw\"ock}},\ and\ \bibinfo {author} {\bibfnamefont {S.~R.}\ \bibnamefont
  {White}},\ }\href {https://doi.org/10.1103/PhysRevB.79.245101} {\bibfield
  {journal} {\bibinfo  {journal} {Phys. Rev. B}\ }\textbf {\bibinfo {volume}
  {79}},\ \bibinfo {pages} {245101} (\bibinfo {year} {2009})}\BibitemShut
  {NoStop}%
\bibitem [{\citenamefont {Haegeman}\ \emph {et~al.}(2011)\citenamefont
  {Haegeman}, \citenamefont {Cirac}, \citenamefont {Osborne}, \citenamefont
  {Pi\ifmmode~\check{z}\else \v{z}\fi{}orn}, \citenamefont {Verschelde},\ and\
  \citenamefont {Verstraete}}]{Haegeman2011}%
  \BibitemOpen
  \bibfield  {author} {\bibinfo {author} {\bibfnamefont {J.}~\bibnamefont
  {Haegeman}}, \bibinfo {author} {\bibfnamefont {J.~I.}\ \bibnamefont {Cirac}},
  \bibinfo {author} {\bibfnamefont {T.~J.}\ \bibnamefont {Osborne}}, \bibinfo
  {author} {\bibfnamefont {I.}~\bibnamefont {Pi\ifmmode~\check{z}\else
  \v{z}\fi{}orn}}, \bibinfo {author} {\bibfnamefont {H.}~\bibnamefont
  {Verschelde}},\ and\ \bibinfo {author} {\bibfnamefont {F.}~\bibnamefont
  {Verstraete}},\ }\href {https://doi.org/10.1103/PhysRevLett.107.070601}
  {\bibfield  {journal} {\bibinfo  {journal} {Phys. Rev. Lett.}\ }\textbf
  {\bibinfo {volume} {107}},\ \bibinfo {pages} {070601} (\bibinfo {year}
  {2011})}\BibitemShut {NoStop}%
\bibitem [{\citenamefont {Haegeman}\ \emph {et~al.}(2016)\citenamefont
  {Haegeman}, \citenamefont {Lubich}, \citenamefont {Oseledets}, \citenamefont
  {Vandereycken},\ and\ \citenamefont {Verstraete}}]{Haegeman2016}%
  \BibitemOpen
  \bibfield  {author} {\bibinfo {author} {\bibfnamefont {J.}~\bibnamefont
  {Haegeman}}, \bibinfo {author} {\bibfnamefont {C.}~\bibnamefont {Lubich}},
  \bibinfo {author} {\bibfnamefont {I.}~\bibnamefont {Oseledets}}, \bibinfo
  {author} {\bibfnamefont {B.}~\bibnamefont {Vandereycken}},\ and\ \bibinfo
  {author} {\bibfnamefont {F.}~\bibnamefont {Verstraete}},\ }\href
  {https://doi.org/10.1103/PhysRevB.94.165116} {\bibfield  {journal} {\bibinfo
  {journal} {Phys. Rev. B}\ }\textbf {\bibinfo {volume} {94}},\ \bibinfo
  {pages} {165116} (\bibinfo {year} {2016})}\BibitemShut {NoStop}%
\bibitem [{\citenamefont {Bulla}\ \emph {et~al.}(1998)\citenamefont {Bulla},
  \citenamefont {Hewson},\ and\ \citenamefont {Pruschke}}]{Bulla1998}%
  \BibitemOpen
  \bibfield  {author} {\bibinfo {author} {\bibfnamefont {R.}~\bibnamefont
  {Bulla}}, \bibinfo {author} {\bibfnamefont {A.~C.}\ \bibnamefont {Hewson}},\
  and\ \bibinfo {author} {\bibfnamefont {T.}~\bibnamefont {Pruschke}},\ }\href
  {https://doi.org/10.1088/0953-8984/10/37/021} {\bibfield  {journal} {\bibinfo
   {journal} {J. Phys. Condens. Matter}\ }\textbf {\bibinfo {volume} {10}},\
  \bibinfo {pages} {8365} (\bibinfo {year} {1998})}\BibitemShut {NoStop}%
\bibitem [{\citenamefont {de~Vega}\ \emph {et~al.}(2015)\citenamefont
  {de~Vega}, \citenamefont {Schollw\"ock},\ and\ \citenamefont
  {Wolf}}]{deVega2015}%
  \BibitemOpen
  \bibfield  {author} {\bibinfo {author} {\bibfnamefont {I.}~\bibnamefont
  {de~Vega}}, \bibinfo {author} {\bibfnamefont {U.}~\bibnamefont
  {Schollw\"ock}},\ and\ \bibinfo {author} {\bibfnamefont {F.~A.}\ \bibnamefont
  {Wolf}},\ }\href {https://doi.org/10.1103/PhysRevB.92.155126} {\bibfield
  {journal} {\bibinfo  {journal} {Phys. Rev. B}\ }\textbf {\bibinfo {volume}
  {92}},\ \bibinfo {pages} {155126} (\bibinfo {year} {2015})}\BibitemShut
  {NoStop}%
\bibitem [{\citenamefont {Strand}\ \emph {et~al.}(2019)\citenamefont {Strand},
  \citenamefont {Zingl}, \citenamefont {Wentzell}, \citenamefont {Parcollet},\
  and\ \citenamefont {Georges}}]{Strand2019}%
  \BibitemOpen
  \bibfield  {author} {\bibinfo {author} {\bibfnamefont {H.~U.~R.}\
  \bibnamefont {Strand}}, \bibinfo {author} {\bibfnamefont {M.}~\bibnamefont
  {Zingl}}, \bibinfo {author} {\bibfnamefont {N.}~\bibnamefont {Wentzell}},
  \bibinfo {author} {\bibfnamefont {O.}~\bibnamefont {Parcollet}},\ and\
  \bibinfo {author} {\bibfnamefont {A.}~\bibnamefont {Georges}},\ }\href
  {https://doi.org/10.1103/PhysRevB.100.125120} {\bibfield  {journal} {\bibinfo
   {journal} {Phys. Rev. B}\ }\textbf {\bibinfo {volume} {100}},\ \bibinfo
  {pages} {125120} (\bibinfo {year} {2019})}\BibitemShut {NoStop}%
\bibitem [{\citenamefont {Blaha}\ \emph {et~al.}(2018)\citenamefont {Blaha},
  \citenamefont {Schwarz}, \citenamefont {Madsen}, \citenamefont {Kvasnicka},
  \citenamefont {Luitz} \emph {et~al.}}]{wien2k_a}%
  \BibitemOpen
  \bibfield  {author} {\bibinfo {author} {\bibfnamefont {P.}~\bibnamefont
  {Blaha}}, \bibinfo {author} {\bibfnamefont {K.}~\bibnamefont {Schwarz}},
  \bibinfo {author} {\bibfnamefont {G.~K.}\ \bibnamefont {Madsen}}, \bibinfo
  {author} {\bibfnamefont {D.}~\bibnamefont {Kvasnicka}}, \bibinfo {author}
  {\bibfnamefont {J.}~\bibnamefont {Luitz}}, \emph {et~al.},\ }\href@noop {}
  {\emph {\bibinfo {title} {WIEN2k, An Augmented Plane Wave + Local Orbitals
  Program for Calculating Crystal Properties}}}\ (\bibinfo  {publisher}
  {Karlheinz Schwarz, Techn. Universit\"at Wien, Austria},\ \bibinfo {year}
  {2018})\BibitemShut {NoStop}%
\bibitem [{\citenamefont {Blaha}\ \emph {et~al.}(2020)\citenamefont {Blaha},
  \citenamefont {Schwarz}, \citenamefont {Tran}, \citenamefont {Laskowski},
  \citenamefont {Madsen},\ and\ \citenamefont {Marks}}]{wien2k_b}%
  \BibitemOpen
  \bibfield  {author} {\bibinfo {author} {\bibfnamefont {P.}~\bibnamefont
  {Blaha}}, \bibinfo {author} {\bibfnamefont {K.}~\bibnamefont {Schwarz}},
  \bibinfo {author} {\bibfnamefont {F.}~\bibnamefont {Tran}}, \bibinfo {author}
  {\bibfnamefont {R.}~\bibnamefont {Laskowski}}, \bibinfo {author}
  {\bibfnamefont {G.~K.~H.}\ \bibnamefont {Madsen}},\ and\ \bibinfo {author}
  {\bibfnamefont {L.~D.}\ \bibnamefont {Marks}},\ }\href
  {https://doi.org/10.1063/1.5143061} {\bibfield  {journal} {\bibinfo
  {journal} {J. Chem. Phys.}\ }\textbf {\bibinfo {volume} {152}},\ \bibinfo
  {pages} {074101} (\bibinfo {year} {2020})}\BibitemShut {NoStop}%
\bibitem [{\citenamefont {Mostofi}\ \emph {et~al.}(2014)\citenamefont
  {Mostofi}, \citenamefont {Yates}, \citenamefont {Pizzi}, \citenamefont {Lee},
  \citenamefont {Souza}, \citenamefont {Vanderbilt},\ and\ \citenamefont
  {Marzari}}]{wannier90}%
  \BibitemOpen
  \bibfield  {author} {\bibinfo {author} {\bibfnamefont {A.~A.}\ \bibnamefont
  {Mostofi}}, \bibinfo {author} {\bibfnamefont {J.~R.}\ \bibnamefont {Yates}},
  \bibinfo {author} {\bibfnamefont {G.}~\bibnamefont {Pizzi}}, \bibinfo
  {author} {\bibfnamefont {Y.-S.}\ \bibnamefont {Lee}}, \bibinfo {author}
  {\bibfnamefont {I.}~\bibnamefont {Souza}}, \bibinfo {author} {\bibfnamefont
  {D.}~\bibnamefont {Vanderbilt}},\ and\ \bibinfo {author} {\bibfnamefont
  {N.}~\bibnamefont {Marzari}},\ }\href
  {https://doi.org/https://doi.org/10.1016/j.cpc.2014.05.003} {\bibfield
  {journal} {\bibinfo  {journal} {Comput. Phys. Commun.}\ }\textbf {\bibinfo
  {volume} {185}},\ \bibinfo {pages} {2309 } (\bibinfo {year}
  {2014})}\BibitemShut {NoStop}%
\bibitem [{\citenamefont {Saberi}\ \emph {et~al.}(2008)\citenamefont {Saberi},
  \citenamefont {Weichselbaum},\ and\ \citenamefont {von Delft}}]{Saberi2008}%
  \BibitemOpen
  \bibfield  {author} {\bibinfo {author} {\bibfnamefont {H.}~\bibnamefont
  {Saberi}}, \bibinfo {author} {\bibfnamefont {A.}~\bibnamefont
  {Weichselbaum}},\ and\ \bibinfo {author} {\bibfnamefont {J.}~\bibnamefont
  {von Delft}},\ }\href {https://doi.org/10.1103/PhysRevB.78.035124} {\bibfield
   {journal} {\bibinfo  {journal} {Phys. Rev. B}\ }\textbf {\bibinfo {volume}
  {78}},\ \bibinfo {pages} {035124} (\bibinfo {year} {2008})}\BibitemShut
  {NoStop}%
\bibitem [{\citenamefont {Deng}\ \emph {et~al.}(2019)\citenamefont {Deng},
  \citenamefont {Stadler}, \citenamefont {Haule}, \citenamefont {Weichselbaum},
  \citenamefont {von Delft},\ and\ \citenamefont {Kotliar}}]{Deng2019}%
  \BibitemOpen
  \bibfield  {author} {\bibinfo {author} {\bibfnamefont {X.}~\bibnamefont
  {Deng}}, \bibinfo {author} {\bibfnamefont {K.~M.}\ \bibnamefont {Stadler}},
  \bibinfo {author} {\bibfnamefont {K.}~\bibnamefont {Haule}}, \bibinfo
  {author} {\bibfnamefont {A.}~\bibnamefont {Weichselbaum}}, \bibinfo {author}
  {\bibfnamefont {J.}~\bibnamefont {von Delft}},\ and\ \bibinfo {author}
  {\bibfnamefont {G.}~\bibnamefont {Kotliar}},\ }\href
  {https://doi.org/10.1038/s41467-019-10257-2} {\bibfield  {journal} {\bibinfo
  {journal} {Nat. Commun.}\ }\textbf {\bibinfo {volume} {10}},\ \bibinfo
  {pages} {2721} (\bibinfo {year} {2019})}\BibitemShut {NoStop}%
\bibitem [{\citenamefont {Bergemann}\ \emph {et~al.}(2000)\citenamefont
  {Bergemann}, \citenamefont {Julian}, \citenamefont {Mackenzie}, \citenamefont
  {NishiZaki},\ and\ \citenamefont {Maeno}}]{Bergemann2000}%
  \BibitemOpen
  \bibfield  {author} {\bibinfo {author} {\bibfnamefont {C.}~\bibnamefont
  {Bergemann}}, \bibinfo {author} {\bibfnamefont {S.~R.}\ \bibnamefont
  {Julian}}, \bibinfo {author} {\bibfnamefont {A.~P.}\ \bibnamefont
  {Mackenzie}}, \bibinfo {author} {\bibfnamefont {S.}~\bibnamefont
  {NishiZaki}},\ and\ \bibinfo {author} {\bibfnamefont {Y.}~\bibnamefont
  {Maeno}},\ }\href {https://doi.org/10.1103/PhysRevLett.84.2662} {\bibfield
  {journal} {\bibinfo  {journal} {Phys. Rev. Lett.}\ }\textbf {\bibinfo
  {volume} {84}},\ \bibinfo {pages} {2662} (\bibinfo {year}
  {2000})}\BibitemShut {NoStop}%
\bibitem [{\citenamefont {Kim}\ and\ \citenamefont
  {Khaliullin}(2017)}]{Kim2017}%
  \BibitemOpen
  \bibfield  {author} {\bibinfo {author} {\bibfnamefont {B.~J.}\ \bibnamefont
  {Kim}}\ and\ \bibinfo {author} {\bibfnamefont {G.}~\bibnamefont
  {Khaliullin}},\ }\href {https://doi.org/10.1103/PhysRevB.96.085108}
  {\bibfield  {journal} {\bibinfo  {journal} {Phys. Rev. B}\ }\textbf {\bibinfo
  {volume} {96}},\ \bibinfo {pages} {085108} (\bibinfo {year}
  {2017})}\BibitemShut {NoStop}%
\bibitem [{\citenamefont {Suzuki}\ \emph {et~al.}(2019)\citenamefont {Suzuki},
  \citenamefont {Gretarsson}, \citenamefont {Ishikawa}, \citenamefont {Ueda},
  \citenamefont {Yang}, \citenamefont {Liu}, \citenamefont {Kim}, \citenamefont
  {Kukusta}, \citenamefont {Yaresko}, \citenamefont {Minola}, \citenamefont
  {Sears}, \citenamefont {Francoual}, \citenamefont {Wille}, \citenamefont
  {Nuss}, \citenamefont {Takagi}, \citenamefont {Kim}, \citenamefont
  {Khaliullin}, \citenamefont {Yava{\c{s}}},\ and\ \citenamefont
  {Keimer}}]{Suzuki2019}%
  \BibitemOpen
  \bibfield  {author} {\bibinfo {author} {\bibfnamefont {H.}~\bibnamefont
  {Suzuki}}, \bibinfo {author} {\bibfnamefont {H.}~\bibnamefont {Gretarsson}},
  \bibinfo {author} {\bibfnamefont {H.}~\bibnamefont {Ishikawa}}, \bibinfo
  {author} {\bibfnamefont {K.}~\bibnamefont {Ueda}}, \bibinfo {author}
  {\bibfnamefont {Z.}~\bibnamefont {Yang}}, \bibinfo {author} {\bibfnamefont
  {H.}~\bibnamefont {Liu}}, \bibinfo {author} {\bibfnamefont {H.}~\bibnamefont
  {Kim}}, \bibinfo {author} {\bibfnamefont {D.}~\bibnamefont {Kukusta}},
  \bibinfo {author} {\bibfnamefont {A.}~\bibnamefont {Yaresko}}, \bibinfo
  {author} {\bibfnamefont {M.}~\bibnamefont {Minola}}, \bibinfo {author}
  {\bibfnamefont {J.~A.}\ \bibnamefont {Sears}}, \bibinfo {author}
  {\bibfnamefont {S.}~\bibnamefont {Francoual}}, \bibinfo {author}
  {\bibfnamefont {H.-C.}\ \bibnamefont {Wille}}, \bibinfo {author}
  {\bibfnamefont {J.}~\bibnamefont {Nuss}}, \bibinfo {author} {\bibfnamefont
  {H.}~\bibnamefont {Takagi}}, \bibinfo {author} {\bibfnamefont {B.~J.}\
  \bibnamefont {Kim}}, \bibinfo {author} {\bibfnamefont {G.}~\bibnamefont
  {Khaliullin}}, \bibinfo {author} {\bibfnamefont {H.}~\bibnamefont
  {Yava{\c{s}}}},\ and\ \bibinfo {author} {\bibfnamefont {B.}~\bibnamefont
  {Keimer}},\ }\href {https://doi.org/10.1038/s41563-019-0327-2} {\bibfield
  {journal} {\bibinfo  {journal} {Nat. Mater.}\ }\textbf {\bibinfo {volume}
  {18}},\ \bibinfo {pages} {563} (\bibinfo {year} {2019})}\BibitemShut
  {NoStop}%
\bibitem [{\citenamefont {Braden}\ \emph
  {et~al.}(2002{\natexlab{a}})\citenamefont {Braden}, \citenamefont {Sidis},
  \citenamefont {Bourges}, \citenamefont {Pfeuty}, \citenamefont {Kulda},
  \citenamefont {Mao},\ and\ \citenamefont {Maeno}}]{Braden2002a}%
  \BibitemOpen
  \bibfield  {author} {\bibinfo {author} {\bibfnamefont {M.}~\bibnamefont
  {Braden}}, \bibinfo {author} {\bibfnamefont {Y.}~\bibnamefont {Sidis}},
  \bibinfo {author} {\bibfnamefont {P.}~\bibnamefont {Bourges}}, \bibinfo
  {author} {\bibfnamefont {P.}~\bibnamefont {Pfeuty}}, \bibinfo {author}
  {\bibfnamefont {J.}~\bibnamefont {Kulda}}, \bibinfo {author} {\bibfnamefont
  {Z.}~\bibnamefont {Mao}},\ and\ \bibinfo {author} {\bibfnamefont
  {Y.}~\bibnamefont {Maeno}},\ }\href
  {https://doi.org/10.1103/PhysRevB.66.064522} {\bibfield  {journal} {\bibinfo
  {journal} {Phys. Rev. B}\ }\textbf {\bibinfo {volume} {66}},\ \bibinfo
  {pages} {064522} (\bibinfo {year} {2002}{\natexlab{a}})}\BibitemShut
  {NoStop}%
\bibitem [{\citenamefont {Braden}\ \emph
  {et~al.}(2002{\natexlab{b}})\citenamefont {Braden}, \citenamefont {Friedt},
  \citenamefont {Sidis}, \citenamefont {Bourges}, \citenamefont {Minakata},\
  and\ \citenamefont {Maeno}}]{Braden2002b}%
  \BibitemOpen
  \bibfield  {author} {\bibinfo {author} {\bibfnamefont {M.}~\bibnamefont
  {Braden}}, \bibinfo {author} {\bibfnamefont {O.}~\bibnamefont {Friedt}},
  \bibinfo {author} {\bibfnamefont {Y.}~\bibnamefont {Sidis}}, \bibinfo
  {author} {\bibfnamefont {P.}~\bibnamefont {Bourges}}, \bibinfo {author}
  {\bibfnamefont {M.}~\bibnamefont {Minakata}},\ and\ \bibinfo {author}
  {\bibfnamefont {Y.}~\bibnamefont {Maeno}},\ }\href
  {https://doi.org/10.1103/PhysRevLett.88.197002} {\bibfield  {journal}
  {\bibinfo  {journal} {Phys. Rev. Lett.}\ }\textbf {\bibinfo {volume} {88}},\
  \bibinfo {pages} {197002} (\bibinfo {year} {2002}{\natexlab{b}})}\BibitemShut
  {NoStop}%
\bibitem [{\citenamefont {Ng}\ and\ \citenamefont {Sigrist}(2000)}]{Ng2000}%
  \BibitemOpen
  \bibfield  {author} {\bibinfo {author} {\bibfnamefont {K.~K.}\ \bibnamefont
  {Ng}}\ and\ \bibinfo {author} {\bibfnamefont {M.}~\bibnamefont {Sigrist}},\
  }\href {https://doi.org/10.1209/epl/i2000-00173-x} {\bibfield  {journal}
  {\bibinfo  {journal} {EPL}\ }\textbf {\bibinfo {volume} {49}},\ \bibinfo
  {pages} {473} (\bibinfo {year} {2000})}\BibitemShut {NoStop}%
\bibitem [{\citenamefont {Scaffidi}\ \emph {et~al.}(2014)\citenamefont
  {Scaffidi}, \citenamefont {Romers},\ and\ \citenamefont
  {Simon}}]{Scaffidi2014}%
  \BibitemOpen
  \bibfield  {author} {\bibinfo {author} {\bibfnamefont {T.}~\bibnamefont
  {Scaffidi}}, \bibinfo {author} {\bibfnamefont {J.~C.}\ \bibnamefont
  {Romers}},\ and\ \bibinfo {author} {\bibfnamefont {S.~H.}\ \bibnamefont
  {Simon}},\ }\href {https://doi.org/10.1103/PhysRevB.89.220510} {\bibfield
  {journal} {\bibinfo  {journal} {Phys. Rev. B}\ }\textbf {\bibinfo {volume}
  {89}},\ \bibinfo {pages} {220510(R)} (\bibinfo {year} {2014})}\BibitemShut
  {NoStop}%
\bibitem [{\citenamefont {Klett}\ \emph {et~al.}(2020)\citenamefont {Klett},
  \citenamefont {Wentzell}, \citenamefont {Sch\"afer}, \citenamefont
  {Simkovic}, \citenamefont {Parcollet}, \citenamefont {Andergassen},\ and\
  \citenamefont {Hansmann}}]{Klett2020}%
  \BibitemOpen
  \bibfield  {author} {\bibinfo {author} {\bibfnamefont {M.}~\bibnamefont
  {Klett}}, \bibinfo {author} {\bibfnamefont {N.}~\bibnamefont {Wentzell}},
  \bibinfo {author} {\bibfnamefont {T.}~\bibnamefont {Sch\"afer}}, \bibinfo
  {author} {\bibfnamefont {F.}~\bibnamefont {Simkovic}}, \bibinfo {author}
  {\bibfnamefont {O.}~\bibnamefont {Parcollet}}, \bibinfo {author}
  {\bibfnamefont {S.}~\bibnamefont {Andergassen}},\ and\ \bibinfo {author}
  {\bibfnamefont {P.}~\bibnamefont {Hansmann}},\ }\href
  {https://doi.org/10.1103/PhysRevResearch.2.033476} {\bibfield  {journal}
  {\bibinfo  {journal} {Phys. Rev. Research}\ }\textbf {\bibinfo {volume}
  {2}},\ \bibinfo {pages} {033476} (\bibinfo {year} {2020})}\BibitemShut
  {NoStop}%
\bibitem [{\citenamefont {Sch\"afer}\ \emph {et~al.}(2021)\citenamefont
  {Sch\"afer}, \citenamefont {Wentzell}, \citenamefont {\ifmmode~\check{S}\else
  \v{S}\fi{}imkovic}, \citenamefont {He}, \citenamefont {Hille}, \citenamefont
  {Klett}, \citenamefont {Eckhardt}, \citenamefont {Arzhang}, \citenamefont
  {Harkov}, \citenamefont {Le~R\'egent}, \citenamefont {Kirsch}, \citenamefont
  {Wang}, \citenamefont {Kim}, \citenamefont {Kozik}, \citenamefont {Stepanov},
  \citenamefont {Kauch}, \citenamefont {Andergassen}, \citenamefont {Hansmann},
  \citenamefont {Rohe}, \citenamefont {Vilk}, \citenamefont {LeBlanc},
  \citenamefont {Zhang}, \citenamefont {Tremblay}, \citenamefont {Ferrero},
  \citenamefont {Parcollet},\ and\ \citenamefont {Georges}}]{Schaefer2021}%
  \BibitemOpen
  \bibfield  {author} {\bibinfo {author} {\bibfnamefont {T.}~\bibnamefont
  {Sch\"afer}}, \bibinfo {author} {\bibfnamefont {N.}~\bibnamefont {Wentzell}},
  \bibinfo {author} {\bibfnamefont {F.}~\bibnamefont {\ifmmode~\check{S}\else
  \v{S}\fi{}imkovic}}, \bibinfo {author} {\bibfnamefont {Y.-Y.}\ \bibnamefont
  {He}}, \bibinfo {author} {\bibfnamefont {C.}~\bibnamefont {Hille}}, \bibinfo
  {author} {\bibfnamefont {M.}~\bibnamefont {Klett}}, \bibinfo {author}
  {\bibfnamefont {C.~J.}\ \bibnamefont {Eckhardt}}, \bibinfo {author}
  {\bibfnamefont {B.}~\bibnamefont {Arzhang}}, \bibinfo {author} {\bibfnamefont
  {V.}~\bibnamefont {Harkov}}, \bibinfo {author} {\bibfnamefont {F.~m. c.-M.}\
  \bibnamefont {Le~R\'egent}}, \bibinfo {author} {\bibfnamefont
  {A.}~\bibnamefont {Kirsch}}, \bibinfo {author} {\bibfnamefont
  {Y.}~\bibnamefont {Wang}}, \bibinfo {author} {\bibfnamefont {A.~J.}\
  \bibnamefont {Kim}}, \bibinfo {author} {\bibfnamefont {E.}~\bibnamefont
  {Kozik}}, \bibinfo {author} {\bibfnamefont {E.~A.}\ \bibnamefont {Stepanov}},
  \bibinfo {author} {\bibfnamefont {A.}~\bibnamefont {Kauch}}, \bibinfo
  {author} {\bibfnamefont {S.}~\bibnamefont {Andergassen}}, \bibinfo {author}
  {\bibfnamefont {P.}~\bibnamefont {Hansmann}}, \bibinfo {author}
  {\bibfnamefont {D.}~\bibnamefont {Rohe}}, \bibinfo {author} {\bibfnamefont
  {Y.~M.}\ \bibnamefont {Vilk}}, \bibinfo {author} {\bibfnamefont {J.~P.~F.}\
  \bibnamefont {LeBlanc}}, \bibinfo {author} {\bibfnamefont {S.}~\bibnamefont
  {Zhang}}, \bibinfo {author} {\bibfnamefont {A.-M.~S.}\ \bibnamefont
  {Tremblay}}, \bibinfo {author} {\bibfnamefont {M.}~\bibnamefont {Ferrero}},
  \bibinfo {author} {\bibfnamefont {O.}~\bibnamefont {Parcollet}},\ and\
  \bibinfo {author} {\bibfnamefont {A.}~\bibnamefont {Georges}},\ }\href
  {https://doi.org/10.1103/PhysRevX.11.011058} {\bibfield  {journal} {\bibinfo
  {journal} {Phys. Rev. X}\ }\textbf {\bibinfo {volume} {11}},\ \bibinfo
  {pages} {011058} (\bibinfo {year} {2021})}\BibitemShut {NoStop}%
\bibitem [{\citenamefont {Wietek}\ \emph {et~al.}()\citenamefont {Wietek},
  \citenamefont {Rossi}, \citenamefont {Šimkovic IV}, \citenamefont {Klett},
  \citenamefont {Hansmann}, \citenamefont {Ferrero}, \citenamefont
  {Stoudenmire}, \citenamefont {Schäfer},\ and\ \citenamefont
  {Georges}}]{Wietek2021}%
  \BibitemOpen
  \bibfield  {author} {\bibinfo {author} {\bibfnamefont {A.}~\bibnamefont
  {Wietek}}, \bibinfo {author} {\bibfnamefont {R.}~\bibnamefont {Rossi}},
  \bibinfo {author} {\bibfnamefont {F.}~\bibnamefont {Šimkovic IV}}, \bibinfo
  {author} {\bibfnamefont {M.}~\bibnamefont {Klett}}, \bibinfo {author}
  {\bibfnamefont {P.}~\bibnamefont {Hansmann}}, \bibinfo {author}
  {\bibfnamefont {M.}~\bibnamefont {Ferrero}}, \bibinfo {author} {\bibfnamefont
  {E.~M.}\ \bibnamefont {Stoudenmire}}, \bibinfo {author} {\bibfnamefont
  {T.}~\bibnamefont {Schäfer}},\ and\ \bibinfo {author} {\bibfnamefont
  {A.}~\bibnamefont {Georges}},\ }\href@noop {} {}\Eprint
  {https://arxiv.org/abs/2102.12904} {arXiv:2102.12904} \BibitemShut {NoStop}%
\bibitem [{\citenamefont {Hubig}\ \emph {et~al.}(2015)\citenamefont {Hubig},
  \citenamefont {McCulloch}, \citenamefont {Schollw\"ock},\ and\ \citenamefont
  {Wolf}}]{Hubig2015}%
  \BibitemOpen
  \bibfield  {author} {\bibinfo {author} {\bibfnamefont {C.}~\bibnamefont
  {Hubig}}, \bibinfo {author} {\bibfnamefont {I.~P.}\ \bibnamefont
  {McCulloch}}, \bibinfo {author} {\bibfnamefont {U.}~\bibnamefont
  {Schollw\"ock}},\ and\ \bibinfo {author} {\bibfnamefont {F.~A.}\ \bibnamefont
  {Wolf}},\ }\href {https://doi.org/10.1103/PhysRevB.91.155115} {\bibfield
  {journal} {\bibinfo  {journal} {Phys. Rev. B}\ }\textbf {\bibinfo {volume}
  {91}},\ \bibinfo {pages} {155115} (\bibinfo {year} {2015})}\BibitemShut
  {NoStop}%
\end{thebibliography}%

\end{document}